\documentstyle [12pt,epsf,rotate] {article}

\newcommand{\nwc}{\newcommand}
\nwc{\cl}  {$\clubsuit$}

\nwc{\hyp} {\hyphenation} 
\nwc{\be}  {\begin{equation}}
\nwc{\ee}  {\end{equation}}
\nwc{\ba}  {\begin{array}}
\nwc{\ea}  {\end{array}}
\nwc{\bdm} {\begin{displaymath}}
\nwc{\edm} {\end{displaymath}}
\nwc{\bea} {\be\ba{rcl}}
\nwc{\eea} {\ea\ee}
\nwc{\ben} {\begin{eqnarray}}
\nwc{\een} {\end{eqnarray}}
\nwc{\bda} {\bdm\ba{lcl}}
\nwc{\eda} {\ea\edm}
\nwc{\bc}  {\begin{center}}
\nwc{\ec}  {\end{center}}
\nwc{\ds}  {\displaystyle}
\nwc{\bmat}{\left(\ba}
\nwc{\emat}{\ea\right)}
\nwc{\non} {\nonumber}
\nwc{\bib} {\bibitem}
\nwc{\lra} {\longrightarrow}
\nwc{\Llra}{\Longleftrightarrow}
\nwc{\ra}  {\rightarrow}
\nwc{\Ra}  {\Rightarrow}
\nwc{\lmt} {\longmapsto}
\nwc{\prl} {\partial}
\nwc{\iy}  {\infty}
\nwc{\ol}  {\overline}
\nwc{\hm}  {\hspace{3mm}}
\nwc{\lf}  {\left}
\nwc{\ri}  {\right}
\nwc{\lm}  {\limits}
\nwc{\lb}  {\lbrack}
\nwc{\rb}  {\rbrack}
\nwc{\ov}  {\over}
\nwc{\pri}  {\prime}
\nwc{\nnn} {\nonumber \vspace{.2cm} \\ }
\nwc{\Sc}  {{\cal S}}
\nwc{\Lc}  {{\cal L}}
\nwc{\Rc}  {{\cal R}}
\nwc{\Dc}  {{\cal D}}
\nwc{\Oc}  {{\cal O}}
\nwc{\Cc}  {{\cal C}}
\nwc{\Pc}  {{\cal P}}
\nwc{\Mc}  {{\cal M}}
\nwc{\Ec}  {{\cal E}}
\nwc{\Fc}  {{\cal F}}
\nwc{\Hc}  {{\cal H}}
\nwc{\Kc}  {{\cal K}}
\nwc{\Xc}  {{\cal X}}
\nwc{\Gc}  {{\cal G}}
\nwc{\Zc}  {{\cal Z}}
\nwc{\Nc}  {{\cal N}}
\nwc{\fca} {{\cal f}}
\nwc{\xc}  {{\cal x}}
\nwc{\Ac}  {{\cal A}}
\nwc{\Bc}  {{\cal B}}
\nwc{\Uc}  {{\cal U}}
\nwc{\Vc}  {{\cal V}}
\nwc{\Th} {\Theta}
\nwc{\th} {\theta}
\nwc{\vth} {\vartheta}
\nwc{\eps}{\epsilon}
\nwc{\si} {\sigma}
\nwc{\Gm} {\Gamma}
\nwc{\gm} {\gamma}
\nwc{\bt} {\beta}
\nwc{\La} {\Lambda}
\nwc{\la} {\lambda}
\nwc{\om} {\omega}
\nwc{\Om} {\Omega}
\nwc{\dt} {\delta}
\nwc{\Si} {\Sigma}
\nwc{\Dt} {\Delta}
\nwc{\al} {\alpha}
\nwc{\vp} {\varphi}
\nwc{\kp} {\kappa}
\def\tr{\mathop{\rm tr}}
\def\Tr{\mathop{\rm Tr}}

\def\pr#1{#1^\prime}
\def\ltap{\raisebox{-.4ex}{\rlap{$\sim$}} \raisebox{.4ex}{$<$}}
\def\gtap{\raisebox{-.4ex}{\rlap{$\sim$}} \raisebox{.4ex}{$>$}}
\nwc{\Id}  {{\bf 1}}
\nwc{\diag} {{\rm diag}}
\nwc{\inv}  {{\rm inv}}
\nwc{\mod}  {{\rm mod}}
\nwc{\hal} {\frac{1}{2}}
\nwc{\tpi}  {2\pi i}

\def\ijmpc#1{Int.\ J.\ Mod.\ Phys.\ {\bf C#1}}
\def\jpa#1{J.\ Phys.\ {\bf A#1}}

\def\npb#1{Nucl.\ Phys.\ {\bf B#1}}

\def\plb#1{Phys.\ Lett.\ {\bf B#1}}
\def\pr#1{Phys.\ Rev.\ {\bf #1}}
\def\pra#1{Phys.\ Rev.\ {\bf A#1}}
\def\prb#1{Phys.\ Rev.\ {\bf B#1}}

\def\prd#1{Phys.\ Rev.\ {\bf D#1}}
\def\prle#1{Phys.\ Rev.\ Lett.\ {\bf #1}}

\def\zpb#1{Z.\ Phys.\ {\bf B#1}}
\def\zpc#1{Z.\ Phys.\ {\bf C#1}}

\newsavebox{\nnin} \sbox{\nnin}{$\hspace{1mm}\in\kern -.8em /
                   \hspace{1mm}$}

\newcommand{\sub}{\subset}
\newsavebox{\nnsub} \sbox{\nnsub}{$\hspace{1mm}\sub\kern -.9em /
            \hspace{1mm}$}

\def\KK{{\rm I\kern -.2em  K}}
\def\NN{{\rm I\kern -.16em N}}
\def\RR{{\rm I\kern -.2em  R}}
\def\ZZ{Z \kern -.43em Z}
\def\QQ{{\rm \kern .25em
             \vrule height1.4ex depth-.12ex width.06em\kern-.31em Q}}
\def\CC{{\rm \kern .25em
             \vrule height1.4ex depth-.12ex width.06em\kern-.31em C}}
\def\ZZZ{Z\kern -0.31em Z}

\nwc{\olnu}  {\ol{\nu}}
\nwc{\olla}  {\ol{\la}}
\nwc{\olm}   {\ol{m}}
\nwc{\olmu}  {\ol{\mu}}
\nwc{\olh}   {\ol{h}}
\nwc{\olpsi} {\ol{\psi}}
\nwc{\olsi}  {\ol{\sigma}}
\nwc{\olgm}  {\ol{\gm}}
\nwc{\prlt}  {\frac{\prl}{\prl t}}
\nwc{\ttau}  {\tilde{\tau}}
\nwc{\trho}  {\tilde{\rho}}
\nwc{\tP}    {\tilde{P}}
\nwc{\tU}    {\tilde{U}}
\nwc{\teps}  {\tilde{\eps}}
\nwc{\tla}   {\tilde{\la}}
\nwc{\tit}    {\tilde{t}}
\nwc{\iddq}  {\int\frac{d^dq}{(2\pi)^d}}
\nwc{\prpr}  {\prime\prime}
\nwc{\rN}    {\left(\frac{\rho}{N}\right)}
\nwc{\rNt}    {\left(\frac{\rho}{N}\right)^{\frac{N-2}{2}}}
\nwc{\rnN}   {\left(\frac{\rho_0}{N}\right)}
\nwc{\rnNt}    {\left(\frac{\rho_0}{N}\right)^{\frac{N-2}{2}}}
\nwc{\rnNf}    {\left(\frac{\rho_0}{N}\right)^{\frac{N-4}{2}}}
\nwc{\rNs}    {\left(\frac{\rho_0}{N}\right)^{\frac{N-6}{2}}}
\nwc{\kNt}    {\left(\frac{\kappa}{N}\right)^{\frac{N-2}{2}}}
\nwc{\kNf}    {\left(\frac{\kappa}{N}\right)^{\frac{N-4}{2}}}
\nwc{\kNs}    {\left(\frac{\kappa}{N}\right)^{\frac{N-6}{2}}}

\newcounter{app}
\def\app{\par
 \addtocounter{app}{1}
 \def\thesection{\Alph{app}}
 \def\ksection{\Alph{app}}}

\def\appendix#1{\app\sect{#1}}

\newcommand{\sect}[1]{ \section{#1} \setcounter{equation}{0} }

\begin{document}

\begin{titlepage}

\title{ Equation of state and \\ 
coarse grained free
energy \\
for matrix models }

\author{{\sc J.\ Berges\thanks{Email:
J.Berges@thphys.uni-heidelberg.de}} \\
 \\ and \\ \\
{\sc C.\ Wetterich\thanks{Email: C.Wetterich@thphys.uni-heidelberg.de}}
\\ \\ \\
{\em Institut f\"ur Theoretische Physik} \\
{\em Universit\"at Heidelberg} \\
{\em Philosophenweg 16} \\
{\em 69120 Heidelberg, Germany}}

\date{}
\maketitle

\begin{picture}(5,2.5)(-350,-450)
\put(12,-115){HD--THEP--96--37}
\end{picture}

\thispagestyle{empty}

\begin{abstract}

We investigate phase transitions in three dimensional
scalar matrix models, with special emphasis on complex
$2 \times 2$ matrices. The universal equation of state 
for weak first order phase transitions is computed.
We also study the coarse grained free energy. Its
dependence on the coarse graining scale gives a 
quantitative criterion for the validity of the standard
treatment of bubble nucleation. 

\end{abstract}

\end{titlepage}

\sect{Introduction \label{intro}}

Matrix models are extensively discussed in statistical
physics. Beyond the $O(N)$ symmetric
Heisenberg models ('vector models') they correspond
to the simplest scalar field theories. 
There is a wide set of different applications as 
the metal insulator transition \cite{Weg}
or liquid crystals \cite{deG} or strings and
random surfaces \cite{FGZ} .\ .\ .  
The universal behavior of these models in the vicinity
of a second order or weak first order phase transition
is determined by the symmetries and the field content
of the corresponding field theories. We will consider
here models with $U(N) \times U(N)$ symmetry with
a scalar field in the $(\bar{N},N)$ representation,
described by an arbitrary complex $N \times N$ matrix
$\vp$. We do not impose nonlinear constraints for
$\vp$ a priori  but rather use a 'classical'
potential.  This enforces nonlinear constraints
in certain limiting cases. Among those, our model
describes a nonlinear matrix model for unitary matrices
or one for singular $2 \times 2$ matrices. The universal
critical behavior does not depend on the details of the 
classical potential and there is no difference
between the linear and nonlinear models in the 
vicinity of  the limiting cases. We concentrate in
this paper on three dimensions, relevant for statistical
physics and critical phenomena in high temperature field 
theory.

The cases $N=2$, $3$ have a relation to high
temperature strong interaction physics. At vanishing
temperature the four dimensional models can be used
for a description of the pseudoscalar and scalar mesons
for $N$ quark flavors. For $N=3$ the antihermitean
part of $\vp$ describes here the pions, kaons, $\eta$
and $\eta^{\pri}$ whereas the hermitean part accounts for the 
nonet of scalar $0^{++}$ mesons.\footnote{See \cite{JuWet96}
for a recent phenomenological analysis.} For nonzero
temperature $T$ the effects of fluctuations with momenta
$p^2 \, \ltap \, (2\pi T)^2$ are described by the 
corresponding three dimensional models. These models
account for the long distance physics and obtain after
integrating out the short distance fluctuations by
virtue of dimensional reduction \cite{DR,TetWet,DR3}.
In particular, the three dimensional models embody
the essential dynamics in the immediate vicinity
of a second order or weak first order chiral phase transition
\cite{PRW}. The four dimensional models at nonvanishing
temperature have also been used for investigations of
the temperature dependence of meson masses \cite{mes2,mes}.
The simple model investigated in this paper is not yet 
realistic -- it neglects the effect of the axial anomaly
which reduces the chiral flavor symmetry to 
$SU(N) \times SU(N)$. It also accounts only for meson
fluctuations and ignores, for example, the temperature 
dependence of the binding mechanism which gives rise
to mesons as quark-antiquark bound states. Nevertheless,
it should serve as an interesting starting point
for a later analysis of more realistic effective three
dimensional meson models arising in high temperature 
QCD. For simplicity we will concentrate here on
$N=2$, but our methods can be generalized to $N=3$
and the inclusion of the axial anomaly.

The case $N=2$ also has a relation to the 
electroweak phase transition in models with two
Higgs doublets. Our model corresponds here
to the critical behavior in a special class
of left-right symmetric theories in the limit where 
the gauge couplings are neglected. Even though
vanishing gauge couplings are not a good 
approximation for typical realistic models
one would like to understand this limiting
case reliably.

We are mainly interested in the character and
the detailed physics of the phase transition.
Standard (linear) high temperature perturbation
theory can give a reliable description only
if the relevant dimensionless couplings remain small near 
the phase transition. This requires two conditions: First,
the effective three dimensional couplings evaluated for
typical momenta $p^2 \simeq (2 \pi T)^2$ must be
small in units of temperature. For the example of a
quartic coupling this means $\bar{\la} (2 \pi T) /T
\ll 1$. In high temperature field theory this ratio
corresponds to the four dimensional coupling,
e.g.\ $\bar{\la} (2 \pi T) = \la_4 T$. The above 
condition is fulfilled whenever the four dimensional
(or zero temperature) couplings remain perturbative.
On the other hand, it never holds for nonlinear matrix
models where the corresponding ratios diverge. The 
second condition requires the phase transition
to be sufficiently strong first order. If we denote by
$m_{c}$ the relevant mass at the critical temperature,
standard perturbation theory turns out to be actually
an expansion in $\bar{\la}/m_{c}$ rather than
$\bar{\la}/T$. For a given value of  $\bar{\la}$ a
perturbative expansion can therefore only converge
if $m_{c}$ is sufficiently large, $m_{c} \gg \la_4 T$.
In consequence, perturbation theory is not applicable
for the interesting cases of second order 
$(m_{c}=0)$ or weak first order $(m_{c}\ll \la_4 T)$
phase transitions.

This property can also be understood in terms of 
running couplings. One may consider the model
in presence of an infrared cutoff with typical scale
$k$. Here $k$ may be given by external conditions 
(as a finite volume) or induced by the problem studied
(e.g.\ typical external momenta in vertices or a 
characteristic scale of critical bubbles in problems
of bubble formation). One may also associate $k$ with
a relevant mass away from the critical temperature 
or for nonvanishing sources or external fields.
Or else the infrared cutoff can be introduced as a
purely technical tool. One can define running
renormalized couplings $\bar{\la}_R(k)$ by appropriate
renormalized n-point functions in presence of the
infrared cutoff and study the flow of 
$\bar{\la}_R(k)$ as $k$ is lowered. Often an originally
small dimensionless ratio $\la(k)=\bar{\la}_R(k)/k$
increases in the course of the evolution. Perturbation
theory breaks down when $\la(k)$ reaches values of order
one. One needs nonperturbative methods to follow
the flow for larger values of $\la(k)$ if $k$ becomes small.
This is typically what happens for a second order phase 
transition where one needs at the critical temperature
the behavior for $k \to 0$. In this case the 
dimensionless coupling $\la(k)$ reaches a fixed
point $\la_{\star}$ which is for most simple models
substantially larger than one. For the present matrix models
one wants to know if the phase transition becomes second
order in certain regions of parameter space. This is equivalent 
to the question if the system of running couplings admits
a fixed point which is infrared stable (except one
relevant direction corresponding to $T-T_{c}$).
Clearly, this is a nonperturbative question. It
can be addressed by an expansion in $\eps=4-d$
\cite{WilFis72,ZJ}. However, the reliability of this expansion
is not guaranteed a priori since for three dimensions
$\eps=1$ is not a small parameter. We will employ here
a method which is based on the general concepts of
the Wilsonian renormalization group \cite{Ka,RG}.
More precisely, we study an approximate solution 
to an exact flow equation for a coarse grained free energy.
We find that the phase transition for the investigated
matrix models with $N=2$ and symmetry breaking
pattern $U(2) \times U(2) \to U(2)$ is always 
(fluctuation induced) first order,
except for a boundary case with enhanced 
$O(8)$ symmetry. For a large part
of parameter space the transition is weak and one finds 
large renormalized dimensionless couplings near the critical
temperature. If the running of the couplings towards approximate
fixed points (there are no exact fixed points) is 
sufficiently fast the large distance physics looses 
memory of the details of the short distance or classical
action. In this case the physics near the phase transition
is described by an universal equation of state.

Besides the possible practical applications it is
a theoretical challenge to find the universal equation of 
state for weak first order phase transitions. For
the second order phase transition in $O(N)$ symmetric
vector models (Heisenberg models) this has only 
recently been achieved \cite{BTW95} by the method employed here
and confirmed by numerical lattice simulations \cite{Tsy}.
Whereas for second order transitions the universal equation 
of state can be expressed as a function of only one scaling 
variable, the additional scale $m_{c}$ in a first order 
transition induces a second dimensionless ratio. 
The equation of state or, equivalently,
the effective potential or free energy therefore
depends on two different scaling variables in the
universal region. We have succeeded to solve this
problem for the present matrix models and present
the universal equation of state in sect.\ \ref{sce}.

Another old and challenging problem for first order phase 
transitions concerns the question of the validity of Langer's 
theory of bubble formation \cite{Langer}. This requires first
a meaningful definition of a coarse grained free energy
with a coarse graining scale $k$ and second the 
validity of a saddle point approximation for the 
treatment of fluctuations around the critical bubble.
Here only fluctuations with momenta smaller than $k$
must be included. We will see that the two issues
are closely related. The validity of the saddle point 
approximation typically requires small dimensionless
couplings $\la(k)$. On the other hand we observe
for large $\la(k)$ that the form of the coarse grained
effective potential $U_k$ depends strongly on $k$
even for scales $k$ where the location of the minima
of $U_k$
is essentially independent of $k$. This means that
the lowest order in the saddle point approximation
(classical contribution) depends strongly on the details of
the coarse graining procedure. Since the final results
as tunneling rates etc.\ must be independent of the
coarse graining prescription this is only compatible
with a large contribution from the higher orders of the
saddle point expansion. Section \ref{coarse} deals with
this issue in a quantitative way.

Our paper is organized as follows. The nonperturbative
method we employ relies on an exact flow equation for
the effective average action $\Gm_k$ which is introduced
in section \ref{averact}. In section \ref{model}
we define the $U(2) \times U(2)$ symmetric matrix model
and we establish the connection to a matrix model
for unitary matrices and to one for
singular complex $2 \times 2$ matrices. There we also give
an interpretation of the model as the coupled system
of two $SU(2)$-doublets for the weak interaction Higgs sector.
Section \ref{ps} is devoted to an overview over the phase 
structure and the coarse grained effective potential $U_k$
for the three dimensional theory. The evolution equation
for $U_k$ and its scaling form is computed in section 
\ref{scale}. A method for its numerical solution is discussed 
in section \ref{solv}
which also contains the flow equation for the wave 
function renormalization constant or the anomalous dimension
$\eta$. A detailed account on the renormalization group flow
is presented in section \ref{rg}. We compute the 
universal form of the equation of state 
for weak first order phase transitions 
in section \ref{sce} and we extract critical exponents
and the corresponding index relations.
The dependence of the coarse grained effective potential
on the coarse graining scale is studied in detail
in section \ref{coarse}. The quantitative analysis
is applied to Langer's theory of bubble formation.
Section \ref{con} contains the conclusions and an outlook.

\sect{Effective average action \label{averact}}

The effective average action $\Gm_k$ \cite{Wet91-1,Wet93-2} 
is a coarse
grained free energy with an infrared cutoff 
or a coarse graining scale $\sim k$. More precisely,
for a theory described by a classical action $S$, 
the effective average action results from the 
integration of degrees of freedom with characteristic momenta larger
than $k$.
To be explicit 
we consider a $k$-dependent
generating functional with $N^2$ complex scalar fields 
$\chi_{ab}$ $(a,b=1 \ldots N)$
\be
W_k[j]=\ln \int D \chi \exp\left(-S[\chi]-\Dt_k S[\chi]+
\hal \int d^dx \left(j^*_{ab}(x)\chi^{ab}(x)+
\chi^*_{ab}(x) j^{ab}(x)\right)\right)
\ee
with $S$ the classical action and $j_{ab}$ arbitrary sources.
The scale dependence arises from the 
introduction of an additional infrared cutoff term
\be
\Dt_k S[\chi]=\int 
\frac{d^dq}{(2\pi)^d} R_k(q)\chi^*_{ab}(q)\chi^{ab}(q).
\ee
Without this term $W_k$ becomes the usual generating functional
for the connected Green functions. Here the 
infrared cutoff function $R_k$ is required to vanish
for $k \to 0$ and to diverge for $k \to \infty$
and fixed $q^2$.
This can be achieved, for example, by the choice
\be
 R_k(q)=\frac{Z_k q^2 e^{-q^2/k^2}}{1- e^{-q^2/k^2}}
 \label{Rk(q)}
\ee
where $Z_k$ denotes an appropriate wave function renormalization
constant which will be defined below.
For fluctuations with small momenta
$q^2\ll k^2$ the cutoff $R_k\simeq Z_k k^2$ acts like an
additional mass term and prevents their propagation. 
For $q^2\gg k^2$ the infrared cutoff vanishes 
such that
the functional integration of the high momentum modes
is not disturbed. 
The expectation value of $\chi$ in the presence of $\Dt_k S[\chi]$
and $j$ reads $\vp^{ab} \equiv <\chi^{ab}> = 
2 \, \dt W_k[j]/\dt j^*_{ab}$.
We define the effective average action via a Legendre
transform
\be
\Gm_k[\vp]=-W_k[j]+\hal \int d^dx \left(j^*_{ab}(x)\chi^{ab}(x)+
\chi^*_{ab}(x) j^{ab}(x)\right)-\Dt_k S[\vp]. 
\ee
As a consequence, as the scale $k$ is lowered
$\Gm_k$ interpolates
from the classical action $S=\lim_{k \to \infty}\Gm_k$ to the 
standard effective action $\Gm=\lim_{k \to 0}\Gm_k$, i.e.\
the generating functional of $1PI$ Green functions \cite{Wet93-2}. 
Lowering $k$ results in 
a successive inclusion of fluctuations with momenta
$q^2\, \gtap \, k^2$ and therefore permits to explore the theory on 
larger and larger length scales. One can view 
$\Gm_k$ as the effective action for averages of fields
over a volume of size $\sim k^{-d}$ and the approach is similar
in spirit to the block spin action 
\cite{Ka,RG} on the lattice.
The interpolation property of $\Gm_k$ can be used to 'start' 
at some high momentum
scale $\La$ where $\Gm_{\La}$ can be taken as the classical or
short distance action and to solve the theory by following
$\Gm_k$ to $k \to 0$.
The scale dependence of $\Gm_k$ can be
described by an exact nonperturbative evolution 
equation \cite{Wet93-2,BAM,El,Mor}
\be
 \prlt \Gm_k [\vp] =
 \hal\Tr\left\{\left(
 \Gm_k^{(2)}[\vp]+R_k\right)^{-1}
 \frac{\prl R_k}{\prl t}\right\}
 \label{ERGE}
\ee
where $t=\ln(k/\La)$.
The evolution is described in terms of the {\em exact} inverse
propagator $\Gm_k^{(2)}$ as given by
the second functional derivative of $\Gm_k$ with respect 
to the fields. The trace involves a momentum 
integration as well as a summation
over the internal indices which count the $2 N^2$ real
scalar fields contained in $\vp$. 
The additional cutoff function $R_k$ with a form like
the one given above renders the momentum integration both 
infrared (IR) and ultraviolet 
(UV) finite. In particular, the direct implementation of
the additional mass term $R_k \simeq Z_k k^2$ for $q^2 \ll k^2$
into the inverse average propagator makes the formulation suitable
for dealing with theories which are plagued by infrared problems
in perturbation theory. The flow equation (\ref{ERGE})
is compatible with all symmetries of the model and can be 
generalized to include possible local gauge symmetries
\cite{gauge}. The exact renormalization group equation
can be formulated in many different but formally
equivalent ways 
\cite{RG,RG2} and it may be interpreted as a differential
form of the Schwinger-Dyson equations \cite{SD}.

From its construction, i.e.\ the
inclusion of fluctuations with characteristic momenta larger
than a given infrared cutoff $\sim k$, the effective average
action is the appropriate quantity for the study of physics at
a scale $k$. It therefore realizes the concept of a coarse
grained free energy in the sense of ref.\ \cite{Langer}.
Such a quantity becomes especially desirable for 
a description of first order phase transitions. 
In contrast to a second order phase transition there
is an inherent length scale $l_0$ due to a finite 
correlation length at a first order phase
transition. This length scale acts as a physical
infrared cutoff. The coarse grained effective action
$\Gm_k$ with $k \sim l_0^{-1}$ accounts for all
fluctuations with momenta larger than this physical
infrared cutoff and it is the appropriate quantity
for the study of the physics at the scale $l_0$. 
To be explicit, one may consider the coarse grained effective
potential $U_k$. It
is obtained from the coarse grained effective
action $\Gm_k$ for a constant field $\vp$.
At a first order phase transition there is a nonzero
difference between the field expectation value 
(order parameter) in
the symmetric (disordered) 
and in the spontaneously broken (ordered) phase.
The two phases correspond to two different minima
of the coarse grained
effective potential $U_k$ and both minima are 
separated by a potential barrier. 
The coarse grained potential $U_k$ is therefore a
nonconvex function whereas the standard effective
potential $U=\lim_{k \to 0} U_k$ has to be convex 
by its definition as a Legendre transform. The
convexity is due to the effect of fluctuations
with characteristic length scales larger than
$l_0$ \cite{RTW}.
The study of physical processes such as tunneling
or inflation usually
relies on the nonconvex part of the potential
which is discussed in section \ref{coarse}. 
 
Though the evolution equation (\ref{ERGE})
for the effective average
action is exact, it remains a complicated functional
differential equation. In practice one has to find
a truncation for $\Gm_k$ in order to obtain approximate solutions.
An important feature of the exact flow equation is therefore
its simple and intuitive form which helps to find a 
nonperturbative approximation scheme. The r.h.s.\ of eq.\
(\ref{ERGE}) expresses the scale dependence of $\Gm_k$
in terms of the exact propagator. Known properties
of the propagator can be used as a guide to find an
appropriate truncation for the effective average action.
For a scalar theory the propagator is a matrix 
characterized by mass terms and kinetic terms $\sim Z q^2$.
The mass matrix is given by the second derivative of 
the potential $U_k$ with respect to the fields. In general
$Z$ can be a complicated function of the fields and momenta.
We may exploit the fact that the function $Z$ plays the role of
a field and momentum dependent wave function 
renormalization. For second order phase transitions
and approximately for weak first order phase transitions
the behavior of $Z$ is governed by the anomalous 
dimension $\eta$. Typically for three and four dimensional
scalar theories $\eta$ is small. (In our case the relevant
value is $\eta \simeq 0.022$ as given by the corresponding
index in the $O(8)$ symmetric Heisenberg model). 
One therefore expects a weak
dependence of $Z$ on the fields and momenta. 
We will exploit this in the following.

\sect{\bf $U(2) \times U(2)$ symmetric scalar matrix model
\label{model}}

We consider a $U(2) \times U(2)$ symmetric effective 
action for a scalar field
$\vp$ which transforms in the $(2,2)$ representation
with respect to the subgroup $SU(2) \times SU(2)$. Here $\vp$
is represented by a complex $2 \times 2$ matrix
and the transformations are
\bea
 \vp &\ra& U^{ }\vp V^\dagger\,\, , \nnn
 \vp^\dagger &\ra& V^{ }\vp^\dagger U^\dagger
 \label{Transformations}
\eea
where $U$ and $V$ are unitary $2 \times 2$ matrices 
corresponding to the two distinct $U(2)$ factors.

We classify the invariants for the construction of the effective 
average action by the number of derivatives. The lowest order
in a systematic derivative expansion \cite{TW94-1,Mor}
of $\Gm_k$ is given by
\be
 \Gm_k = \ds{\int d^d x\left\{U_k(\vp,\vp^\dagger )+
 Z_k \prl_\mu \vp^*_{ab} \prl^\mu \vp^{ab}
  \right\}}\qquad  (a,b=1,2).
 \label{Ansatz}
\ee
The term with no derivatives defines the scalar potential $U_k$
which is an arbitrary function of traces of powers of 
$\vp^\dagger \vp$. The most general $U(2) \times U(2)$ symmetric
scalar potential can be expressed as a function of only two 
independent invariants,
\bea
 \rho &=& \ds{\tr\left(\vp^\dagger \vp \right) }\non\\
 \tau &=& \ds{2 \tr\left(\vp^\dagger \vp - \frac{1}{2} \rho \right)^2}
 = \ds{2 \tr\left(\vp^\dagger \vp \right)^2 - \rho^2 }.
 \label{Invariants}
\eea
Here we have used for later convenience the traceless matrix 
$\vp^\dagger \vp - \frac{1}{2} \rho$ to construct the second
invariant.
Higher invariants, $\tr\left(\vp^\dagger \vp 
- \frac{1}{2} \rho \right)^n$ for
$n > 2$, can be expressed as functions of $\rho$ and $\tau$ 
\cite{Ju95-7}.

For the derivative part we consider a standard kinetic term with a 
scale dependent wave function renormalization constant $Z_k$. 
The first correction to the kinetic term would include field 
dependent wave function
renormalizations $Z_k(\rho,\tau)$ plus functions not
specified in eq.\ (\ref{Ansatz}) which account for a
different index structure of invariants with two
derivatives. These wave function renormalizations may be
defined at zero momentum. The next level involves 
invariants with four derivatives and so on.
We define $Z_k$ 
at the minimum $\rho_0$, $\tau_0$ of $U_k$ and
at vanishing momenta $q^2$,
\be
Z_k=Z_k(\rho=\rho_0,\tau=\tau_0;q^2=0). \label{zet}
\ee
The factor $Z_k$ appearing
in the definition of the infrared cutoff $R_k$ in eq.\ (\ref{Rk(q)})
is identified with (\ref{zet}).
The $k$-dependence of this function is given by the anomalous
dimension
\be
\eta(k)=-\frac{\mbox{d}}{\mbox{d}t}\mbox{ln} Z_k .
\label{Eta} 
\ee 

If the ansatz (\ref{Ansatz}) is inserted into the flow equation
for the effective average action (\ref{ERGE}) one obtains flow equations  
for the effective average potential $U_k(\rho,\tau)$ and for the
wave function renormalization constant $Z_k$ (or equivalently the 
anomalous dimension $\eta$).
This is done in sections
\ref{scale} and \ref{solv}. These flow equations 
have to be integrated starting from some short distance scale 
$\La$ and one has to specify $U_{\La}$ and $Z_{\La}$
as initial conditions. At the scale $\La$, where 
$\Gm_{\La}$ can be taken as the classical 
or short distance action, no integration 
of fluctuations 
has been performed.
The short distance potential is taken to be a quartic potential
which is parametrized by two quartic couplings $\bar{\la}_{1\La}$,
$\bar{\la}_{2\La}$ and a mass term. We start
in the spontaneously broken regime where the minimum
of the potential occurs at a nonvanishing field value 
and there is a negative mass term at
the origin of the potential $(\bar{\mu}_{\La}^2 > 0)$, 
\be
U_{\La}(\rho,\tau)=-\bar{\mu}_{\La}^2 \rho + \hal \bar{\la}_{1\La}
\rho^2 +\frac{1}{4} \bar{\la}_{2\La} \tau \quad
\label{uinitial}
\ee 
and $Z_{\La}=1$. 
The potential is bounded from
below provided 
$\bar{\la}_{1\La} > 0$ and 
$\bar{\la}_{2\La} > - 2 \bar{\la}_{1\La}$.
For $\bar{\la}_{2\La} > 0$ one observes 
the potential minimum for the configuration
$\vp_{ab}=\vp \delta_{ab}$ corresponding
to the spontaneous symmetry breaking 
down to the diagonal $U(2)$ subgroup of $U(2) \times U(2)$.
For negative $\bar{\la}_{2\La}$ the potential is minimized
by the configuration $\vp_{ab}=\vp \delta_{a1} \delta_{ab}$
which corresponds to the symmetry breaking pattern 
$U(2) \times U(2) \lra U(1) \times U(1) \times U(1)$. 
In the special case $\bar{\la}_{2\La}=0$ the theory
exhibits an enhanced $O(8)$ symmetry. This constitutes
the boundary between two phases with different symmetry
breaking patterns.

The limits of infinite couplings correspond to nonlinear
constraints in the matrix model. For 
$\bar{\la}_{1\La} \to \infty$ with fixed ratio 
$\bar{\mu}_{\La}^2/\bar{\la}_{1\La}$ one finds the constraint
$\tr(\vp^{\dagger}\vp)=2 \bar{\mu}_{\La}^2/\bar{\la}_{1\La}$.
By a convenient choice of $Z_{\La}$ (rescaling of $\vp$)
this can be brought to the form $\tr(\vp^{\dagger}\vp)=2$.
On the other hand, the limit $\bar{\la}_{2\La} \to +\infty$
enforces the constraint 
$\vp^{\dagger}\vp=\hal \tr(\vp^{\dagger}\vp)$.
Combining the limits $\bar{\la}_{1\La} \to \infty$,
$\bar{\la}_{2\La} \to \infty$ the constraint reads
$\vp^{\dagger}\vp=1$ and we deal with a matrix model for 
unitary matrices. (These considerations generalize to arbitrary
$N$.) Another interesting limit obtains for 
$\bar{\la}_{1\La}=-\hal \bar{\la}_{2\La} + \Dt_{\la}$,
$\Dt_{\la} > 0$ if $\bar{\la}_{2\La} \to - \infty$. In this 
case the nonlinear constraint reads
$(\tr\vp^{\dagger}\vp)^2=\tr(\vp^{\dagger}\vp)^2$ which
implies for $N=2$ that $\det \vp =0$. This is a matrix model
for singular complex $2 \times 2$ matrices.

One can also interpret our model as the coupled 
system of two
$SU(2)$-doublets for the weak interaction Higgs
sector. This is simply done by decomposing the 
matrix $\vp_{ab}$ into two two-component complex 
fundamental representations of one of the $SU(2)$
subgroups, $\vp_{ab} \to \vp_{1b},\vp_{2b}$. The
present model corresponds to a particular
left-right symmetric model with interactions specified
by
\bea
\rho&=&\vp_1^{\dagger}\vp_1+\vp_2^{\dagger}\vp_2\\
\tau&=&\left(\vp_1^{\dagger}\vp_1-\vp_2^{\dagger}\vp_2
\right)^2+4\left(\vp_1^{\dagger}\vp_2\right)
\left(\vp_2^{\dagger}\vp_1\right) \label{th}\, .
\eea
We observe that for a typical weak interaction symmetry
breaking pattern the expectation values of $\vp_1$ and
$\vp_2$ should be aligned in the same direction or one
of them should vanish. In the present model this 
corresponds to the choice $\bar{\la}_{2\La} < 0$.
The phase structure of a related model without the term
$\sim (\vp_1^{\dagger}\vp_2)(\vp_2^{\dagger}\vp_1)$
has been investigated previously \cite{TwoHig}
and shows second or first order 
transitions\footnote{First order phase transitions and 
coarse graining have also been discussed in a 
multi-scalar model with $Z_2$ symmetry \cite{AlMr}.
}. Combining
these results with the outcome of this work leads 
already to a detailed qualitative overview over the phase 
pattern in a more general setting with three independent
couplings for the quartic invariants 
$(\vp_1^{\dagger}\vp_1+\vp_2^{\dagger}\vp_2)^2$, 
$(\vp_1^{\dagger}\vp_1-\vp_2^{\dagger}\vp_2)^2$
and $(\vp_1^{\dagger}\vp_2)(\vp_2^{\dagger}\vp_1)$. 
We also note that the special case 
$\bar{\la}_{2\La}=2\bar{\la}_{1\La}$ corresponds to two
Heisenberg models interacting only by a term sensitive
to the alignment between $\vp_1$ and $\vp_2$, i.e.\
a quartic interaction of the form 
$(\vp_1^{\dagger}\vp_1)^2+(\vp_2^{\dagger}\vp_2)^2
+2(\vp_1^{\dagger}\vp_2)(\vp_2^{\dagger}\vp_1)$.

The model is now completely specified and it remains to extract
the flow equations for $U_k$ and $Z_k$. Before this is carried
out in sections \ref{scale} and \ref{solv}
we present an overview over the phase structure in the next
section. These results are obtained from a numerical solution
of the evolution equations.

\sect{Phase structure \label{ps}}

In this section we consider the
$U(2) \times U(2)$ symmetric model in three space dimensions.
The aim is to give an overview of our results concerning
the phase structure and
the effective average potential.  
We concentrate here on the spontaneous 
symmetry breaking with a residual $U(2)$ symmetry
group. 
This symmetry breaking can be observed for a
configuration which is proportional to the 
identity and with (\ref{Invariants}) one finds $\tau = 0$. 
In this case we shall use
an expansion of $U_k(\rho,\tau)$ around $\tau=0$ keeping only the
linear term in $\tau$. This amounts to 
assuming
\be
\ds{\frac{\prl^n U_k}{\prl \tau^n}}
(\rho,\tau=0)=0 \quad \mbox{for}
\quad n \ge 2.
\label{Truncationdim}
\ee
We will motivate this truncation in section \ref{solv} where
we present a more detailed analysis. 
We make no expansion of $U_k(\rho,\tau)$ in terms of $\rho$.
This allows the description
of a first order phase transition where a second local minimum
of $U_k(\rho) \equiv U_k(\rho,\tau=0)$ appears. The 
$\rho$-dependence also gives information about the
equation of state of the system. 
 
For the considered symmetry breaking pattern 
the short distance potential $U_{\La}$
given in eq.\ (\ref{uinitial}) is parametrized by positive 
quartic couplings, 
\be
\bar{\la}_{1\La},\bar{\la}_{2\La}>0
\ee 
and the location of its minimum is given by 
\be
\rho_{0\La}=\bar{\mu}_{\La}^2/\bar{\la}_{1\La}. 
\ee
To study the phase structure of the model we integrate the
flow equation for the effective average potential $U_k$
(cf.\ sect.\ \ref{scale}, \ref{solv})
for a variety of initial 
conditions $\rho_{0\La},\bar{\la}_{1\La}$ and
$\bar{\la}_{2\La}$. In particular, for general 
$\bar{\la}_{1\La},\bar{\la}_{2\La}>0$ we are able to find a critical
value $\rho_{0\La}=\rho_{0 c}$ for which the system exhibits
a first order phase transition. In this case the evolution of
$U_k$ leads at some scale $k_2 < \La$ to the appearance 
of a second local
minimum at the origin of the effective average potential and both
minima become degenerate in the limit $k \to 0$. If $\rho_0(k)>0$
denotes the $k$-dependent outer minimum of the potential 
($U_k^{\pri}(\rho_0)=0$, where the prime on $U_k$ denotes the
derivative with respect to $\rho$ at fixed $k$) at
a first order phase transition one has
\be
\lim\limits_{k \to 0}( U_k(0)-U_k(\rho_0))=0.
\label{critcon}
\ee
A measure of the distance from the phase transition
is the difference $\dt\kp_{\La}=(\rho_{0\La}-\rho_{0 c})/\La$.
If $\bar{\mu}_{\La}^2$ and therefore
$\rho_{0\La}$ is interpreted as a function of 
temperature, the deviation $\dt\kp_{\La}$ is 
proportional to the deviation from the critical 
temperature $T_{c}$, i.e.\ $\dt\kp_{\La}=A(T)
(T_{c}-T)$ with $A(T_{c}) > 0$. 

\begin{figure}[h]
\unitlength1.0cm
\begin{center}
\begin{picture}(13.,9.)
\put(-0.6,4.4){$\ds{\frac{U_{k_f}}{\vp_{0R}^6}}$}
\put(6.4,-0.5){$\ds{\frac{\vp_R}{\vp_{0R}}}$}
\put(-0.5,0.){
\epsfysize=13.cm
\epsfxsize=9.cm
\rotate[r]{\epsffile{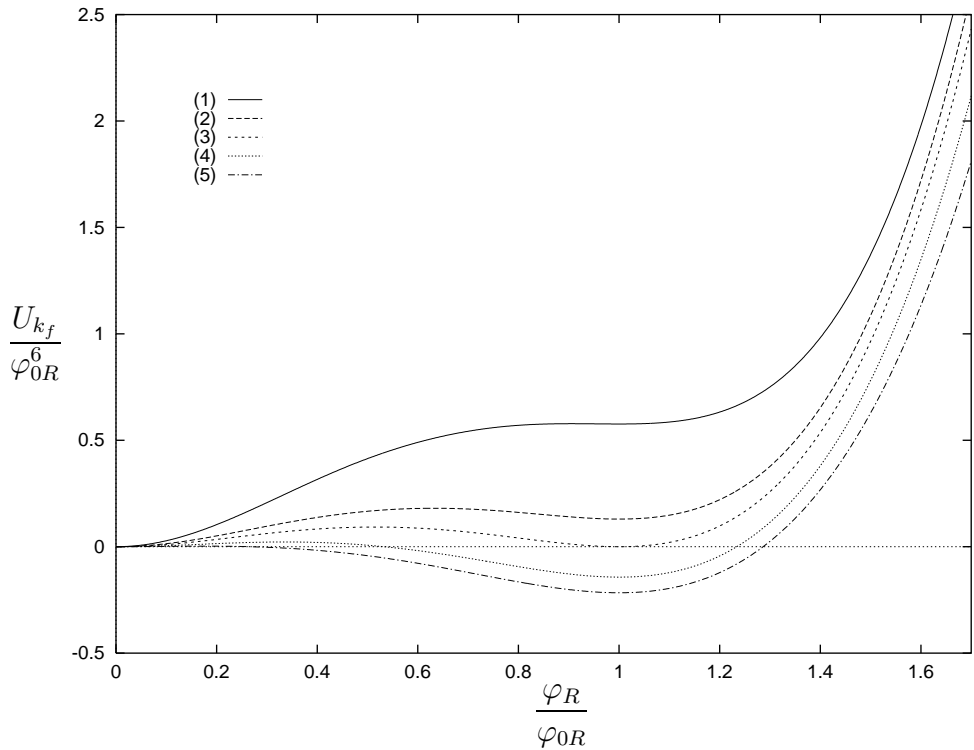}}
}
\end{picture}
\end{center}
\vspace{0.3cm}
\caption{\footnotesize The effective average potential
$U_{k=k_f}$ as a function of the renormalized field 
$\vp_R$. The potential
is shown for various values of $\dt \kp_{\La} \sim T_c-T$.
The parameters for the short distance potential $U_{\La}$ are
(1) $\dt \kp_{\La}=-0.03$, (2) $\dt \kp_{\La}=-0.015$,
(3) $\dt \kp_{\La}=0$, (4) $\dt \kp_{\La}=0.04$,
(5) $\dt \kp_{\La}=0.1$ and
$\la_{1\La}=0.1$, $\la_{2\La}=2$.
\label{tempot}
}
\end{figure}

\begin{figure}[h]
\unitlength1.0cm
\begin{center}
\begin{picture}(13.,9.)
\put(4.5,3.5){\footnotesize $\ds{\frac{m_R}{\La}}$}
\put(4.5,7.){\footnotesize $\ds{\frac{\rho_{0R}}{\La}}$}
\put(6.4,-0.5){$-\dt\kp_{\La}$}
\put(-0.5,0.){
\epsfysize=13.cm
\epsfxsize=9.cm
\rotate[r]{\epsffile{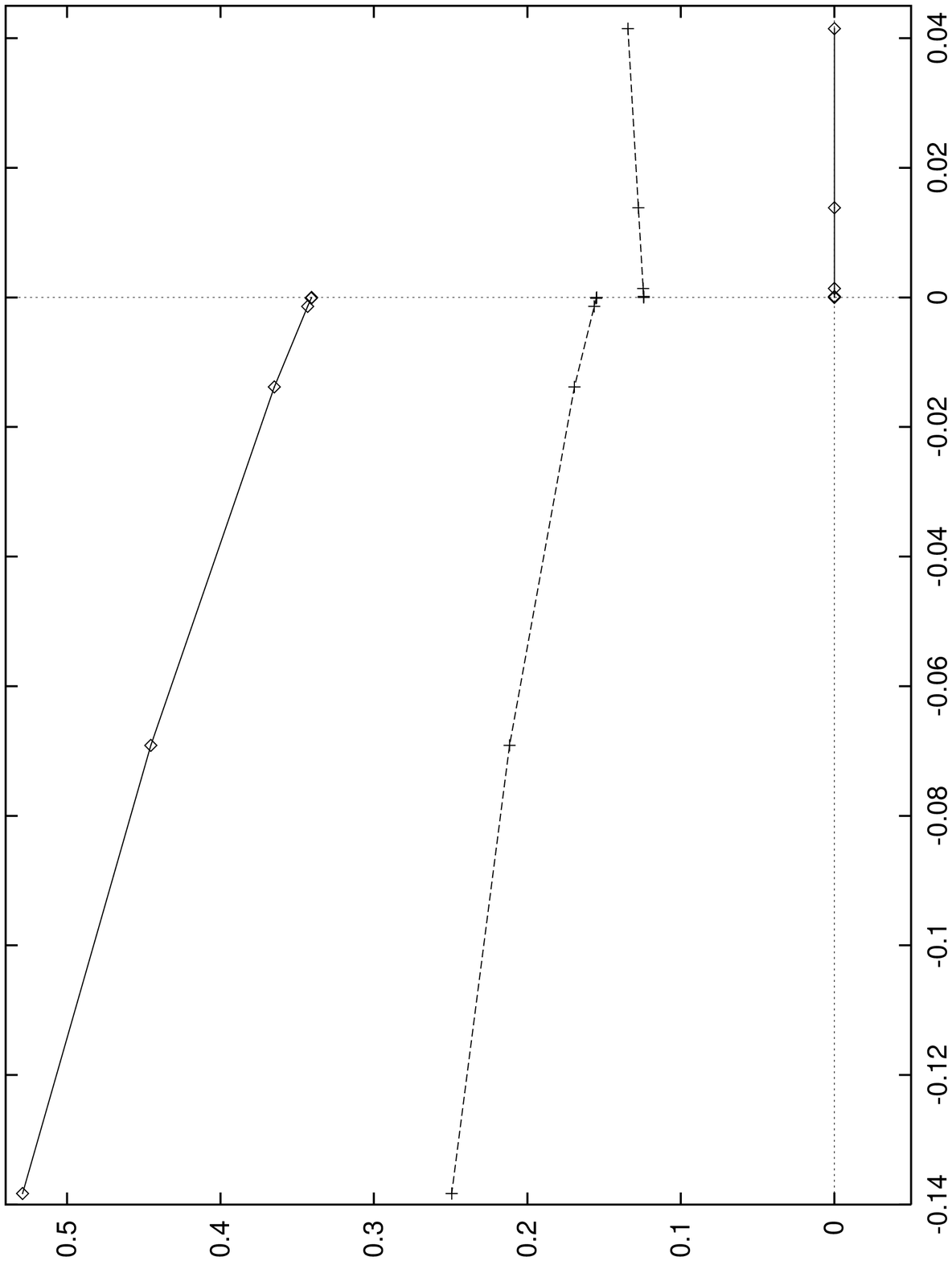}}
}
\end{picture}
\end{center}
\caption[]{\label{temp} 
\footnotesize The
minimum $\rho_{0R}$ 
and the mass term 
$m_R$ in units of the momentum scale $\La$
as a function of $-\dt\kp_{\La}$ or temperature
($\la_{1\La}=0.1$, $\la_{2\La}=2$, $k=k_f$). For 
$\dt\kp_{\La}=0$ one observes the jump in the
renormalized order parameter $\Dt \rho_{0R}$
and mass $\Dt m_R$. 
}
\end{figure}

We consider in the
following the effective average potential $U_k$
for a nonzero scale $k$. This allows
to observe the nonconvex part of the potential (cf.\
sect.\ \ref{coarse}). 
As an example we show in 
fig.\ \ref{tempot} the effective average potential
$U_{k=k_f}$ for $\la_{1\La}= \bar{\la}_{1\La}/\La=0.1$
and $\la_{2\La}= \bar{\la}_{2\La}/\La=2$
as a function of the renormalized field 
$\vp_R=(\rho_R/2)^{1/2}$ with $\rho_{R}=Z_{k=k_f} \rho$. 
The scale $k_f$ is some 
characteristic scale 
below which the location of the minimum $\rho_0(k)$
becomes essentially independent of $k$.  
Its precise definition is given below. We have normalized 
$U_{k_f}$ and $\vp_R$
to powers of the renormalized minimum 
$\vp_{0R}(k_f)= (\rho_{0R}(k_f)/2)^{1/2}$ with
$\rho_{0R}(k_f)=Z_{k_f}\rho_0(k_f)$.
The potential
is shown for various values of deviations from the critical
temperature or $\dt\kp_{\La}$. 
For the given examples $\dt \kp_{\La}=-0.03$, $-0.015$
the minimum at the origin becomes
the absolute minimum and the system is in the
symmetric (disordered) phase. Here $\vp_{0R}$
denotes the minimum in the metastable ordered phase.
In contrast, for $\dt \kp_{\La}=0.04$, $0.1$ the 
absolute minimum is located at $\vp_R/\vp_{0R}=1$ which
characterizes the spontaneously broken phase.
For large enough $\dt \kp_{\La}$ the local minimum at the 
origin vanishes.
For $\dt\kp_{\La}=0$ the two distinct minima are degenerate in 
height\footnote{
We note that the critical 
temperature is determined by condition (\ref{critcon})
in the limit $k \to 0$. Nevertheless for the employed
nonvanishing scale $k=k_f$ the minima of $U_k$
become almost degenerate at the critical temperature.
}.
As a consequence the order parameter makes a discontinuous jump
at the phase transition which characterizes the transition to be 
first order.
It is instructive to consider some characteristic values
of the effective average potential.
In fig.\ \ref{temp} we consider for
$\la_{1\La}=0.1,\la_{2\La}=2$ the value of the renormalized
minimum $\rho_{0R}(k_f)$ and the radial mass term 
as a function of $-\dt\kp_{\La}$ or temperature.
In the spontaneously
broken phase the renormalized radial mass squared is given by
(cf.\ section \ref{scale})
\be
m^2_R(k_f)=2 Z_{k_f}^{-1} \rho_0 U^{\pri\pri}_{k_f}(\rho_0),
\label{radialmass}
\ee 
in the symmetric phase the renormalized mass term reads
\be
m^2_{0R}(k_f)=Z_{k_f}^{-1} U^{\pri}_{k_f}(0).
\label{originmass}
\ee
At the critical temperature ($\dt\kp_{\La}=0$) one observes
the discontinuity $\Dt \rho_{0R}=\rho_{0R}(k_f)$ and the jump
in the mass term $\Dt m_R=m_R(k_f)-m_{0R}(k_f)
=m_R^c-m_{0R}^c$. (Here the index 'c' denotes
$\dt\kp_{\La}=0$).
The ratio $\Dt \rho_{0R}/\La$ is a rough measure for 
the 'strength' of the first order transition.
For $\Dt \rho_{0R}/\La \ll 1$ the phase transition is
weak in the sense that typical masses are small compared to
$\La$. In consequence, the long-wavelength fluctuations
play a dominant role and the system exhibits universal 
behavior, i.e.\ it becomes largely independent of the
details at the short distance scale $\La^{-1}$.
We will discuss the universal behavior in more 
detail below.     

\begin{figure}[h]
\unitlength1.0cm
\begin{center}
\begin{picture}(13.,9.)
\put(-0.9,4.6){$\ds{\ln\left(\la_{2\La}\right)}$}
\put(6.4,-0.5){$\ds{\ln\left(\la_{1\La}\right)}$}
\put(9.5,6.5){\footnotesize $(1)$}
\put(9.5,5.4){\footnotesize $(2)$}
\put(9.5,3.){\footnotesize $(3)$}
\put(9.5,1.5){\footnotesize $(4)$}
\put(-0.5,0.){
\epsfysize=13.cm
\epsfxsize=9.cm
\rotate[r]{\epsffile{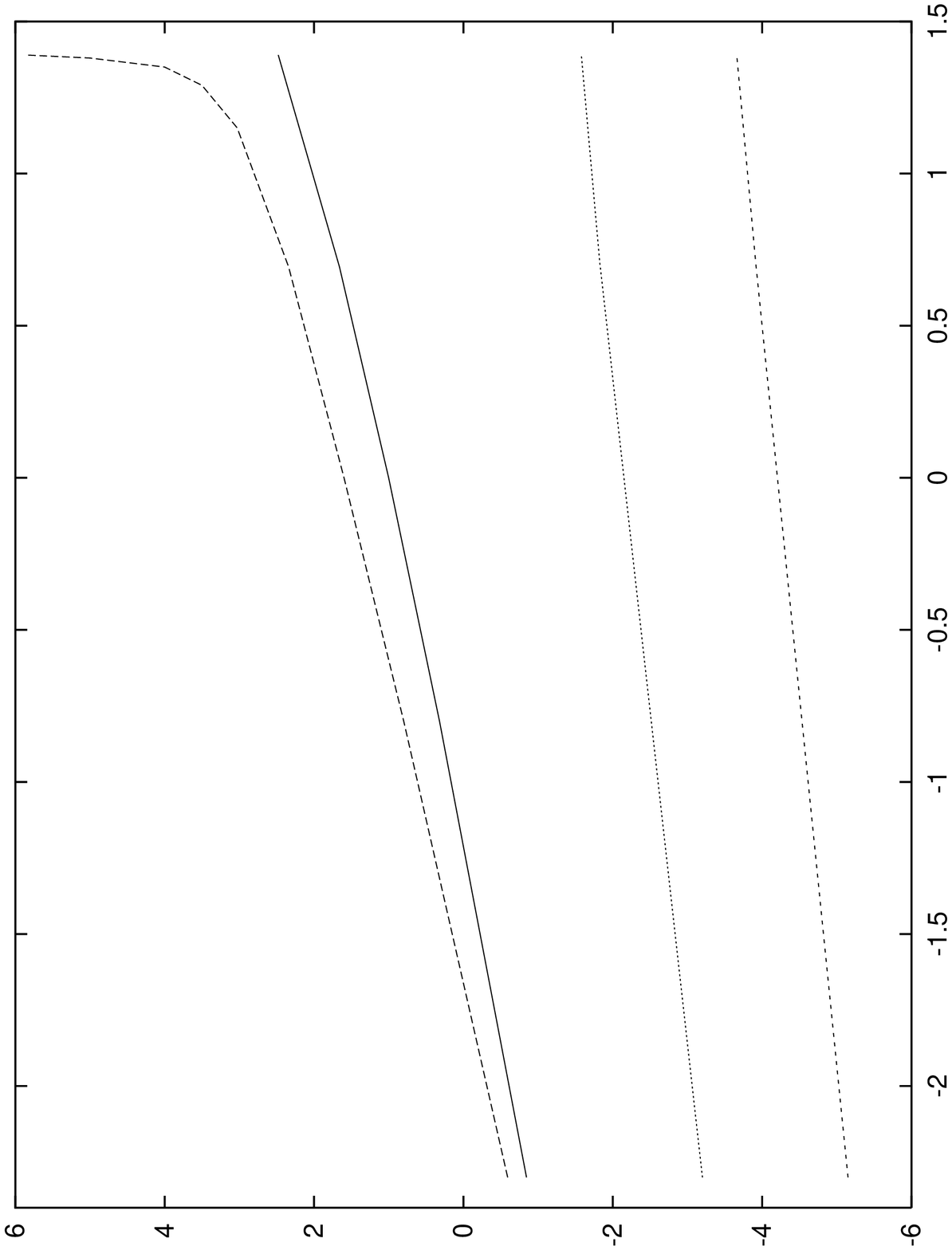}}
}
\end{picture}
\end{center}
\caption[]{\label{la1la2} \footnotesize
Lines of constant jump of the 
renormalized order parameter
$\Dt \rho_{0R}/\La$ at the phase transition in the
$\ln(\la_{1\La}) , \ln(\la_{2\La})$ plane. The curves
correspond to (1) $\ln(\Dt \rho_{0R}/\La)=-4.0$,
(2) $\ln(\Dt \rho_{0R}/\La)=-4.4$,
(3) $\ln(\Dt \rho_{0R}/\La)=-10.2$,
(4) $\ln(\Dt \rho_{0R}/\La)=-14.3$.    
}
\end{figure}

In order to characterize the strength of 
the phase transition for 
arbitrary positive values of $\la_{1\La}$ and $\la_{2\La}$
we consider lines of constant $\Dt \rho_{0R}/\La$ in the
$\la_{1\La} , \la_{2\La}$ plane. In fig.\ \ref{la1la2} this is
done for the logarithms of these quantities. 
For fixed
$\la_{2\La}$ one observes that the discontinuity
at the phase transition weakens with increased $\la_{1\La}$.
On the other hand for given $\la_{1\La}$ one finds a larger jump in the
order parameter for increased $\la_{2\La}$. This is true up to a 
saturation point where $\Dt \rho_{0R}/\La$ becomes independent
of $\la_{2\La}$. In the plot this can be observed from the
vertical part of the line of constant ln$(\Dt \rho_{0R}/\La)$.
This phenomenon occurs for arbitrary nonvanishing   
$\Dt \rho_{0R}/\La$ in the strong $\la_{2\La}$ coupling limit
and is discussed in section \ref{rg}. 

\begin{figure}[h]
\unitlength1.0cm
\begin{center}
\begin{picture}(13.,9.)
\put(-1.2,4.6){$\ds{\ln\left(\frac{\Dt \rho_{0R}}{\La}\right)}$}
\put(6.2,-0.5){$\ds{\ln\left(\la_{2\La}\right)}$}
\put(8.9,8.5){\footnotesize $(1)$}
\put(10.5,7.7){\footnotesize $(2)$}
\put(10.5,6.7){\footnotesize $(3)$}
\put(-0.2,0.){
\epsfysize=13.cm
\epsfxsize=9.cm
\rotate[r]{\epsffile{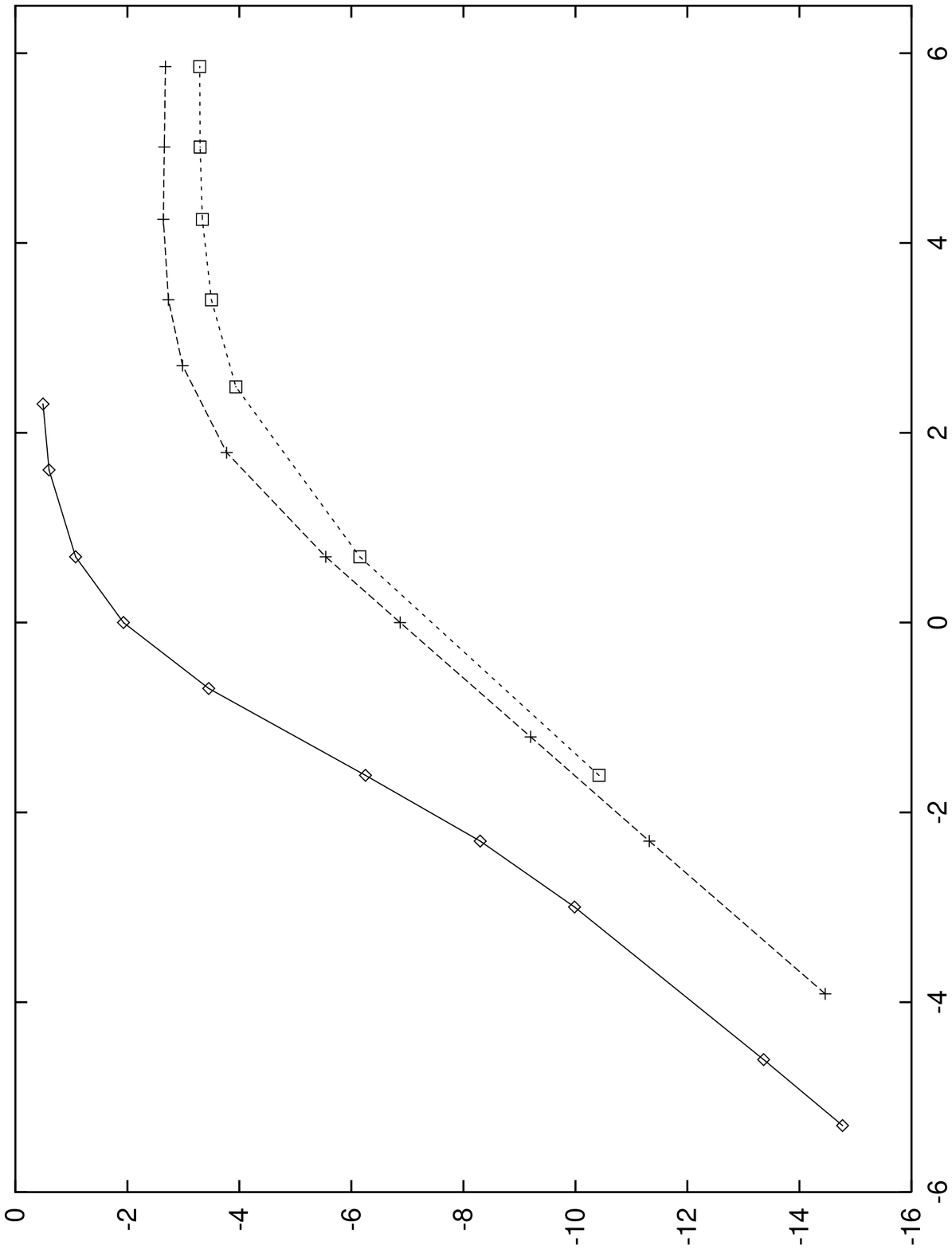}}
}
\end{picture}
\end{center}
\caption[]{\footnotesize The logarithm of the discontinuity 
of the renormalized order parameter
$\Dt \rho_{0 R}/\La$ 
as a function of $\ln(\la_{2\La})$.
Data points
for fixed (1) $\la_{1\La}=0.1$,
(2) $\la_{1\La}=2$, (3) $\la_{1\La}=4$ are
connected by straight lines. 
\label{phase}
}
\end{figure}

In the following we give a detailed quantitative
description of the first order phase
transitions and a separation in weak and strong transitions.
We consider some characteristic quantities for the effective average 
potential in dependence on the short distance parameters 
$\la_{1\La}$ and
$\la_{2\La}$ for $\dt \kp_{\La}=0$.  
We consider the discontinuity
in the renormalized order parameter $\Dt \rho_{0R}$ 
and the inverse correlation
lengths (mass terms) 
$m_R^c$ and $m_{0R}^c$ in the ordered and the 
disordered phase respectively. 
Fig.\ \ref{phase}
shows the logarithm of  
$\Dt \rho_{0 R}$ in units of 
$\La$ as a function of the logarithm of the initial 
coupling $\la_{2\La}$. 
We have connected the calculated values
obtained for various fixed $\la_{1\La}=0.1$, 
$2$ and $\la_{1\La}=4$
by straight lines. The values are listed in table \ref{table1}. 
\begin{table} [h]
\renewcommand{\arraystretch}{1.5}
\hspace*{\fill}
\begin{tabular}{|c|c|c|c|c||c|c|c|c|c|}     \hline

$\la_{1 \La}$
&$\la_{2 \La}$
&$\ds{\frac{\Dt \rho_{0 R}}{\La}}$
&$\ds{\frac{m_R^c}{\Dt \rho_{0 R}}}$
&$\ds{\frac{m_{0R}^c}{\Dt \rho_{0 R}}}$
&$\la_{1 \La}$
&$\la_{2 \La}$
&$\ds{\frac{\Dt \rho_{0 R}}{\La}}$
&$\ds{\frac{m_R^c}{\Dt \rho_{0 R}}}$
&$\ds{\frac{m_{0R}^c}{\Dt \rho_{0 R}}}$
\\ \hline \hline
$0.1$
&0.005
&$0.386 \times 10^{-6}$
&$1.69$
&$1.26$
&$2$
&2
&$ 0.392\times 10^{-2}$
&1.66
&1.24
\\ \hline
$0.1$
&0.01
&$0.158 \times 10^{-5}$
& $1.68$
&1.26
&2
&6
&$0.230\times 10^{-1}$
&1.68
&1.25 
\\ \hline
$0.1$
&0.05
&$0.461 \times 10^{-4}$
& $1.66$
&1.23
&2
&15 
&$0.505 \times 10^{-1}$
&1.80
&1.35 
\\ \hline
$0.1$
&0.1
&$0.249 \times 10^{-3}$
& $1.58$
&1.17
&2
&30
&$0.649 \times 10^{-1}$
&1.91
&1.45
\\ \hline
$0.1$
&0.2
&$0.193 \times 10^{-2}$
& $1.34$
&0.992
&2
&70
&$0.712 \times 10^{-1}$
&2.01
&1.60
\\ \hline
$0.1$
&0.5
&$0.316 \times 10^{-1}$
& $0.772$
&0.571
&2
&150
&$0.699 \times 10^{-1}$
&2.02 
&1.65
\\ \hline
$0.1$
&1
&0.145
& $0.527$
&0.395
&2
&350
&$0.685 \times 10^{-1}$
&2.03
&1.68
\\ \hline
$0.1$
&2
&0.341
& $0.455$
&0.360
&4
&0.2
&$0.298 \times 10^{-4}$
&1.69 
&1.26
\\ \hline
$0.1$
&5
&0.547
& $0.450$
&0.414
&4
&2
&$0.213 \times 10^{-2}$
&1.70
&1.27 
\\ \hline
$0.1$
&10
&0.610
& $0.462$
&0.490
&4
&12
&$0.195 \times 10^{-1}$ 
&1.80
&1.35
\\ \hline
$2$
&0.02
&$ 0.523 \times 10^{-6}$
&$ 1.69 $
&1.26
&4
&30
&$0.302 \times 10^{-1}$
&1.89 
&1.43
\\ \hline
$2$
&0.1
&$0.121 \times 10^{-4}$
&$ 1.69 $
&1.26
&4
&70
&$0.355 \times 10^{-1}$
&1.96
&1.49
\\ \hline
$2$
&0.3
&$0.101 \times 10^{-3}$
&$ 1.69 $
&1.25
&4
&150
&$0.369 \times 10^{-1}$
&1.98
&1.55
\\ \hline
$2$
&1
&$ 0.104\times 10^{-2}$
&1.66
&1.25
&4
&350
&$0.372 \times 10^{-1}$
&1.97
&1.57
\\ \hline 
\end{tabular}
\hspace*{\fill}
\caption{\footnotesize
The discontinuity in the renormalized order parameter 
$\Dt \rho_{0R}$ and the critical inverse correlation
lengths $m_R^c$ and $m_{0R}^c$ in the ordered and the 
disordered phase respectively. 
For small $\la_{2\La}/\la_{1\La}$ the ratios 
$m_R^c/\Dt \rho_{0R}$ and $m_{0R}^c/\Dt \rho_{0R}$ become
universal. 
\label{table1}
}
\end{table}
For 
$\la_{2\La}/\la_{1\La}\,\, \ltap \,\,1$ the curves show constant 
positive slope. In this range $\Dt \rho_{0 R}$ follows a 
power law behavior
\be
\Dt \rho_{0 R}=R \, (\la_{2\La})^{\th}, \quad  \th=1.93 
\label{Powrho}.
\ee
The critical exponent $\th$ is obtained from the slope
of the curve in fig.\ \ref{phase} for 
$\la_{2\La}/\la_{1\La}\ll 1$. The exponent is
universal and, therefore, does not depend on the specific 
value for $\la_{1\La}$.
On the other hand, the amplitude $R$ grows with
decreasing $\la_{1\La}$. For vanishing $\la_{2\La}$
the order parameter changes continuously at the transition
point and one observes a second order phase transition
as expected for the $O(8)$ symmetric vector model. As 
$\la_{2\La}/\la_{1\La}$ becomes larger than one the
curves deviate substantially
from the linear behavior. The deviation
depends on the specific choice of the short distance
potential. For $\la_{2\La}/\la_{1\La}\gg 1$ the curves
flatten. In this range
$\Dt \rho_{0 R}$ becomes insensitive to a variation
of the quartic coupling $\la_{2\La}$.

\begin{figure}[h]
\unitlength1.0cm
\begin{center}
\begin{picture}(13.,9.)
\put(5.5,-0.8){$\ds{\ln\left(\frac{\la_{2 \La}}
{\la_{1 \La}}\right)}$}
\put(10.6,1.3){\footnotesize $(1)$}
\put(10.6,8.28){\footnotesize $(2)$}
\put(10.6,7.55){\footnotesize $(3)$}
\put(10.6,6.65){\footnotesize $(2)$}
\put(10.6,5.7){\footnotesize $(3)$}
\put(2.2,6.05){\footnotesize $\ds{\frac{m_R^c}
{\Dt \rho_{0R}}}$}
\put(2.2,4.1){\footnotesize $\ds{\frac{m_{0R}^c}
{\Dt \rho_{0R}}}$}
\put(-0.5,0.){
\epsfysize=13.cm
\epsfxsize=9.cm
\rotate[r]{\epsffile{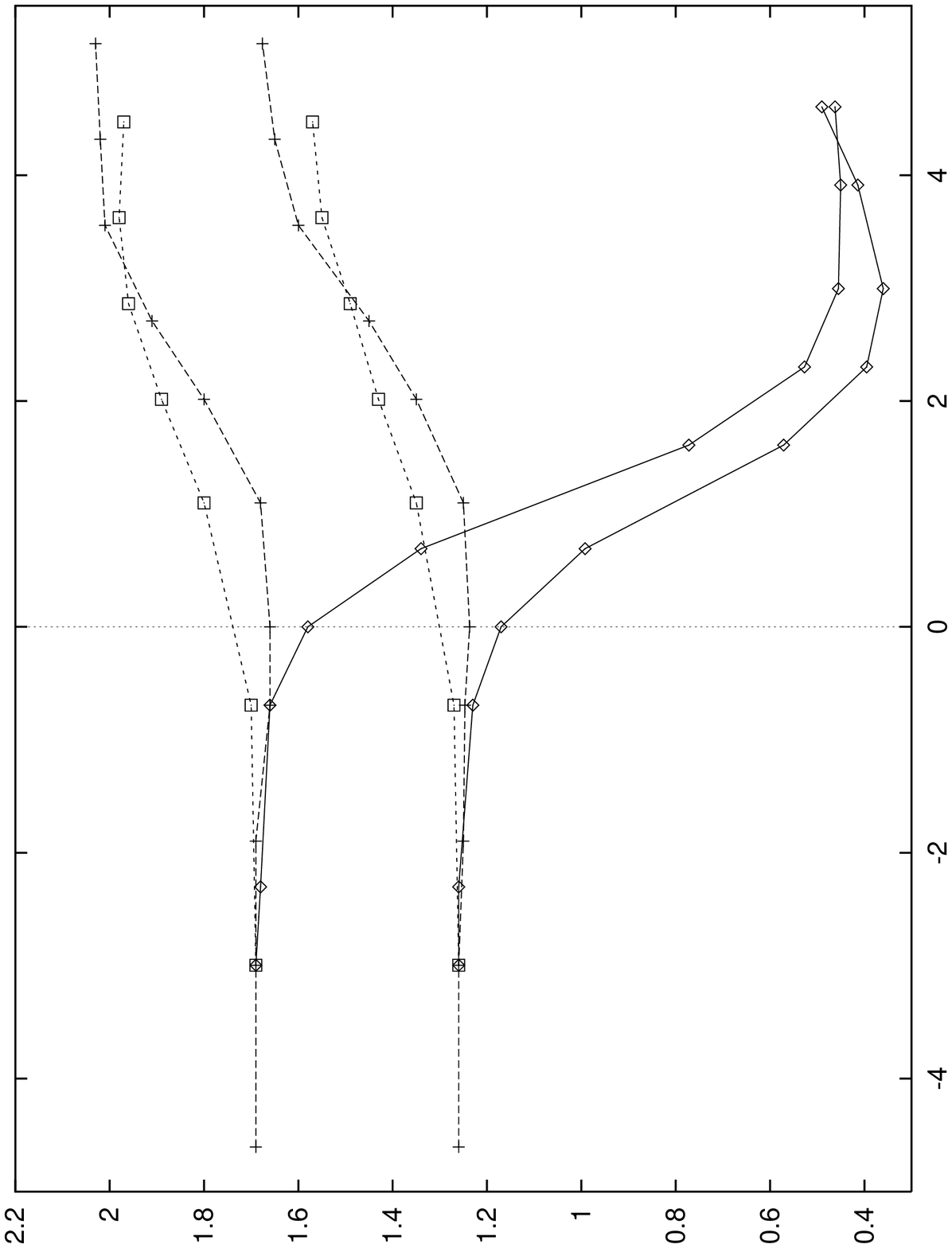}}
}
\end{picture}
\end{center}
\vspace{0.5cm}
\caption[]{\label{ratio} \footnotesize
The inverse correlation
lengths $m_R^c$ and $m_{0R}^c$ in the ordered and the 
disordered phase respectively. They are normalized to
$\Dt \rho_{0R}$ and given as a function
of $\ln(\la_{2\La}/\la_{1\La})$. Data points
for fixed (1) $\la_{1\La}=0.1$,
(2) $\la_{1\La}=2$, (3) $\la_{1\La}=4$ are
connected by straight lines. 
}
\end{figure}

In addition to the jump in the order parameter 
we present the mass terms $m_R^c$ 
and $m_{0R}^c$ which we
normalize to $\Dt \rho_{0 R}$.
In fig.\ \ref{ratio} these ratios are
plotted versus the logarithm of the ratio of the initial
quartic couplings $\la_{2\La}/\la_{1\La}$. 
Again values
obtained for fixed $\la_{1\La}=0.1$, $2$ and $\la_{1\La}=4$
are connected by straight lines. The universal range is set
by the condition $m_R^c/ \Dt \rho_{0 R} \simeq \mbox{const}$
(equivalently for $m_{0R}^c/ \Dt \rho_{0 R}$).  
The universal ratios are 
$m_R^c/ \Dt \rho_{0 R}=1.69$
and $m_{0R}^c/ \Dt \rho_{0 R}=1.26$
as can be seen from table \ref{table1}. 
For the given curves universality holds 
approximately for
$\la_{2\La}/\la_{1\La} \, \ltap \, 1/2$ 
and becomes 'exact' in the limit 
$\la_{2\La}/\la_{1\La} \to 0$. In this range we obtain
\be
m_R^c = S (\la_{2\La})^{\th},\qquad 
m_{0R}^c = \tilde{S} (\la_{2\La})^{\th} 
\label{Powm}.
\ee
The 
universal features of the system are not restricted to the weak coupling 
region of $\la_{2\La}$. This is demonstrated in fig.\ \ref{ratio}
for values up to  $\la_{2\La} \simeq 2$. 
The ratios $m_R^c/ \Dt \rho_{0 R}$ and $m_{0R}^c/ \Dt \rho_{0 R}$
deviate from the universal
values as $\la_{2\La}/\la_{1\La}$ is increased. For fixed
$\la_{2\La}$
a larger $\la_{1\La}$ results in a weaker transition
concerning $\Dt \rho_{0 R}/\La$.
The ratio $m_R^c/ \Dt \rho_{0 R}$ increases with $\la_{1\La}$
for small fixed $\la_{2\La}$ whereas in the asymptotic
region, $\la_{2\La}/\la_{1\La} \gg 1$, one observes from
fig.\ \ref{ratio} that this tendency is reversed 
and $m_R^c/ \Dt \rho_{0 R}$,
$m_{0R}^c/ \Dt \rho_{0 R}$ start to decrease at about 
$\la_{1\La} \simeq 2$.  

In summary,
the above results show that though the short distance 
potential $U_{\La}$ indicates a second order phase transition,
the transition becomes first order once fluctuations are taken into
account. This fluctuation induced first order phase transition
is known in four dimensions as the Coleman-Weinberg phenomenon
\cite{Col73}. Previous studies of the three dimensional 
$U(2) \times U(2)$ symmetric model using
the $\eps$-expansion \cite{Pisa84} already indicated that the
phase transition should be a fluctuation induced first order
transition. The validity of the $\eps$-expansion for weak 
first order transitions is, however, not clear a priori
since the expansion parameter is not small
-- there are known cases where it fails 
to predict correctly the order of the transition \cite{AbeHig}.
The question of the order of the phase transition has been
addressed also in lattice studies \cite{Dreher92Shen94} and in
high-temperature expansion \cite{Khl95}. All studies are 
consistent with the first order nature of the 
transition and with the absence of nonperturbative infrared
stable fixed points. It is, however, notoriously
difficult to distinguish by these methods between weak
first order and second order transitions. Our method
gives here a clear and unambiguous answer and allows
a detailed quantitative
description of the phase transition. The universal form
of the equation of state for weak first order phase
transitions is presented in section \ref{sce}.

In the following we specify the scale $k_f$ 
for which we have given
the effective average potential $U_{k}$.
We observe that $U_k$ depends
strongly on the infrared cutoff $k$ as long as $k$ is larger
than the scale $k_2$ where 
the second minimum of the potential
appears. Below $k_2$ the two minima start to become
almost degenerate for $T$ near $T_{c}$ and the running
of $\rho_0(k)$ stops rather soon. The
nonvanishing value of $k_2$ induces
a physical infrared cutoff and represents a characteristic
scale for the first order phase transition.  
We stop the integration of the flow equation
for the effective average potential at a scale
$k_f < k_2$ which is determined in terms of the 
curvature (mass term) at the top
of the potential barrier that separates the two local minima
of $U_k$ at the origin and at $\rho_{0}(k)$.
The top of the potential barrier at
$\rho_B(k)$ is determined
by
\be
U_k^{\pri}(\rho_B)=0 \label{bar}
\ee
for $0 < \rho_B(k) < \rho_0(k)$ and for the 
renormalized mass term at 
$\rho_B(k)$ one obtains 
\be
m_{B,R}^2(k)=2 Z_k^{-1} \rho_B U_k^{\pri\pri}(\rho_B) < 0. 
\label{mbr}
\ee  
We fix our final value for the running by 
\be
\ds{\frac{k_f^2-|m_{B,R}^2(k_f)|}{k_f^2}}=0.01
\label{fix}
\ee 
For this choice the coarse grained effective potential $U_{k_f}$
essentially includes all fluctuations with momenta larger than the
mass $|m_{B,R}|$ at the top of the potential barrier. It is a
nonconvex function which is the appropriate quantity for the 
study of physical
processes such as tunneling or inflation.
The nonconvex part of $U_k$ is
considered in detail in section \ref{coarse}. 
There we will also discuss that 
the appropriate choice for a coarse graining
scale $k$ is often far from obvious.\\

\sect{
$\!\!\!\!\!$Scale dependence of the effective average potential 
\label{scale}}

In this section we derive the flow equation for
the effective average potential $U_k$.
We point out a related
investigation of the four dimensional
$SU(N) \times SU(N)$ symmetric
linear sigma model coupled to fermions that has been
studied previously within the framework of the effective
average action \cite{Ju95-7}. 
There the flow equations for a polynomial approximation of
$U_k$ can be found for arbitrary dimension $d$. 
To study the equation of state of the three dimensional
theory we keep here the most general form for $U_k$.

We define $U_k$ by evaluating the average action 
for a constant field with $\Gm_k=\Omega U_k$ where
$\Omega$ denotes the  
volume. To evaluate the r.h.s.\ of (\ref{ERGE}) with the
ansatz (\ref{Ansatz}) we expand $\Gm_k$  
around a constant background configuration.
With the help of the $U(2)\times
U(2)$ transformations the matrix field $\vp$ can be turned  
into a standard diagonal form with real nonnegative eigenvalues.
Without loss of generality the evolution
equation for the effective
potential can therefore be obtained by calculating the trace 
in (\ref{ERGE}) for small field fluctuations 
$\chi_{ab}$ around a constant
background configuration which is real and
diagonal, 
\be
\vp_{ab}=\vp_{a} \delta_{ab} 
\,\, , \quad \vp^*_{a}=\vp_{a}\label{ConstConfig}.
\ee
We separate the fluctuation field into its real and
imaginary part, $\chi_{ab}=\frac{1}{\sqrt{2}}
(\chi_{Rab}+i\chi_{Iab})$ and perform the second functional
derivatives of $\Gm_k$ with respect to the
eight real components.
For the constant configuration
(\ref{ConstConfig}) it turns out that $\Gm_k^{(2)}$ has a 
block diagonal form because mixed derivatives with 
respect to real and imaginary parts of the field vanish.
The remaining submatrices 
$\dt^2\Gm_k/\dt\chi_R^{ab}\dt\chi_R^{cd}$ and 
$\dt^2\Gm_k/\dt\chi_I^{ab}\dt\chi_I^{cd}$ can be
diagonalized in order to find
the inverse of $\Gm_k^{(2)} + R_k$ under the trace
occuring in eq.\ (\ref{ERGE}).
Here the momentum independent part
of $\Gm_k^{(2)}$ defines the mass matrix
by the second functional derivatives
of $U_k$. 
The eight eigenvalues of the mass matrix are
\bea
(M_1^{\pm})^2 &=& U_k^{\pri}+2\left(\rho \pm
(\rho^2-\tau)^{1/2}\right)\prl_{\tau}U_k \,,\nnn
(M_2^{\pm})^2 &=& U_k^{\pri}\pm 2 \tau^{1/2} \prl_{\tau}U_k
\label{MassEigena} 
\eea
corresponding to second derivatives with respect to $\chi_I$
and
\bea
(M_3^{\pm})^2 &=& (M_1^{\pm})^2 \,,\nnn
(M_4^{\pm})^2 &=& U_k^{\pri}+\rho U_k^{\pri\pri}
+2 \rho \prl_{\tau} U_k
+4 \tau \prl_{\tau} U_k^{\pri} + 4 \rho \tau \prl_{\tau}^2 U_k 
\nnn &&
\pm \left\{ \tau \left( U_k^{\pri\pri} + 4 \prl_{\tau} U_k
+ 4 \rho \prl_{\tau} U_k^{\pri} + 4 \tau \prl_{\tau}^2 U_k \right)^2 
\right.\nnn
&& \left.
+ \left(\rho^2-\tau\right) \left( U_k^{\pri\pri} 
- 2 \prl_{\tau} U_k 
-4 \tau \prl_{\tau}^2 U_k \right)^2 \right\}^{1/2}
\label{MassEigen}
\eea
corresponding to second derivatives with respect to $\chi_R$.
Here the eigenvalues are expressed in terms of the 
invariants $\rho$ and $\tau$ using
\be
\vp_1^2=\hal(\rho+\tau^{1/2}),\quad
 \vp_2^2=\hal(\rho-\tau^{1/2})
\ee
and we adopt the convention that a prime on $U_k(\rho,\tau)$
denotes the derivative with respect to $\rho$ at fixed $\tau$ 
and $k$ and $\prl_{\tau}^n U_k \equiv \prl^nU_k/(\prl\tau)^n$.

The flow equation for the effective average potential is
simply expressed in terms of the mass eigenvalues
\bea
\lefteqn{ \ds{\prlt U_k(\rho,\tau)} = \ds{
 \hal\iddq\prlt R_k(q)}}\nnn
 &&\ds{ \left\{\frac{2}{P_k(q)+(M_1^+(\rho,\tau))^2}+
 \frac{2}{P_k(q)+(M_1^-(\rho,\tau))^2}
+\frac{1}{P_k(q)+(M_2^+(\rho,\tau))^2}\right.} \nnn
 &&+ \ds{ \left.
 \frac{1}{P_k(q)+(M_2^-(\rho,\tau))^2}+
 \frac{1}{P_k(q)+(M_4^+(\rho,\tau))^2}+
 \frac{1}{P_k(q)+(M_4^-(\rho,\tau))^2} \right\}  }.
 \label{UkEvol}
\eea
On the right hand side of the evolution equation appears the 
(massless) inverse average propagator 
\be
 P_k(q)=Z_k q^2+R_k(q)=\frac{Z_k q^2}{1-e^{-q^2/k^2}}
 \label{Propagator}
\ee
which incorporates the infrared cutoff function $R_k$ given
by eq.\ (\ref{Rk(q)}).
The only approximation so far is due to the derivative 
expansion (\ref{Ansatz}) of $\Gm_k$ which
enters into the flow equation 
(\ref{UkEvol}) through the form of $P_k$.
The mass eigenvalues (\ref{MassEigena}) and (\ref{MassEigen}) 
appearing in the above flow equation are exact since we have kept
for the potential the most general form $U_k(\rho,\tau)$.\\

{\bf Spontaneous symmetry breaking and mass spectra}

In the following we consider spontaneous symmetry 
breaking patterns
and the corresponding mass spectra for
a few special cases.
For the origin at $\vp_{ab}=0$ all eigenvalues equal 
$U_k^{\pri}(0,0)$. If the origin is the absolute minimum of the
potential we are in the symmetric regime where all excitations have
mass squared $U_k^{\pri}(0,0)$. 

Spontaneous symmetry breaking to the diagonal $U(2)$ subgroup
of $U(2) \times U(2)$ 
can be observed for a field configuration which is proportional
to the identity matrix, i.e. $\vp_{ab}=\vp \delta_{ab}$.
The invariants $(\ref{Invariants})$ take on values  
$\rho=2 \vp^2$ and $\tau=0$. The relevant
information for this symmetry breaking pattern is contained
in $U_k(\rho) \equiv U_k(\rho,\tau=0)$.
In case of spontaneous symmetry breaking there is a 
nonvanishing value for the minimum $\rho_0$ of the potential.
With $U_k^{\pri}(\rho_0)=0$ one finds the expected four 
massless Goldstone bosons with 
$(M_1^-)^2=(M_2^{\pm})^2=(M_3^-)^2=0$.
In addition there are three massive scalars in the 
adjoint representation of the unbroken diagonal
$SU(2)$ with mass squared 
$(M_1^+)^2=(M_3^+)^2=(M_4^-)^2=
4 \rho_0 \prl_{\tau} U_k$ and one singlet with mass squared
$(M_4^+)^2=2 \rho_0 U_k^{\pri\pri}$. The situation
corresponds to chiral symmetry breaking in two flavor
QCD in absence of quark masses and the chiral anomaly.
The Goldstone modes are the pseudoscalar pions and the
$\eta$ (or $\eta^{\prime}$), the scalar triplet has the
quantum numbers of $a_0$ and the singlet is the so-called
$\sigma$-field.

Another interesting case is the spontaneous symmetry breaking
down to a residual
$U(1) \times U(1) \times U(1)$ subgroup of $U(2) \times U(2)$
which can be observed for the configuration 
$\vp_{ab} =\vp \dt_{a1} \dt_{ab}$ ($\rho=\vp^2$, 
$\tau=\vp^4=\rho^2$). 
Corresponding to the number of broken generators 
one observes the five massless Goldstone bosons 
$(M_1^{\pm})^2=(M_2^+)^2=(M_3^{\pm})^2=0$ for the minimum 
of the potential at 
$U_k^{\pri}+2 \rho_0 \prl_{\tau} U_k = 0$. In addition there are two
scalars with mass squared 
$(M_2^-)^2=(M_4^-)^2=U_k^{\pri}-2 \rho_0 \prl_{\tau} U_k$
and one with 
$(M_4^+)^2=U_k^{\pri}+2 \rho_0 U_k^{\pri\pri} + 6 \rho_0
\prl_{\tau} U_k + 8 \rho_0^2 \prl_{\tau} U_k^{\pri} + 8 \rho_0^3
\prl_{\tau}^2 U_k$. 

We finally point out the special case where 
the potential is independent of the second invariant
$\tau$. In this case there is an enhanced 
$O(8)$ symmetry instead of 
$U(2) \times U(2)$. With $\prl^n_{\tau}U_k \equiv 0$
and $U_k^{\pri}(\rho_0)=0$ one observes the expected
seven massless Goldstone bosons and one massive mode with mass
squared $2 \rho_0 U_k^{\pri\pri}$.\\

{\bf Scaling form of the flow equation}

For the $O(8)$ symmetric  
model in the limit $\bar{\la}_{2\La}=0$
one expects a region of
the parameter space which is characterized by renormalized
masses much smaller than the ultraviolet cutoff
or inverse microscopic length scale of the theory. 
In particular, in the absence of a mass scale one
expects a scaling behavior of the effective average
potential $U_k$. The behavior of $U_k$ at or near a second
order phase transition is most conveniently studied
using the scaling form of the evolution equation.
This form is also appropriate for an investigation that has 
to deal with weak first order phase transitions as 
encountered in the present model for 
$\bar{\la}_{2\La} > 0$. The remaining
part of this section is devoted to the derivation of
the scaling form (\ref{DlessEvol}) of 
the flow equation (\ref{UkEvol}).

In the present form of eq.\ (\ref{UkEvol}) the r.h.s.\
shows an explicit dependence on the scale $k$ once the
momentum integration is performed.
By a proper choice of 
variables we cast the evolution equation into a form
where the scale no longer appears explicitly. We
introduce a dimensionless potential $u_k=k^{-d}  U_k$
and express it in terms of dimensionless 
renormalized fields 
\bea
\trho &=& Z_k k^{2-d} \rho\,,\nnn
\ttau &=& Z_k^2 k^{4-2 d} \tau\,.  
\label{dlessfield}
\eea
The derivatives of $u_k$ are given by
\be
\prl_{\ttau}^n u_k^{(m)}(\trho,\ttau)=Z_k^{-2 n-m} 
k^{(2 n+m-1) d-4 n-2 m} \prl_{\tau}^n U_k^{(m)}
\left(\rho,\tau\right)\,.
\label{dlessder}
\ee  
(Note that $u_k^{(m)}$ denotes $m$ derivatives
with respect to $\trho$ at fixed $\ttau$ and $k$, while 
$U_k^{(m)}$ denotes $m$ derivatives
with respect to $\rho$ at fixed $\tau$ and $k$). With
\bea
\ds{\prlt u_k(\trho,\ttau)_{|\trho,\ttau}} &=& 
\ds{-d u_k(\trho,\ttau) +
(d-2+\eta) \trho u_k^{\pri}(\trho,\ttau)
 + (2 d-4+2 \eta) \ttau
\prl_{\ttau} u_k(\trho,\ttau)}\nnn 
&&+ \ds{k^{-d} \prlt 
U_k\left(\rho(\trho),\tau(\ttau)\right)_{|\rho,\tau}}
\label{dlesstrans}    
\eea
one obtains from (\ref{UkEvol})
the evolution equation for the 
dimensionless potential. Here the anomalous dimension 
$\eta$ arises from the $t$-derivative acting on $Z_k$ and 
is given by eq.\ (\ref{Eta}). It is convenient to introduce
dimensionless integrals by 
\be
\ds{\hal\iddq\frac{\prl R_k}{\prl t}}\frac{1}{P_k+Z_k k^2 \om}
=2 v_d k^d \left[ l^d_0(\om) - \eta \hat{l}^d_0(\om) \right]
\ee
where 
\be
v_d^{-1} = 2^{d+1} \pi^\frac{d}{2} \Gm\left(\frac{d}{2}\right). 
\ee
The explicit form of $l^d_0$ and $\hat{l}^d_0$
reads
\bea
\ds{l^d_0(\om)} &=&
\ds{-  \int^{\infty}_0 dy y^{\frac{d}{2}+1} 
\frac{\partial r(y)}{\partial y}
\left[ y(1+r(y)) + \om \right]^{-1}}, 
\nnn 
\ds{\hat{l}^d_0(\om)}
&= &\ds{ \frac{1}{2} \int^{\infty}_0 dy y^{\frac{d}{2}} }
r(y)
\left[ y(1+r(y)) + \om \right]^{-1}
\label{threshld0} 
\eea
with the dimensionless infrared cutoff function
\be
r(y) = \frac{e^{-y}}{1 - e^{-y}}.
\label{ry} 
\ee
The behavior of these functions is discussed in section 
\ref{solv}
where we also consider
their derivatives with respect to $\om$. Using the notation
\be
l^d_0(\om;\eta )=l^d_0(\om) - \eta \hat{l}^d_0(\om)
\ee  
one obtains the flow equation for the dimensionless potential
\bea
\lefteqn{\ds{\prlt u_k(\trho,\ttau)} = 
\ds{-d u_k(\trho,\ttau) +
(d-2+\eta) \trho u_k^{\pri}(\trho,\ttau)
 + (2 d-4+2 \eta) \ttau
\prl_{\ttau} u_k(\trho,\ttau)}}\nnn
&&+ \ds{4 v_d l^d_0\left((m_1^+(\trho,\ttau))^2;\eta\right)
+ 4 v_d l^d_0\left((m_1^-(\trho,\ttau))^2;\eta\right)
+ 2 v_d l^d_0\left((m_2^+(\trho,\ttau))^2;\eta\right)}\nnn
&&+ \ds{2 v_d l^d_0\left((m_2^-(\trho,\ttau))^2;\eta\right)
+2 v_d l^d_0\left((m_3^+(\trho,\ttau))^2;\eta\right)
+2 v_d l^d_0\left((m_3^-(\trho,\ttau))^2;\eta\right)
}
\label{DlessEvol}    
\eea
where the dimensionless mass terms are related to 
(\ref{MassEigen}) according to 
\be
(m_i^{\pm}(\trho,\ttau))^2=\frac
{\left(M_i^{\pm}(\rho(\trho),\tau(\ttau))\right)^2}{Z_k k^2}.
\label{massrel}
\ee
Eq.\ (\ref{DlessEvol}) is the scaling form of the
flow equation we are looking for. For a 
$\ttau$-independent potential it reduces to the 
evolution equation for the $O(8)$ symmetric model
\cite{TW94-1,Mor}. The potential $u_k$ at
a second order phase transition is given by
a $k$-independent (scaling) solution 
$\prl u_k/\prl t=0$ \cite{TW94-1,Mor}. For this solution 
all the $k$-dependent functions
on the r.h.s.\ of eq.\
(\ref{DlessEvol}) become independent
of $k$. For a weak
first order phase transition these functions
will show a weak $k$-dependence for $k$
larger than the inherent mass scale of the
system (cf.\ section \ref{rg}). There is no
particular advantage of the scaling form of
the flow equation for strong first order phase transitions.\\

\sect{Solving the flow equation \label{solv}}

Eq.\ (\ref{DlessEvol}) describes the scale dependence
of the effective average potential $u_k$
by a nonlinear partial differential equation for the
three variables $t$, $\trho$ and $\ttau$. 
A major difficulty for its analytical
study is that the integral $l^d_0(\om;\eta)$
(cf.\ eq.\ (\ref{threshld0}))
can be done analytically only for certain limits
of the arguments. The complicated form of the equation
therefore suggests a numerical solution. 
We will use a method that relies on a simultaneous
expansion of the potential around a number 
of field values $\trho_i,\ttau_j$, $i=1,\ldots,l$,
$j=1,\ldots,l^{\pri}$ for given numbers $l,l^{\pri}$. 
The expansions around
different points are matched to obtain the general field
dependence of the potential. As a consequence we
cast the partial differential equation (\ref{DlessEvol})
into a system of ordinary differential equations.
The method is developed in \cite{ABBTW95} for a computation of 
the critical equation of state for $O(N)$ symmetric
models \cite{BTW95}. The generalization to the present model
is described below. This approach has some
favorable aspects. The main advantage is that it 
allows the integration of the differential equations using
a standard Runge-Kutta algorithm without the occurance
of numerical instabilities. (See e.g.\ \cite{ABBTW95} for 
instability problems with standard discretised versions
of partial differential equations in the context of flow 
equations). The coupled set of
ordinary differential equations describes the
flow of couplings defined as derivatives of the potential
at given points, e.g.\ at the minimum of the potential.
These couplings often allow direct physical interpretation 
and some of their properties can be read off from the
analytic structure of their flow equations.
We will exploit this fact to explain the 
results we obtain from the numerical solution
in section \ref{rg}.

We concentrate in the following on spontaneous 
symmetry breaking with a residual $U(2)$ symmetry
group. As we have already pointed out in section \ref{model}
this symmetry breaking can be observed for a
configuration which is proportional to the 
identity and we have $\ttau=0$. In this case
the eigenvalues (\ref{MassEigena}) and
(\ref{MassEigen}) of the mass matrix 
with (\ref{massrel}) are given by
\bea
(m_1^-)^2 &=& (m_2^{\pm})^2 = (m_3^-)^2 = u_k^{\pri}\,,\nnn 
(m_1^+)^2 &=& (m_3^+)^2 = (m_4^-)^2 = 
u_k^{\pri}+4 \trho \prl_{\ttau}u_k\,,\nnn
(m_4^+)^2 &=& u_k^{\pri} + 2 \trho u_k^{\pri\pri}\,
\eea
and on the r.h.s.\ of the partial differential equation 
(\ref{DlessEvol}) for $u_k(\trho) \equiv u_k(\trho,\ttau=0)$
only the functions $u_k^{\pri}(\trho)$, $u_k^{\pri\pri}(\trho)$
and $\prl_{\ttau}u_k(\trho)$ appear. At fixed 
$\trho=\trho_i$ the $k$-dependence of $u_k$ is
then determined by the couplings 
$u_k^{\pri}(\trho_i)$, $u_k^{\pri\pri}(\trho_i)$
and $\prl_{\ttau}u_k(\trho_i)$. We determine these
couplings through the use of flow equations which
are obtained by taking the derivative in eq.\ 
(\ref{DlessEvol}) with respect to $\trho$ and
$\ttau$ evaluated at $\trho=\trho_i$, $\ttau=0$.
These flow equations for $u_k^{\pri}$, $u_k^{\pri\pri}$
and $\prl_{\ttau}u_k$ involve also
higher derivatives of the
potential as $u_k^{\pri\pri\pri}$,
$u_k^{(4)}$, $\prl_{\ttau}u_k^{\pri}$
and $\prl_{\ttau}u_k^{\pri\pri}$. The procedure
will be to evaluate the flow
equations for $u_k^{\pri}$ and $u_k^{\pri\pri}$
at different points $\trho_i,\ttau=0$ for $i= 1,\ldots ,l$
and to estimate the higher $\trho$-derivatives 
appearing on the right hand side of the flow equations
by imposing matching conditions. The same procedure
can be applied to $\prl_{\ttau}u_k$ and in order to obtain
an equivalent matching we
also consider the flow equation for $\prl_{\ttau}u_k^{\pri}$.
One could proceed in a similar way for $\prl_{\ttau}^2u_k$ which
appears on the right hand side of the evolution equation
of $\prl_{\ttau}u_k$ and so on.
However, since we are 
interested in the $\trho$-dependence of the potential at 
$\ttau=0$ we shall use a truncated expansion\footnote{
In principle one could also consider points with $\ttau \not = 0$ 
in the neighborhood
of $\ttau=0$ and use the additional information to estimate
the higher $\ttau$-derivatives as it is done for the
higher $\trho$-derivatives.} in $\ttau$ with 
\be
\prl_{\ttau}^n u_k(\trho,\ttau=0)=0 \quad \mbox{for}
\quad n \ge 2.
\label{Truncation}
\ee
In three space dimensions the neglected  
($\trho$-dependent) operators have 
negative canonical mass dimension.
We make no expansion in terms of $\trho$
since the general $\trho$-dependence allows a description
of a first order phase transition where a second local minimum
of $u_k(\trho)$ appears. 
The approximation (\ref{Truncation}) only affects  
the flow equations for $\prl_{\ttau}u_k$ and 
$\prl_{\ttau}u_k^{\pri}$ (cf.\ eqs.\ (\ref{La2}) and 
(\ref{La2Pri})). The form of the flow equations for
$u_k^{\pri}$ and $u_k^{\pri\pri}$ is not affected by the 
truncation (cf.\ eqs.\ (\ref{Epsi}) and 
(\ref{La1})). 
From $u_k^{\pri}$ we obtain the effective average
potential $u_k$ by simple integration.
We have tested the sensitivity of our results
for $u_k^{\pri}$ to a change in $\prl_{\ttau}u_k$ by neglecting
the $\trho$-dependence of the $\ttau$-derivative. We observed
no qualitative change of the results. We expect that the main
truncation error is due to the derivative expansion (\ref{Ansatz}) for
the effective average action.  

To simplify notation we introduce 
\bea
\eps(\trho)&=&u_k^{\pri}(\trho,\ttau=0),\nnn
\la_1(\trho)&=&u_k^{\pri\pri}(\trho,\ttau=0),\nnn
\la_2(\trho)&=&4 \prl_{\ttau} u_k(\trho,\ttau=0).
\label{notation}
\eea
Higher derivatives are denoted by primes on the 
$\trho$-dependent quartic
'couplings',
i.e.\ $\la_1^{\pri}=u_k^{\pri\pri\pri}$,
$\la_2^{\pri}=\prl_{\ttau}u_k^{\pri}$ etc. It is
convenient to introduce functions $l^d_n(\om;\eta)$
that can be related
to $l^d_0(\om;\eta)$
(\ref{threshld0}) by differentiation with respect to the
mass argument: 
\bea
\ds{l^d_1(\om;\eta)} 
&=& l^d_1(\om) - \eta \hat{l}^d_1(\om)\nnn
&=& \ds{- \frac{\prl}{\prl \om} l^d_0(\om;\eta),} \nnn
l^d_{n+1}(\om;\eta) &=& \ds{-\frac{1}{n} 
\frac{\prl}{\prl \om} l^d_n(\om;\eta)}\quad
\mbox{for}\quad n \ge 1.
\eea
The explicit form of $l^d_n$ and $\hat{l}^d_n$ is given in
eq.\ (\ref{threshldn}) in the appendix. 
For $\om \ge -1$ they are positive,
monotonically decreasing functions of $\om$.
In leading order $l^d_n$ and $\hat{l}^d_n$ vanish
$\sim \om^{-(n+1)}$ for arguments $\om \gg 1$. They 
introduce a 'threshold' behavior that accounts for
the decoupling of modes with mass squared larger
than the infrared cutoff $Z_k k^2$. For 
vanishing argument they are of order unity. As $\om \to -1$
the functions $l^d_n$, $\hat{l}^d_n$ exhibit
a pole for $d < 2(n+1)$. The pole structure is discussed
in the appendix. We also use two-parameter functions
$l^d_{n_1,n_2}(\om_1,\om_2;\eta)$ \cite{Ju95-7}. 
For $n_1=n_2=1$ their relation to the functions
$l^d_n(\om;\eta)$ can be expressed as
\bea
l^d_{1,1}(\om_1,\om_2;\eta) &=&
\ds{\frac{1}{\om_2 - \om_1} \left[ l^d_1(\om_1;\eta)
- l^d_1(\om_2;\eta) \right]} \quad 
\mbox{for}\quad \om_1 \ne \om_2, \nnn
l^d_{1,1}(\om ,\om ;\eta)&=& l^d_2(\om ;\eta)
\eea
and
\be
l^d_{n_1 + 1, n_2}(\om_1,\om_2;\eta) =\ds{
-\frac{1}{n_1}  \frac{\prl}{\prl \om_1}
l^d_{n_1,n_2}(\om_1,\om_2;\eta)},\quad
l^d_{n_1,n_2}(\om_1,\om_2;\eta) =  
l^d_{n_2,n_1}(\om_2,\om_1;\eta).                         
\ee

With the help of these functions the scale dependence of 
$\eps$ is described by
\bea
\ds{\frac{\prl \eps}{\prl t}} &=& 
\ds{(-2+\eta) \eps +
(d-2+\eta) \trho \la_1 - 6 v_d (\la_1+\la_2+\trho \la_2^{\pri}) 
l^d_1(\eps+\trho \la_2;\eta)}\nnn
&&- \ds{2 v_d (3 \la_1+2 \trho \la_1^{\pri}) 
l^d_1(\eps+2\trho \la_1;\eta)   
-8 v_d \la_1 l^d_1(\eps ;\eta)}
\label{Epsi}
\eea
and for $\la_1$ one finds
\bea
\ds{\frac{\prl \la_1}{\prl t}} &=& 
\ds{(d-4+2\eta) \la_1 +
(d-2+\eta) \trho \la_1^{\pri} }\nnn
&&+\ds{6 v_d \left[ (\la_1+\la_2+\trho \la_2^{\pri})^2
l^d_2(\eps+\trho \la_2;\eta)
-(\la_1^{\pri}+2\la_2^{\pri}+\trho \la_2^{\pri\pri})
l^d_1(\eps+\trho \la_2;\eta)\right]}\nnn
&&+\ds{2 v_d \left[(3\la_1+2\trho \la_1^{\pri})^2
l^d_2(\eps+2\trho \la_1;\eta)
-(5\la_1^{\pri}+2\trho \la_1^{\pri\pri})
l^d_1(\eps+2\trho \la_1;\eta)\right] }\nnn
&&+\ds{8 v_d \left[(\la_1)^2 l^d_2(\eps;\eta)
-\la_1^{\pri}l^d_1(\eps;\eta)\right]}.
\label{La1}                      
\eea
Similarly the scale dependence of $\la_2$ 
is given by
\bea
\ds{\frac{\prl \la_2}{\prl t}} &=& 
\ds{(d-4+2\eta) \la_2 +
(d-2+\eta) \trho \la_2^{\pri}       
- 4 v_d (\la_2)^2 l^d_{1,1}(\eps+\trho \la_2,\eps;\eta) }\nnn
&&+\ds{2 v_d \left[ 3 (\la_2)^2+12\la_1\la_2+8 \trho
\la_2^{\pri}(\la_1+\la_2)+4\trho^2 (\la_2^{\pri})^2\right]
l^d_{1,1}(\eps+\trho \la_2,\eps+2\trho \la_1;\eta) }\nnn
&&- \ds{14 v_d \la_2^{\pri} l^d_1(\eps+\trho \la_2;\eta)
-2 v_d (5 \la_2^{\pri}+2\trho \la_2^{\pri\pri})
l^d_1(\eps+2\trho \la_1;\eta) }\nnn
&&+\ds{2 v_d \left[ (\la_2)^2 l^d_2(\eps;\eta)-4 \la_2^{\pri}
l^d_1(\eps;\eta)\right] }
\label{La2} 
\eea
and for $\la_2^{\pri}$ it reads
\bea
\ds{\frac{\prl \la_2^{\pri}}{\prl t}} &=&
(2 d-6+3\eta) \la_2^{\pri} + 
(d-2+\eta) \trho \la_2^{\pri \pri} \nnn
&&
+4 v_d (\la_2)^2 \Big[(\la_1+\la_2+\trho \la_2^{\pri})
l^d_{2,1}(\eps+\trho \la_2,\eps;\eta)+
\la_1 l^d_{1,2}(\eps+\trho \la_2,\eps;\eta)\Big]\nnn
&&
-2 v_d \Big[3 \la_2 (4 \la_1+\la_2) + 4\trho \la_2^{\pri}
(2\la_1+2\la_2+\trho \la_2^{\pri})\Big]
\Big[(\la_1+\la_2+\trho \la_2^{\pri})\nnn
&&
l^d_{2,1}(\eps+\trho \la_2,\eps+2\trho \la_1;\eta)+
(3 \la_1+ 2 \trho \la_1^{\pri})
l^d_{1,2}(\eps+\trho \la_2,\eps+2\trho \la_1;\eta)\Big]\nnn
&&
+4 v_d \Big[\la_2^{\pri}(7 \la_2  +10 \la_1) + 
6 \la_2 \la_1^{\pri} + 4 \trho \Big(\la_2^{\pri \pri}
(\la_1+\la_2+\trho \la_2^{\pri})
+\la_2^{\pri} (2\la_2^{\pri}+\la_1^{\pri})\Big)\Big]\nnn
&&
l^d_{1,1}(\eps+\trho \la_2,\eps+2\trho \la_1;\eta)
-8 v_d \la_2 \la_2^{\pri} l^d_{1,1}(\eps+\trho \la_2,\eps;\eta)
\nnn &&
+2 v_d (3 \la_1 + 2\trho \la_1^{\pri})(5 \la_2^{\pri}+
2\trho\la_2^{\pri \pri})
l^d_{2}(\eps+2\trho \la_1;\eta)
+14 v_d \la_2^{\pri} (\la_1+\la_2+\trho \la_2^{\pri})\nnn
&&
l^d_{2}(\eps+\trho \la_2;\eta)-2 v_d
(7 \la_2^{\pri \pri} + 2 \trho \la_2^{\pri \pri \pri})
l^d_{1}(\eps+2\trho \la_1;\eta)-14 v_d
\la_2^{\pri \pri} l^d_{1}(\eps+\trho \la_2;\eta)\nnn
&&
-4 v_d (\la_2)^2 \la_1 l^d_{3}(\eps;\eta)
+4 v_d \la_2^{\pri} (2 \la_1+\la_2) l^d_{2}(\eps;\eta)
-8 v_d \la_2^{\pri \pri} l^d_{1}(\eps;\eta)
\label{La2Pri} .
\eea
We evaluate the above flow equations
at different points $\trho_i$ for $i= 1,\ldots ,l$ 
and use a set of 
matching conditions that has been proposed in ref.\ \cite{ABBTW95}.
The generalization
of these conditions to the present model is obtained by
considering fourth order polynomial expansions of 
$\eps (\trho)$ and $\la_2 (\trho)$ around some arbitrary
point $\trho_i$,
\bea
(\eps)_i (\trho) &=& \ds{ \eps_i+\la_{1,i}
(\trho-\trho_i)+\hal \la_{1,i}^{\pri} (\trho-\trho_i)^2
+\frac{1}{6} \la_{1,i}^{\pri \pri} (\trho-\trho_i)^3}
,\nnn
(\la_2)_i (\trho) &=&\ds{ \la_{2,i} +\la_{2,i}^{\pri}
(\trho-\trho_i) + \hal \la_{2,i}^{\pri \pri} (\trho-\trho_i)^2
+\frac{1}{6} \la_{2,i}^{\pri \pri \pri} (\trho-\trho_i)^3}
\eea
with $\eps_i=\eps (\trho_i)$, $\la_{2,i}=\la_2 (\trho_i)$ etc.
Using similar expressions for
\be
(\la_1)_i (\trho) = \frac{\prl}{\prl \trho}
(\eps)_i (\trho)\quad,
\qquad  (\la_2^{\pri})_i (\trho) = \frac{\prl}{\prl \trho} 
(\la_2)_i (\trho)
\ee
the matching is done by imposing continuity 
at half distance between neighboring expansion points,
\bea 
\ds{ (\eps)_i \left( \frac{\trho_i+\trho_{i+1}}{2} \right)} =   
\ds{ (\eps)_{i+1} \left( \frac{\trho_i+\trho_{i+1}}{2} \right) 
}&,&
\ds{ (\la_1)_i \left( \frac{\trho_i+\trho_{i+1}}{2} \right)} =
\ds{ (\la_1)_{i+1} \left( \frac{\trho_i+\trho_{i+1}}{2} \right), 
}\nnn
\ds{ (\la_2)_i \left( \frac{\trho_i+\trho_{i+1}}{2} \right)} =   
\ds{ (\la_2)_{i+1} \left( \frac{\trho_i+\trho_{i+1}}{2} \right) 
}&,& 
\ds{ (\la_2^{\pri})_i \left( \frac{\trho_i+\trho_{i+1}}{2} \right)} =
\ds{ (\la_2^{\pri})_{i+1} \left( \frac{\trho_i+\trho_{i+1}}{2} \right) 
} \label{match1}
\eea 
for $i=1,\ldots ,l-1$ and 
\bea
\ds{ (\la_1^{\pri})_j \left( \frac{\trho_j+\trho_{j+1}}{2} \right)} =  
\ds{ (\la_1^{\pri})_{j+1} \left( \frac{\trho_j+\trho_{j+1}}{2} \right) 
}&,&
\ds{ (\la_2^{\pri \pri})_j \left( \frac{\trho_j+\trho_{j+1}}{2}
\right)} =
\ds{ (\la_2^{\pri \pri})_{j+1} \left( \frac{\trho_j+\trho_{j+1}}{2} 
\right) 
}\label{match2}
\eea
for the initial and end points, $j=1$ and $j=l-1$. Together these
$4 (l-1)$ conditions (\ref{match1})
for all $l-1 \ge 2$ intermediate points and the four conditions
(\ref{match2})
make up two independent algebraic systems of each
$2 l$ equations. From the first set of equations
one obtains a unique solution 
for $\la_{1,i}^{\pri}$ and $\la_{1,i}^{\pri \pri}$.
The second set is identical in structure and 
$\la_{2,i}^{\pri \pri}$, $\la_{2,i}^{\pri \pri \pri}$
can be obtained
from the solutions for $\la_{1,i}^{\pri}$ and 
$\la_{1,i}^{\pri \pri}$
with the substitutions $\eps_j \to \la_{2,j}$, $\la_{1,j} \to   
\la_{2,j}^{\pri}$, $\la_{1,j}^{\pri} \to \la_{2,j}^{\pri \pri}$,
$\la_{1,j}^{\pri \pri} \to \la_{2,j}^{\pri \pri \pri}$ for 
$j=1,\ldots,l$. With the help of these algebraic solutions
we eliminate $\la_1^{\pri}(\trho)$, $\la_1^{\pri\pri}(\trho)$,
$\la_2^{\pri\pri}(\trho)$ and $\la_2^{\pri\pri\pri}(\trho)$
in the flow equations (\ref{Epsi}) - (\ref{La2Pri}) for all
$l$ points $\trho_i$.\footnote{
The algebraic solutions $\la_{1,i}^{\pri}$ and $\la_{1,i}^{\pri \pri}$
(equivalently $\la_{2,i}^{\pri \pri}$ and 
$\la_{2,i}^{\pri \pri \pri}$) do incorporate information from
the whole range of points $\trho_j$ with $j=1,\ldots,l$. It is
a feature of the matching conditions (\ref{match1}),
(\ref{match2}) that the contributions from points $\trho_{j \not = i}$
to $\la_{1,i}^{\pri}$, $\la_{1,i}^{\pri \pri}$ rapidly
decrease with increasing $|i-j|$. For equal spacings between
neighboring expansion points contributions from points
$\trho_j$ with $j > i+1$ $(j < i-1)$ are typically suppressed
by a factor $\ltap 10^{-|i-j|+1}$ as compared to the contribution
from the nearest neighbor point $\trho_{i+1}$ ($\trho_{i-1}$).
As a consequence solutions $\la_{1,i}^{\pri}$,
$\la_{1,i}^{\pri \pri}$ for inner points with $1 \ll i \ll l$
become independent from boundary points. We observe approximate
translational invariance for inner point solutions, i.e.
$\la_{1,i \pm n}^{\pri}$ and $\la_{1,i \pm n}^{\pri \pri}$
are approximately obtained from the solutions 
$\la_{1,i}^{\pri}$ and $\la_{1,i}^{\pri \pri}$ with the 
substitutions $\eps_j \to \eps_{j \pm n}$, 
$\la_{1,j} \to \la_{1,j \pm n}$ for $1 \le j \le l$
if $i$ and $i\pm n$ are sufficiently far away from
the boundaries.
The decoupling from distant points and the translational 
invariance for inner points can be used to obtain
approximate expressions which become useful if a large
number of expansion points is considered. 
We use the exact algebraic solution 
for $l=10$ points. For $l > 10$
we apply the approximate translational 
invariance to 
generate from $\la_{1,5}^{\pri}$, $\la_{1,6}^{\pri}$
additional solutions $\la_{1,5 + 2i}^{\pri}$, 
$\la_{1,6 + 2i}^{\pri}$ for $i=1,\ldots,(l-10)/2$ 
with $l$ even and equivalently  
for $\la_{1,5}^{\pri\pri}$, $\la_{1,6}^{\pri\pri}$.
With $\la_{1,l-3}^{\pri},\ldots,\la_{1,l}^{\pri}$
and $\la_{1,l-3}^{\pri \pri},\ldots,\la_{1,l}^{\pri \pri}$
from the calculation with 10 points one obtains the 
desired generalization.
We have used runs with different choices of $l$
in order to check the stability of the numerical
results.}
Therefore, equations (\ref{Epsi}) - (\ref{La2Pri})
are turned 
into a closed system of $4l$ ordinary differential equations
for the unknowns $\eps(\trho_i)$, $\la_1(\trho_i)$,
$\la_2(\trho_i)$ and 
$\la_2^{\pri}(\trho_i)$. 

If there is a 
minimum of the potential at nonvanishing
$\kp \equiv \trho_0$
we use expansion points that are proportional to
the minimum, i.e.\ $\trho_i = \frac{i-1}{n} \kappa$ with 
$i=1,\ldots,l$ and fixed integer $n$. 
The condition $\eps(\kp)=0$ can be used to 
obtain the  scale dependence of $\kp(k)$:
\bea
\ds{\frac{\mbox{d} \kp}{\mbox{d} t}}
&=&\ds{-[\la_1(\kp)]^{-1}\frac{\prl \eps}{\prl t}}
|_{\trho=\kp}\nnn
&=&\ds{ -(d-2+\eta) \kp 
+ 6 v_d \left(1+\frac{\la_2(\kp) + \kp \la_2^{\pri}(\kp)}
{\la_1(\kp)} \right) l^d_1\left(\kp \la_2(\kp);\eta \right) }\nnn
&+& \ds{2 v_d \left( 3+\frac{2\kp \la_1^{\pri}(\kp)}{\la_1(\kp)}
\right) l^d_1\left(2\kp \la_1(\kp);\eta \right) 
+8 v_d l^d_1\left(0;\eta\right) }.
\label{kappa}
\eea
To make contact with $\beta$-functions for the couplings 
at the potential minimum $\kp$ we point out the relation
\be
\ds{\frac{\mbox{d} \la_{1,2}^{(m)}(\kp)}{\mbox{d} t}}=
\ds{\frac{\prl \la_{1,2}^{(m)} }{\prl t}|_{\trho=\kp}
+ \la_{1,2}^{(m+1)}(\kp) \frac{\mbox{d} \kp}{\mbox{d} t} }.
\label{lakappa}
\ee
Similar relations hold for $\eps(\trho_i)$,
$\la_{1}(\trho_i)$ etc., e.g.\
\bea
\ds{\frac{\mbox{d} \eps(\trho_i)}{\mbox{d} t}}
&=& \ds{\frac{\prl \eps}{\prl t}|_{\trho=\trho_i}
+\frac{i-1}{n} \la_{1}(\trho_i) \frac{\mbox{d} \kp}
{\mbox{d} t}}, \nnn
\ds{\frac{\mbox{d} \la_1(\trho_i)}{\mbox{d} t}}
&=& \ds{\frac{\prl \la_1}{\prl t}|_{\trho=\trho_i}
+\frac{i-1}{n} \la_{1}^{\pri}(\trho_i) \frac{\mbox{d} \kp}
{\mbox{d} t}}. \label{prop}
\eea
We integrate the $4 l-1$ differential equations
(\ref{Epsi}) - (\ref{La2Pri})
for the couplings $\eps(\trho_i)$, $\la_1(\trho_i)$,
$\la_2(\trho_i)$ and 
$\la_2^{\pri}(\trho_i)$ (with $\prl/\prl t$ replaced by
$\mbox{d} / \mbox{d} t$ according to (\ref{prop}))  
and the one for $\kp$ (\ref{kappa})
with a fifth-order Runge-Kutta algorithm using the
embedded fourth-order method for precision control.
The general $\trho$-dependence is recovered
by patching the simultaneous expansions around
different points at half distance between neighboring 
expansion points.\\ 

It remains
to compute the anomalous dimension $\eta$ defined in (\ref{Eta})
which describes the scale
dependence of the wave function renormalization $Z_k$. 
We consider a space dependent distortion of the constant 
background field configuration (\ref{ConstConfig})
of the form
\be
 \vp_{ab}(x) = \vp_a\dt_{ab} +
 \left[\dt\vp e^{-iQx} + \dt\vp^* e^{iQx}\right] 
\Si_{ab}.
 \label{ConfAnDi}
\ee
Insertion of the above configuration into the parametrization 
(\ref{Ansatz}) of $\Gm_k$ yields
\be
 \ds{
 Z_k }=
\ds{
  Z_k(\rho,\tau,Q^2=0) }
 = \ds{
 \hal\frac{1}{\Si_{ab}^*\Si_{ab}}
 \lim_{Q^2 \ra 0}\frac{\prl}{\prl Q^2}
 \frac{\dt \Gm_k}{\dt (\dt\vp \dt\vp^*)}|_{\dt\vp=0}} .
 \label{Z}
\ee 
\\
To obtain 
the flow equation of the wave function renormalization
one expands the effective average action around a configuration
of the form (\ref{ConfAnDi}) and evaluates the r.h.s.\ of eq.\
(\ref{ERGE}). This computation has been done in ref.\ 
\cite{Ju95-7} for a 'Goldstone' configuration with   
\be
 \Si_{ab}=\dt_{a1}\dt_{b2}-\dt_{a2}\dt_{b1}
\ee
and $\vp_a \dt_{ab}=\vp\dt_{ab}$ corresponding to a 
symmetry breaking pattern
with residual $U(2)$ symmetry. The result of ref.\ \cite{Ju95-7}
can be easily generalized to arbitrary fixed field values 
of $\trho$
and we find
\bea
 \ds{ \eta(k)}
&=&\ds{
 4 \frac{v_d}{d}\trho \left[
 4(\la_1)^2 
m_{2,2}^d(\eps,\eps+
2\trho \la_1;\eta) 
  +(\la_2)^2 
m_{2,2}^d(\eps,\eps+
\trho \la_2;\eta) 
 \right] }.
 \label{EtaSkale}
\eea  
The explicit form of the 'threshold' function
\be 
m_{2,2}^d(\om_1,\om_2;\eta)=
m_{2,2}^d(\om_1,\om_2)-\eta\hat{m}_{2,2}^d(\om_1,\om_2)
\ee
can be found in refs.\ \cite{TW94-1,Ju95-7}. 
For vanishing arguments the 
functions $m_{2,2}^d$ and $\hat{m}_{2,2}^d$ are of order
unity. They are symmetric with respect to their arguments
and in leading order $m_{2,2}^d(0,\om) \sim 
\hat{m}_{2,2}^d(0,\om) \sim \om^{-2}$ for $\om \gg 1$. 
According to eq.\ (\ref{zet})
we use $\trho=\kp$ to define the uniform
wave function renormalization
\be
Z_{k} \equiv Z_{k}(\kp).
\label{wfr}
\ee 
We point out that 
according to our truncation of the
effective average action with eq.\ (\ref{EtaSkale}) 
the anomalous
dimension $\eta$ is exactly zero at $\trho = 0$. 
This is an artefact of the truncation and
we expect the symmetric phase
to be more affected by truncation errors than the
spontaneously broken phase. We typically observe
small values for 
$\eta(k)=-\mbox{d} (\ln Z_k)/ \mbox{d} t$ (of the order of 
a few per cent). The smallness of $\eta$ is crucial
for our approximation of a uniform wave function
renormalization to give quantitatively reliable
results for the equation of state. For the universal
equation of state given in sect.\ \ref{sce}
one has $\eta=0.022$ as given by the corresponding
index of the $O(8)$ symmetric 'vector' model.

\sect{Renormalization group flow \label{rg}}

To understand the detailed picture of the phase structure
presented in section \ref{ps} we will consider 
the flow of some characteristic quantities for the effective
average potential as the infrared
cutoff $k$ is lowered. We will always consider in this section
the trajectories for the critical 'temperature', i.e.\
$\dt \kp_{\La}=0$,
and we follow the flow for different values of the
short distance parameters $\la_{1\La}$
and $\la_{2\La}$. The discussion for sufficiently
small $\dt \kp_{\La}$ is analogous.
In particular, we compare the
renormalization group flow of these quantities for a weak and a
strong first order phase transition. In some limiting cases
their behavior can be studied analytically.
For the discussion we will frequently consider the flow 
equations for the quartic 'couplings' $\la_1(\trho)$,
$\la_2(\trho)$ eqs.\ (\ref{La1}), (\ref{La2}) and for the
minimum $\kp$ eq.\ (\ref{kappa}).    

\begin{figure}[h]
\unitlength1.0cm
\begin{center}
\begin{picture}(17.,10.)
\put(8.5,2.8){\footnotesize $\ds{\frac{U_B}{\La^3}}\, 
2\times 10^{15}$}
\put(2.2,2.3){\footnotesize $\ds{\frac{U_0}{\La^3}}\, 
2\times 10^{15}$}
\put(0.7,6.9){\footnotesize $\ds{\frac{
\rho_{0R}}{\La}} 10^{5}$}
\put(7.2,9.1){\footnotesize $\ds{\frac{
m_{R}^2}{\La^2}} 10^{9}$}
\put(12.2,8.2){\footnotesize $\tilde{m}_2^2 $}
\put(10.1,5.8){\footnotesize $\tilde{m}^2 $}
\put(10.2,4.4){\footnotesize $\kp$}
\put(6.4,-0.5){$t=\ln(k/\La)$}
\put(-0.7,0.){
\epsfysize=11.7cm
\epsfxsize=10.cm
\rotate[r]{\epsffile{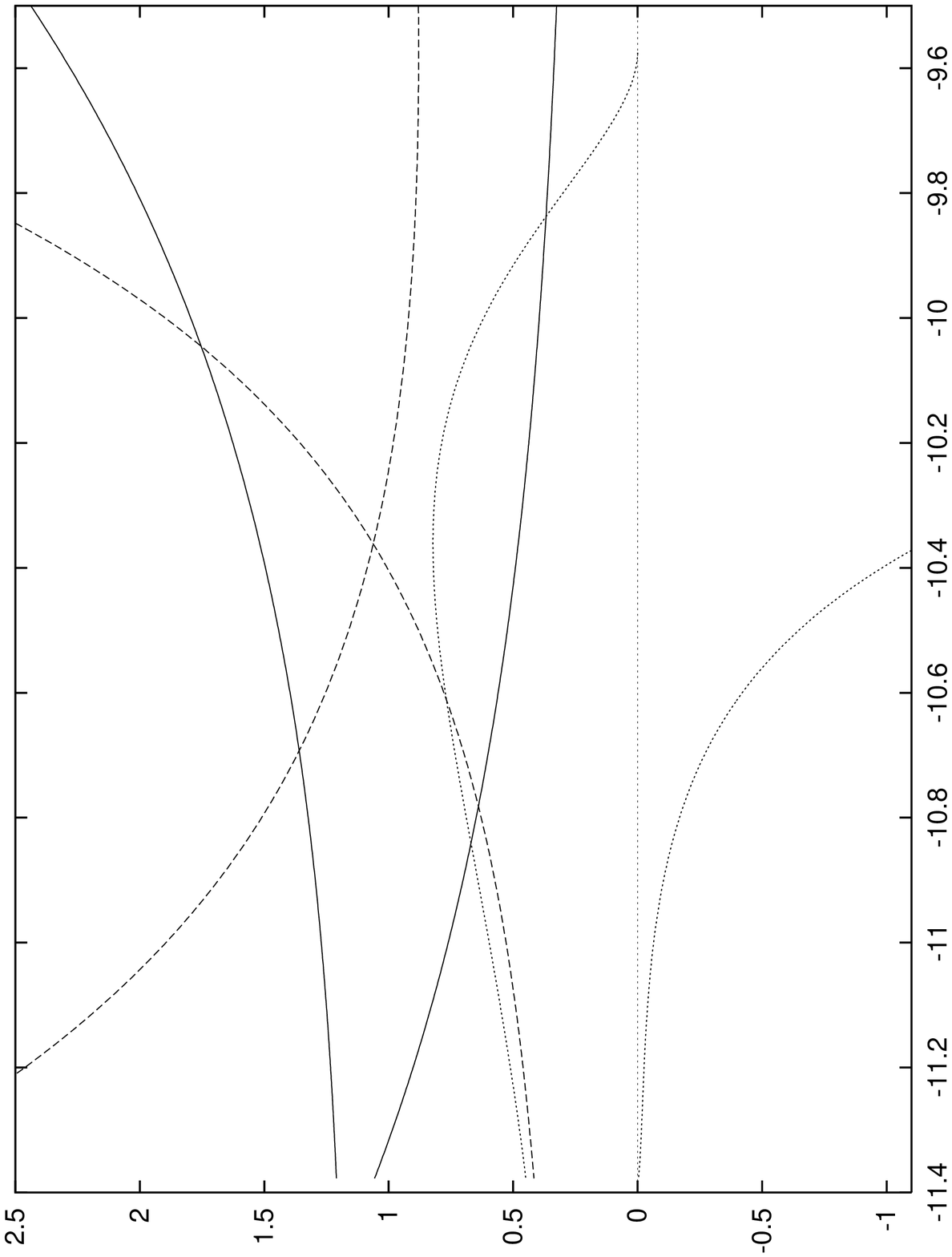}}
}
\put(10.4,0.){
\epsfysize=5.5cm
\epsfxsize=10.cm
\rotate[r]{\epsffile{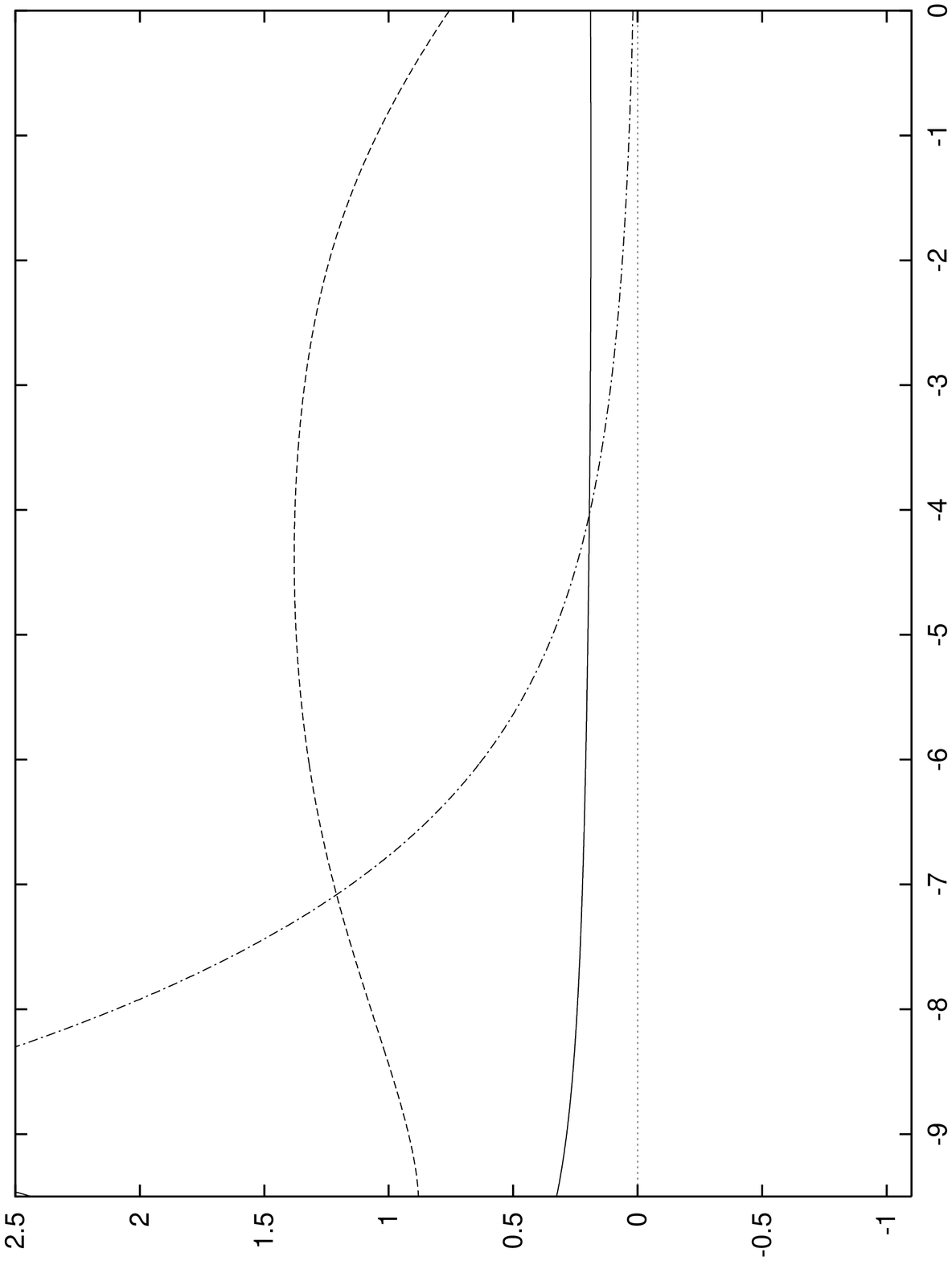}}
}
\end{picture}
\end{center}
\caption[]{\footnotesize
Scale dependence of 
the dimensionless renormalized 
masses $\tilde{m}^2$, $\tilde{m}_2^2$, minimum $\kp$
and dimensionful counterparts 
$m_R^2=k^2 \tilde{m}^2$, $\rho_{0 R}=k \kp$ 
in units of $\La$. We also 
show $U_B(k)$ and
$U_0(k)$, the value of the potential at the top
of the potential barrier and at the minimum
$\rho_{0R}$, respectively. The short distance parameters are  
$\la_{1\La}=2$, $\la_{2\La}=0.1$ and $\dt \kp_{\La}=0$. 
\label{scaledep1}
}
\end{figure}

\begin{figure}[h]
\unitlength1.0cm
\begin{center}
\begin{picture}(15.,10.)
\put(8.,2.15){\footnotesize $\ds{\frac{U_B}{\La^3}}\, 
2\times 10^{3}$}
\put(3.6,1.5){\footnotesize $\ds{\frac{U_0}{\La^3}}\, 
2\times 10^{3}$}
\put(1.1,3.73){\footnotesize $\ds{\frac{
\rho_{0R}}{\La}}$}
\put(6.4,8.6){\footnotesize $\ds{\frac{
m_{R}^2}{\La^2}} 10^{2}$}
\put(11.5,7.8){\footnotesize $\tilde{m}_2^2 $}
\put(13.,3.15){\footnotesize $\tilde{m}^2 $}
\put(13.,4.4){\footnotesize $\kp$}
\put(6.4,-0.5){$t=\ln(k/\La)$}
\put(-0.7,0.){
\epsfysize=15.cm
\epsfxsize=10.cm
\rotate[r]{\epsffile{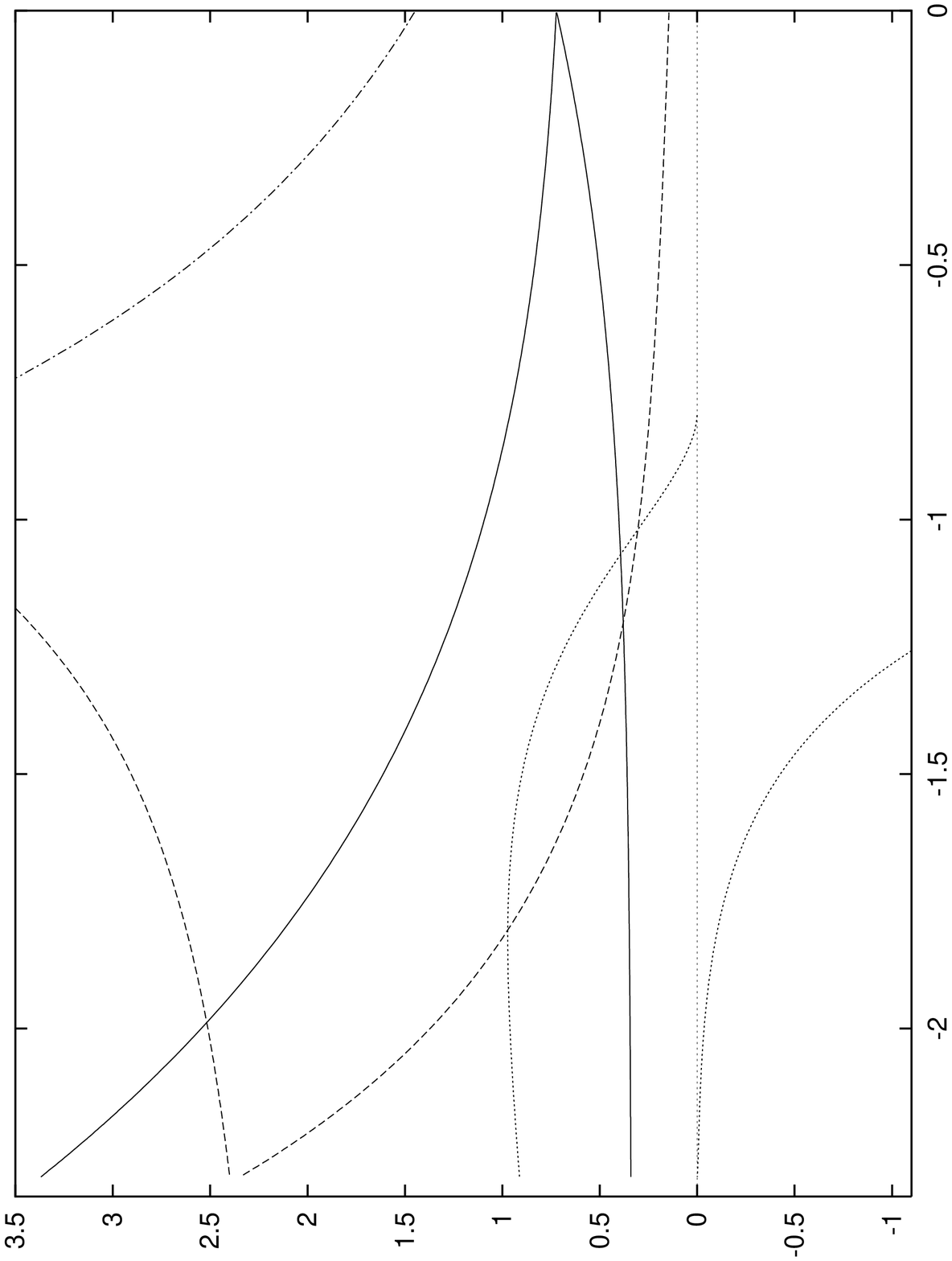}}
}
\end{picture}
\end{center}
\caption[]{\label{scaledep2} \footnotesize
Same as fig.\ \ref{scaledep1}
for $\la_{1\La}=0.1$ and $\la_{2\La}=2$.
}
\end{figure}

In fig.\ \ref{scaledep1}, \ref{scaledep2} we follow the flow
of the dimensionless renormalized minimum $\kp$ and
the radial mass term $\tilde{m}^2=2 \kp \la_{1}(\kp)$ in
comparison to their dimensionful
counterparts $\rho_{0 R}=k \kp$
and $m_R^2=k^2 \tilde{m}^2$ in units of the momentum scale $\La$.
We also consider the dimensionless renormalized mass term
$\tilde{m}_2^2=\kp \la_{2} (\kp)$ corresponding to the curvature of the
potential in the direction of the second
invariant $\ttau$. The height of the potential 
barrier $U_B(k)=k^3 u_k(\trho_B)$ 
with $u_k^{\pri}(\trho_B)=0$, $0 < \trho_B < \kp$,
and the height
of the outer minimum $U_0(k)=k^3 u_k(\kp)$ 
is also displayed and will be discussed
in section \ref{coarse}. Fig.\ \ref{scaledep1} shows these quantities
as a function of $t=\mbox{ln}(k/\La)$ for $\la_{1 \La}=2$, 
$\la_{2 \La}=0.1$.
One observes that
the flow can be separated into two parts.
The first part ranging from $t=0$ to $t \simeq -6$
is characterized by $\kp \simeq \mbox{const}$ and small
$\tilde{m}_2^2$. 
It is instructive to consider what happens in the case
$\tilde{m}_2^2 \equiv 0$. In this case $\la_{2}\equiv 0$
and the flow is governed by the Wilson-Fisher fixed
point of the O(8) symmetric theory.
At the corresponding second order phase transition
the evolution of $u_k$ leads to the scaling
solution of (\ref{DlessEvol}) which obtains for
$\prl u_k/ \prl t=0$. As a consequence $u_k$ becomes a
$k$-independent function that takes on constant (fixed
point) values \cite{TW94-1,Mor}.
In particular, the minimum $\kp$ of the potential  
takes on its fixed point value
$\kp(k)= \kp_{\star}$. 
The fixed point is 
not attractive in the $U(2) \times U(2)$ symmetric theory and
$\la_{2\La}$ is an additional relevant parameter for the system.
For small $\la_2$ the evolution is governed by an anomalous
dimension $d\la_2/ d t=A \la_2$ with $A < 0$, leading to the
increasing $\tilde{m}_2^2$ as $k$ is lowered.

The system exhibits 
scaling behavior only for sufficiently small $\la_{2}$.
As $\tilde{m}_2^2$ increases the quartic coupling $\la_1$ and
therefore the radial mass term $\tilde{m}^2$ is driven to 
smaller values as can be observed from fig.\ \ref{scaledep1}.
For nonvanishing $\la_{2}$ the corresponding 
qualitative change in the flow equation 
(\ref{La1}) for $\la_1$ is the occurance of a term 
$\sim \la_2^2$.
It allows to drive $\la_1$ to negative values in a certain range
of $\trho<\kp$ and, therefore,
to create a potential barrier inducing a first order phase
transition. We observe from the plot that at $t \ltap -9.5$
a second minimum arises $(U_B \not = 0)$. The corresponding value
of $k=\La e^t=k_2$ sets a characteristic scale for the first order
phase transition. Below this scale the dimensionless,
renormalized  quantities approximately scale 
according to their canonical dimension.
The dimensionful quantities like $\rho_{0R}$ or
$m_R^2$ show only a weak scale dependence in this range. 
In contrast to the
above example of a weak first order phase transition 
with characteristic renormalized masses much smaller
than $\La$ fig.\
\ref{scaledep2} shows the flow of the corresponding quantities for
a strong first order phase transition. 
The short distance 
parameters employed are $\la_{1\La}=0.1$,
$\la_{2\La}=2$.
Here the range with 
$\kp \simeq \mbox{const}$ is almost absent and one observes
no approximate scaling behavior.   

With the above examples it becomes easy to understand
the phase structure presented in section \ref{ps}.
For the given curves of figs.\ \ref{phase} and \ref{ratio} we 
distinguished between the range $\la_{2\La}/\la_{1\La} \ll 1$
and $\la_{2\La}/\la_{1\La} \gg 1$ to denote the weak
and the strong first order region.
For $\la_{2\La}/\la_{1\La} \ll 1$ the initial   
renormalization group flow is dominated by the 
Wilson-Fisher fixed point of the $O(8)$ symmetric theory.
In this range 
the irrelevant couplings are driven close to 
the fixed point for some 'time' $|t|=-\mbox{ln}(k/\La)$, loosing
their memory on the initial conditions given by the short
distance potential $u_{\La}$. As a consequence we are able
to observe universal behavior as is 
demonstrated in fig.\ \ref{ratio}.
 
To discuss the case $\la_{2\La}/\la_{1\La} \gg 1$ we consider the
flow equations for the couplings
at the minimum $\kp \not= 0$ of the potential
given by (\ref{kappa}) and (\ref{lakappa})
with (\ref{La1}), (\ref{La2}). 
In the limit of an infinite mass term 
$\tilde{m}_2^2=\kp \la_{2}(\kp) \to \infty$ the $\bt$-functions for
$\la_{1}(\kp)$ and $\kp$ become independent from $\la_{2}(\kp)$ due to
the 'threshold' functions, with 
$l^3_n(\kp \la_{2}) \sim (\kp \la_{2})^{-(n+1)}$ for large
$\kp \la_{2}(\kp)$. 
As a consequence
$\bt_{\la_1}$ and $\bt_{\kp}$ equal the $\beta$-functions for an 
$O(5)$ symmetric model. 
We argue in the following that in this large
coupling limit fluctuations of massless Goldstone bosons
lead to an attractive fixed point for $\la_{2}(\kp)$.
We take the flow equation (\ref{lakappa}),
(\ref{La2}) for $\la_{2} (\kp)$ 
keeping only terms with positive canonical mass dimension
for a qualitative discussion. (This amounts to the 
approximation 
$\la_1^{(n)}(\kp)=\la_2^{(n)}(\kp)=0$ for $n \ge 1$.) 
To be explicit, one may consider the
case for given $\la_{1\La}=2$. The critical cutoff value
for the potential minimum is $\kp_{\La}\simeq 0.2$ for
$\la_{2\La}\gg 1$. For $\kp \la_2(\kp) \gg 1$ and taking 
$\eta \simeq 0$ the $\bt$-function for $\la_2 (\kp)$
is to a good approximation given by $(d=3)$
\be
\frac{\mbox{d} \la_{2} (\kp)}{\mbox{d} t} =
-\la_{2} (\kp)+2 v_3 (\la_{2} (\kp))^2 l^3_2(0).
\label{fpla2}
\ee
The second term on the r.h.s.\ of eq.\ (\ref{fpla2})
is due to massless Goldstone modes which give the
dominant contribution in the considered range. 
The solution of (\ref{fpla2})
implies an attractive fixed point for $\la_{2} (\kp)$ with a value
\be
\la_{2\star} (\kp)=\ds{\frac{1}{2 v_3 l^3_2(0)} } \simeq 4 \pi^2.
\ee 
Starting from $\la_{2\La}$ one finds for the 'time'
$|t|$ necessary
to reach a given $\la_{2} (\kp) > \la_{2\star} (\kp)$ 
\be
|t|= - \mbox{ln}
\ds{\frac{\la_{2} (\kp)-\la_{2\star} (\kp)}
{\la_{2} (\kp) \left(1-\frac{\ds{\la_{2\star} (\kp)}}{
\ds{\la_{2 \La}}} 
\right)} }\quad .
\ee
This converges to a 
finite value for $\la_{2 \La} \to \infty$.  
The further evolution therefore 
becomes insensitive
to the initial value for $\la_{2\La}$ in the large coupling
limit. The flow of $\la_{1}(\kp)$ and $\kp$ is not
affected by the initial running of $\la_{2}(\kp)$
and quantities like $\Dt \rho_{0R}/\La$ or 
$m_R/\Dt \rho_{0R}$ become independent of $\la_{2\La}$
if the coupling is sufficiently large. 
This qualitative
discussion is confirmed by the numerical solution of the full
set of equations presented in figs.\ \ref{phase} and 
\ref{ratio}.
For the fixed point value we obtain $\la_{2\star} (\kp)=38.02$.
We point out that an analogous discussion for the large
coupling region of $\la_{1 \La}$ cannot be made. This can be seen by
considering the mass term at the origin of the short
distance potential 
(\ref{uinitial}) given by 
$u^{\pri}_{\La}(0,0)=-\kp_{\La} \la_{1\La}$. Due to the
pole of $l^3_n(\om,\eta)$ at $\om = -1$ for $n > 1/2$
(cf.\ appendix \ref{poles}) one obtains the constraint
\be
\kp_{\La} \la_{1\La} < 1 \label{constraint}\quad.
\ee  
In the limit $\la_{1\La} \to \infty$
the mass term $2 \kp_{\La} \la_{1\La}$
at the minimum $\kp$ of the potential
at the critical temperature therefore remains finite.

\sect{ Scaling equation of state for weak first order phase
transitions \label{sce}}

We presented in section \ref{ps} some characteristic
quantities for the effective average potential which
become universal at the phase transition 
for a sufficiently small 
quartic coupling $\la_{2\La}=\bar{\la}_{2\La}/\La$
of the short distance potential $U_{\La}$ (\ref{uinitial}). 
The aim of this section is to generalize this observation
and to find a universal scaling form of 
the equation of state for weak first order phase transitions.
The equation of state relates the derivative of the free energy
$U=\lim_{k\to 0}U_k$ to an external source, 
$\prl U/\prl \vp = j$.
Here the derivative has to be evaluated in the outer
convex region of the potential. For instance, for the meson model
of strong interactions the source $j$ is proportional
to the average quark mass \cite{PRW,QuMa} and the 
equation of state
permits to study the quark mass dependence of properties of the
chiral phase transition. We will compute the equation
of state for a nonzero coarse graining scale $k$. It 
therefore contains information for quantities
like the 'classical' bubble surface tension in the
context of Langer's theory of bubble formation which will be
discussed in section \ref{coarse}.
  
In three dimensions the $U(2) \times U(2)$ symmetric 
model exhibits a second order phase transition in
the limit of a vanishing quartic coupling ${\la}_{2\La}$ due to
an enhanced $O(8)$ symmetry. In this case  
there is no scale present in the theory at the critical
temperature. 
In the vicinity of the critical temperature
(small $|\dt \kp_{\La}| \sim |T_c-T|$) and for
small enough $\la_{2\La}$
one therefore expects a scaling behavior of
the effective average potential $U_k$ and
accordingly a universal scaling form of the equation of state. 
At the second order phase transition in the 
$O(8)$ symmetric model there are only two independent
scales that can be related to the deviation from
the critical temperature and to the external source
or $\vp$. 
As a consequence the properly rescaled potential $U/ \rho_R^{3}$
or $U/ \rho^{(\dt +1)/2}$ (with the usual critical exponent
$\dt$) can only depend on one dimensionless ratio.
A possible set of variables
to represent the two independent scales are the renormalized
minimum of the potential $\vp_{0R}=(\rho_{0R}/2)^{1/2}$ 
(or the renormalized mass for the symmetric phase) and
the renormalized field $\vp_R=(\rho_{R}/2)^{1/2}$. The 
rescaled potential will then only depend on the 
scaling variable $z=\vp_R/\vp_{0R}$ \cite{BTW95}. Another
possible choice is the Widom scaling variable
$x=-\dt \kp_{\La}/\vp^{1/\bt}$ \cite{Widom}.  
In the $U(2) \times U(2)$ symmetric 
theory $\la_{2\La}$ is an additional relevant
parameter which renders the phase transition first
order and introduces a new scale, e.g.\ the nonvanishing
value for the jump in the renormalized order parameter
$\Dt \vp_{0R}=(\Dt \rho_{0R}/2)^{1/2}$ at the critical 
temperature or $\dt \kp_{\La}=0$.
In the universal range we therefore observe three
independent scales and the scaling form of the equation
of state will depend on two dimensionless ratios. 
The rescaled potential
$U/\vp_{0R}^6$ can then be written as a universal function $G$
\be
\frac{U}{\vp_{0R}^6}=G(z,v)
\label{scalingeos}
\ee
which depends on the two scaling variables
\be
z=\frac{\vp_R}{\vp_{0R}},\quad v= \frac{\Dt \vp_{0R}}{\vp_{0R}}\,\,.
\ee
The relation (\ref{scalingeos}) is the scaling 
form of the equation of
state we are looking for. At a second order phase
transition the variable $v$ vanishes identically and 
$G(z,0)$ describes the scaling equation of state
for the model with $O(8)$ symmetry \cite{BTW95}.
The variable $v$ accounts for the additional scale 
present at the first order phase transition.
We note that $z=1$ corresponds to a vanishing source
and $G(1,v)$ describes the temperature dependence
of the free energy for $j=0$. In this case $v=1$ denotes the 
critical temperature $T_c$ whereas for $T < T_c$
one has $v < 1$. Accordingly $v > 1$ obtains for
$T > T_c$ and $\vp_{0R}$ describes here the local 
minimum corresponding to the metastable ordered phase. 
The function $G(z,1)$ accounts for the
dependence of the free energy on $j$ for $T=T_c$.

\begin{figure}[h]
\unitlength1.0cm
\begin{center}
\begin{picture}(13.,9.)
\put(-0.6,4.4){$G_{k_f}$}
\put(6.4,-0.5){$z$}
\put(-0.5,0.){
\epsfysize=13.cm
\epsfxsize=9.cm
\rotate[r]{\epsffile{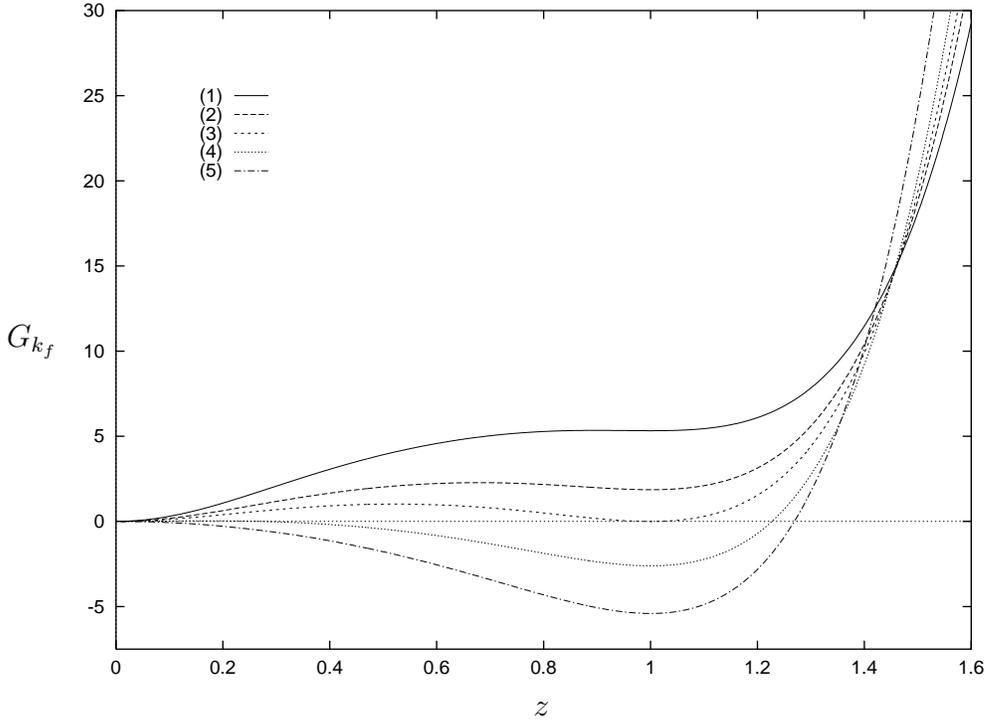}}
}
\end{picture}
\end{center}
\caption[]{\label{scalfunc} \footnotesize
Universal shape of the coarse grained potential $(k=k_f)$
as a function of the scaling variable $z=\vp_R/\vp_{0R}
=(\rho_R/\rho_{0R})^{1/2}$ for different values of
$v=\Dt \vp_{0R}/\vp_{0R}=(\Dt \rho_{0R}/\rho_{0R})^{1/2}$.
The employed values for $v$ are (1) $v=1.18$,
(2) $v=1.07$, (3) $v=1$, (4) $v=0.90$, (5) $v=0.74$. 
For vanishing sources one has $z=1$. In this case $v=1$
denotes the critical temperature $T_c$. Similarly
$v > 1$ corresponds to $T > T_c$ with $\vp_{0R}$
denoting the minimum in the metastable ordered phase.
}
\end{figure}

We consider the scaling form (\ref{scalingeos}) of the
equation of state for a nonzero coarse graining scale $k$.
The renormalized field is given by $\vp_R=Z_k^{1/2} \vp$.
We pointed out in section \ref{ps} that there is a
characteristic scale $k_2$ for the first order phase
transition where the second local minimum of the effective
average potential appears. For weak first order phase
transitions one finds $\rho_{0R} \sim k_2$. To observe
the scaling form of the equation of state the infrared cutoff
$k$ has to run below $k_2$ with $k \ll k_2$. For
the scale $k_f$ defined in eq.\ (\ref{fix})
we observe universal behavior to high
accuracy (cf.\ fig.\ \ref{ratio} and the corresponding universal 
ratios in table \ref{table1} for small $\la_{2\La}/\la_{1\La}$).
The result for the universal function 
$U_{k_f}/\vp_{0R}^6=G_{k_f}(z,v)$ is presented in fig.\
\ref{scalfunc}. 
For $v=1$ one has $\vp_{0R}(k_f)=\Dt \vp_{0R}(k_f)$ 
which denotes the critical temperature.
Accordingly $v > 1$ denotes temperatures above 
and $v < 1$ temperatures below the critical temperature.
One observes that $G_{k_f}(z,1)$ 
shows two almost degenerate minima. (They become exactly
degenerate in the limit $k \to 0$).
For the given examples $v=1.18$, $1.07$
the minimum at the origin becomes
the absolute minimum and the system is in the
symmetric phase. In contrast, for $v = 0.90$, $0.74$ the 
absolute minimum is located at $z=1$ which
characterizes the spontaneously broken phase.
For small enough $v$ the local minimum at the 
origin vanishes.

We explicitly verified that the universal
function $G_{k_f}$ depends only 
on the scaling variables $z$ and $v$ by choosing various
values for $\dt \kp_{\La}$ and for the quartic couplings of
the short distance potential, $\la_{1\La}$ and $\la_{2\La}$.
In section \ref{ps} we observed that the model shows
universal behavior for a certain range of the parameter
space. For given $\la_{1\La}$ and small enough 
$\la_{2\La}$ one always observes universal behavior.
For $\la_{1\La}=0.1$, $2$ and $4$ it is 
demonstrated that (approximate) universality holds
for $\la_{2\La}/ \la_{1\La} \, \ltap \, 1/2$ 
(cf.\ fig.\ \ref{ratio}). 
For $\la_{1\La}$ around $2$ one observes 
from figs.\ \ref{phase}, \ref{ratio} and table \ref{table1} that
the system is to a good accuracy described by
its universal properties for even larger values 
of $\la_{2\La}$. The corresponding phase transitions
cannot be considered as particularly weak first order. 
The universal function $G_{k_f}$
therefore accounts 
for a quite large range of the parameter space.

We emphasize that the universal form of the 
effective potential given in fig.\ \ref{scalfunc} depends on
the scale $k_f$ where the integration of the flow 
equations is stopped (cf.\ eq.\ (\ref{fix})).  
A different prescription
for $k_f$ will, in general, 
lead to a different
form of the effective potential. We may interpret
this as a scheme dependence describing the effect of 
different coarse graining procedures. This is fundamentally
different from nonuniversal corrections since $G_{k_f}$
is completely independent of details of the short distance
or classical action and in this sense universal.
A more quantitative discussion of this scheme
dependence will be presented in the next section.
We note that fluctuations on scales 
$k < k_f$ do not influence substantially the location of
the minima of the coarse grained potential and the
form of $U_k(\vp_R)$ for $\vp_R > \vp_{0R}$ remains almost
$k$-independent. Here
$\prl U_{k_f}/\prl \vp=j(k_f)$ with
$j(k_f) \simeq \lim_{k \to 0} j(k) = j$.\footnote{
The role of massless
Goldstone boson fluctuations for the 
universal form of the effective average potential
in the limit $k \to 0$ has been discussed previously
for the $O(8)$ symmetric model \cite{BTW95}.}

We have established the scaling equation of state using the
renormalized variables $\vp_R$, $\vp_{0R}$ and
$\Dt \vp_{0R}$. Alternatively the scaling equation
of state can be presented in terms of the short
distance parameters $\dt \kp_{\La}$, $\la_{2\La}$
and the unrenormalized field $\vp$. For the 
$O(8)$ symmetric model $(\la_{2\La}=0)$ this is
known as the Widom scaling form of the equation of state
\cite{Widom} and in this case the relation between
$\vp_R$, $\vp_{0R}$ and $\vp$, $\dt \kp_{\La}$
is solely determined by critical exponents 
and amplitudes \cite{BTW95}. For the 
$U(2) \times U(2)$ symmetric model the dependence
of $\vp_R$, $\vp_{0R}$ and $\Dt \vp_{0R}$
on the parameters $\dt \kp_{\La}$, $\la_{2\La}$
and the unrenormalized field $\vp$ can be expressed
in terms of scaling functions. 
In general, only for limiting
cases as $\dt \kp_{\La}=0$ or $\la_{2\La}=0$ the
relation is determined by critical exponents 
and amplitudes. We will consider here these limits. 
The computation of the scaling 
functions for the general case as
well as the corresponding (generalized) Widom scaling
form of the equation of state is the subject of a 
separate work \cite{BW96}.

We consider the
renormalized minimum $\vp_{0R}$ in two limits which are
denoted by  $\Dt \vp_{0R} = \vp_{0R}(\dt \kp_{\La}=0)$
and $\vp^0_{0R} = \vp_{0R}(\la_{2\La}=0)$.  
The behavior of $\Dt \vp_{0R}$ is described in
terms of the exponent $\theta$ according to
eq.\ (\ref{Powrho}),
\be
\Dt \vp_{0R} \sim (\la_{2\La})^{\theta/2}, \qquad \theta=1.93.
\label{scath}
\ee
The dependence of the minimum $\vp^0_{0R}$ of the
$O(8)$ symmetric potential on the temperature is 
characterized by the critical exponent $\nu$,
\be
\vp^0_{0R} \sim (\dt \kp_{\La})^{\nu/2}, \qquad \nu=0.882.
\label{scanu}
\ee
The exponent $\nu$ for the $O(8)$ symmetric model
is determined analogously to $\theta$ as described in section
\ref{phase} \footnote{For the $O(8)$ symmetric model
($\la_{2\La}=0)$ we consider the 
minimum $\vp^0_{0R}$ at $k=0$.}. 
We can also introduce a critical exponent $\zeta$ for the
jump of the unrenormalized order parameter
\be
\Dt \vp_{0} \sim (\la_{2\La})^{\zeta}, 
\qquad \zeta=0.988 \, .
\ee
With
\be
\vp^0_{0} \sim (\dt \kp_{\La})^{\beta}, 
\qquad \beta=0.451
\ee
it is related to $\th$ and $\nu$ by the additional index
relation
\be
\frac{\theta}{\zeta}=\frac{\nu}{\beta}=1.95 \, . 
\ee   
We have verified this numerically. 
For the case $\dt \kp_{\La}=\la_{2\La}=0$ one obtains
\be
j \sim \vp^{\dt}.
\ee
The exponent $\dt$ is related to the 
anomalous dimension $\eta$ via the usual
index relation $\dt=(5-\eta)/(1+\eta)$. From the
scaling solution of eq.\ (\ref{EtaSkale}) we
obtain $\eta=0.0224$.

With the help of the above relations one immediately
verifies that for $\la_{2\La}=0$ 
\be
z \sim (-x)^{-\bt}\, , \qquad v=0
\ee
and for $\dt \kp_{\La}=0$
\be
z \sim y^{-\zeta} \, , \qquad v=1 \, .
\ee
Here we have used the Widom scaling variable $x$ and the 
new scaling variable $y$ given by
\be
x=\frac{-x}{\vp^{1/\bt}}\, , 
\qquad y=\frac{\la_{2\La}}{\vp^{1/\zeta}}\, .
\ee

\sect{Coarse graining and Langer's theory of bubble formation
\label{coarse}}

The coarse grained effective potential $U_k$ results from
the integration of fluctuations with momenta larger
than $\sim k$. It is a nonconvex function whereas the 
standard effective potential $U=\lim_{k\to 0} U_k$
has to be convex by its definition as a Legendre transform.
The difference between $U$ and $U_k$
is due to fluctuations with
characteristic length scales larger than the inverse
coarse graining scale $\sim k^{-1}$. The role of these
fluctuations for the approach to convexity has been
established explicitly \cite{RTW}.
The study of
first order phase transitions usually relies on the
nonconvex 'part' of the potential. As an 
illustration we consider the change of the system
from the high temperature to the low temperature
phase by bubble nucleation as described by Langer's theory
\cite{Langer}. On the one hand, the approach relies on
the definition of a suitable
coarse grained free energy $\Gm_k$ with a coarse graining
scale $k$ and, on the other hand, a saddle point
approximation for the treatment of fluctuations around
the 'critical bubble' is employed.
The problem is therefore separated into two parts:
One part concerns the treatment of fluctuations with 
momenta $q^2 \, \gtap \, k^2$ which are included in the
coarse grained free energy. The second part deals with an 
estimate of fluctuations around the bubble
for which only fluctuations with momenta smaller than
$k$ must be considered. These issues will be discussed 
in this section in a quantitative way. In particular, we will
give a criterion for the validity of Langer's 
formula. 

One may consider a system that starts at some high temperature
$T > T_c$ and investigate what happens as $T$ is lowered
as a function of time as for example during the evolution
of the early universe. For large enough temperature
the origin of the potential $(\vp=0)$ is the only 
minimum and the system is therefore originally in the
symmetric phase. As $T$ approaches $T_c$ a second local
minimum develops at $\vp_0 > 0$. This becomes the
absolute minimum below $T_c$. Nevertheless, the potential
barrier prevents the system to change smoothly to the
ordered phase. For a short while where $T$ is in the vicinity
of $T_c$ but below $T_c$ the system remains therefore 
in a state with higher energy density as
compared to the state corresponding to the absolute minimum
away from the origin. This is the so-called 'false vacuum' in
high energy physics or the metastable state in statistical 
physics. Such a state is unstable with respect to
fluctuations which  penetrate or cross the barrier.
The picture is familiar from the condensation of vapor.
The false vacuum corresponds to the supercooled
vapor phase and the true vacuum to the fluid phase.
Bubbles of the true vacuum (droplets) occur through thermal or
quantum fluctuations\footnote{
In the real world the condensation of vapor is triggered
by impurities but this is not the issue here.}. 
If a bubble is large enough 
so that the decrease in volume energy exceeds the 
surface energy it will grow. The phase transition is 
completed once the whole 
space is filled with the true vacuum. 
On the other hand, small bubbles shrink due to the 
surface tension. The
critical bubble is just large enough that it
does not shrink.
To be explicit we consider a spherical bubble
where the bubble wall with 'thickness' $\Dt$
is thin as compared to the bubble radius $R$, i.e.
$\Dt \ll R$.
In leading order the coarse grained free energy $\Gm_k$ 
for such a bubble configuration
can be decomposed
in a volume and a surface term \cite{ColCal,Lin}
\be
\Gm_k^{(0)} = -\frac{4 \pi}{3}  R^3 \eps + 4 \pi R^2 \sigma_k.
\label{dec}
\ee
In the thin wall approximation one obtains for the 
surface tension $\sigma_k$ in our conventions
\be
\sigma_k=2\int\limits^{\vp_0}_0 d\vp
\sqrt{2 Z_k U_k(\vp)}.
\label{sur}
\ee
For the difference in the free energy density $\eps$ one 
has
\be
\eps=U(0)-U(\vp_0)=-\lim_{k \to 0} U_0(k).
\label{vo}
\ee
We include in $\eps$ fluctuations with arbitrarily
small momenta. In contrast, the long wavelength
contributions to $\si_k$ are effectively cut off by the 
characteristic length scale of the bubble surface
and are described by the 'fluctuation determinant'
$A_k$ (cf.\ eq.\ (\ref{nc})). The determination of a suitable
coarse graining scale $k$ for the computation of 
$\si_k$ is discussed below.

The critical bubble maximizes $\Gm_k^{(0)}$ with respect to the
radius. It minimizes the coarse grained free energy
with respect to other deformations because the spherical 
form is energetically favorable. The critical bubble
therefore represents a saddle point in the space of
possible 'bubble configurations'. 
In Langer's theory of bubble formation
one considers a saddle point expansion 
around the critical bubble. There is exactly one negative 
mode that corresponds to the shrinking or growth
of the bubble and there are infinitely many
positive modes (there are also translational
zero modes) 
\footnote{
Langer's theory is not restricted to the thin wall approximation
which is considered here for simplicity. In particular,
the property of the critical bubble to 
represent a saddle point with exactly one negative mode is 
independent from the thin wall approximation.
}. The bubble nucleation
rate $\bar{\Gm}$, which describes the probability
per unit volume per unit time for the transition
to the new vacuum, 
can be written in the form \cite{ColCal,Lin}
\be
\bar{\Gm} = A_k \, \exp (-\Gm^{(0)}_k [\vp^{(0)}_b]) 
=A_k \, \exp \left(-\frac{16 \pi}{3}\frac{\sigma_k^3}
{\eps^2}\right)
\label{nc}
\ee 
where $\Gm_k^{(0)}$ is evaluated for $\vp^{(0)}_b$ corresponding to
the critical bubble and approximated by (\ref{dec}).
The exponential term with the coarse grained free energy
$\Gm^{(0)}_k$ denotes the lowest order or classical
contribution.
The prefactor $A_k$ contains several
factors that depend on the details of the system
under investigation. 
In particular, $A_k$ accounts for the contribution
to the free energy from the fluctuations
with momenta smaller than $k$.
It depends on $k$ through the effective ultraviolet
cutoff for these fluctuations which is present
since fluctuations with momenta larger than $k$
are already included in 
$\Gm^{(0)}_k[\vp^{(0)}_b]$~\footnote{
The effective average action \cite{Wet91-1} also provides the
formal tool how the ultraviolet cutoff  $\sim k$
is implemented in the remaining functional
integral for large length scale fluctuations.}. 
Langer's formula for
bubble nucleation amounts essentially to a perturbative
one loop estimate of $A_k$.

For a determination of a useful choice of $k$
it is convenient to place the discussion in a more
general context which does not rely on the thin wall 
approximation or a saddle point approximation.
What one is finally interested in is the free energy
$\Gm[\vp_b]$ for bubble configurations of a given shape.
The 'true critical bubble' $\vp_b^c$
corresponds to a saddle point
in the space of 'bubble configurations' which are
characterized by boundary conditions connecting the
false and the true vacuum. The nucleation rate
is then proportional $\exp-\Gm[\vp_b^c]$. The coarse 
graining can be seen as a convenient strategy to
evaluate $\Gm[\vp_b]$ by separating contributions from 
different momentum scales. We propose to choose
the coarse graining scale $k$ somewhat above
but in the vicinity of the inverse thickness $\Dt^{-1}$
of the bubble wall. We will argue below that in
this case the corrections to the effective surface tension
from the prefactor $A_k$ should be best accessible.

In fact, we can write $\bar{\Gm} = B \, \exp-\Gm[\vp_b^c]$
where $B$ contains dynamical factors and $\Gm[\vp_b^c]$
does not include contributions from fluctuations of the
negative mode and the translational modes present for the
critical bubble. The prefactor in eq.\ (\ref{nc}) can then be
written as
\be
A_k=B \exp-(\dt_k+\eta_k)
\ee
where
\be
\dt_k= \Gm[\vp_b^c]-\Gm_k[\vp_b^c]\, ,\qquad 
\eta_k= \Gm_k[\vp_b^c]-\frac{16 \pi}{3}\frac{\sigma_k^3}
{\eps^2}\, .  
\ee
The term $\eta_k$ includes the difference between
the true critical bubble and the configuration used
to estimate $\sigma_k$ as well as a correction term 
to $\eps$ to be discussed below. We first concentrate 
on $\dt_k$ which describes the difference between the
free energy and the coarse grained free energy for
the critical bubble. As mentioned above this is due 
to fluctuations with momenta $q^2\, \ltap \, k^2$
and incorporates the dominant $k$-dependence of $A_k$.
Since the bubble provides for inherent effective
infrared cutoff scales  
$\sim R^{-1}$  
or $\Dt^{-1}$ the
contribution $\dt_k$ is both infrared and 
ultraviolet finite. The larger $k$, the more 
fluctuations are included in $\dt_k$ and from this
point of view one wants to take $k$ as low as possible.
On the other hand, $k$ should not be taken smaller than
$\Dt^{-1}$ if the approximation used for a computation 
of $\Gm_k$ relies on almost constant field configurations
rather than real bubbles, as is usually the case. Only
for $k$ sufficiently large compared to $\Dt^{-1}$ the
difference between an evaluation of the potential
and kinetic terms in $\Gm_k$ for almost constant field 
configurations (e.g.\ by a derivative expansion) 
rather than for bubbles remains small. In this way the 
technique of course graining combines a relatively simple 
treatment of the modes with $q^2 \, \gtap \, k^2$
for which the detailed properties of the bubble are 
irrelevant with an estimate of fluctuations around the
bubble for which the short distance physics
($q^2 \, \gtap \, k^2$) needs not to be considered anymore.
It is clear that $k$ is only a technical construct
and for physical quantities the $k$-dependence of
$\dt_k$ and $\Gm_k^{(0)}$ must cancel. More 
precisely, this concerns the sum 
$\dt_k+\eta_k+16 \pi \sigma_k^3/3 \eps^2$ .
For thin wall bubbles the most important contribution
to $\eta_k$ is easily identified: By our definition
of $\eps$ we have included contributions from fluctuations
with length scales $\gtap \, R$. They should not be
present in the effective action for a bubble with finite 
radius. Therefore $\eta_k$ contains a correction term
$(16 \pi/3)\sigma_k^3(\eps^{-2}(R)-\eps^{-2})$
which replaces effectively $\eps$ by $\eps(R)$ in eq.\ 
(\ref{nc}). We can evaluate $\eps(R)$ in terms of the
coarse grained free energy at a scale $k_R$
\be
\eps(R) \simeq U_{k_R}(0)-U_{k_R}(\vp_0)=-U_0(k_R)\, ,
\qquad
k_R = \ds{\frac{1}{R}} \,\, .
\ee
For $\Dt \ll R$ one should not confound $k_R$
with the coarse graining scale $k$ since one has the
inequality
\be
k_R \ll \frac{1}{\Dt} \,\, \ltap \,\, k\,\, .
\ee
Only for $\Dt \simeq R$ the clear separation between
$k_R$ and $k$ disappears. We note that at the critical
temperature one has $R \to \infty$ and therefore 
$\eps(R)=\eps$\, .
  
Since $\sigma_k$ enters the 
nucleation rate
(\ref{nc}) exponentially even small changes with $k$
will have large effects. If one finds a strong
dependence of $\si_k$ on the
coarse graining scale $k$ this is only compatible with
a large contribution from the higher orders
of the saddle point expansion. The $k$-dependence of
$\si_k$ therefore
gives direct information about the validity of 
Langer's formula. We find a strong scale
dependence of $\si_k$ if the phase transition is characterized
by large effective
dimensionless couplings $\frac{\la_R}{m_R^c}(k)$.
A weak scale dependence is observed for small effective
couplings. This gives a very consistent
picture: The validity of the saddle point
approximation typically requires small dimensionless couplings.
In this case also the details of the
coarse graining are not of crucial importance
within an appropriate range of $k$. The
remaining part of this section provides a detailed
quantitative discussion.

We consider the dependence of
the effective average potential $U_k$ 
and the surface tension $\sigma_k$ on the coarse
graining scale $k$ near the critical temperature $T_{c}$
for three examples in detail.
They are 
distinguished by different choices for the quartic couplings
$\bar{\la}_{1\La}$ and $\bar{\la}_{2\La}$ of the short
distance potential $U_{\La}$ given by eq.\ (\ref{uinitial}).
The choice $\bar{\la}_{1\La}/\La=0.1$ and 
$\bar{\la}_{2\La}/\La=2$ corresponds to a strong first order
phase transition with renormalized masses
not much smaller than the cutoff 
scale $\La$.
The renormalized couplings will turn out
small enough such that the 
notion of a coarse grained potential $U_k$ 
and a surface tension $\si_k$ can be used
without detailed information on the coarse graining
scale within a certain range of $k$. In contrast
we give two examples where the dependence of $U_k$
and $\si_k$ on the coarse graining scale becomes of crucial
importance.
The choice $\bar{\la}_{1\La}/\La=2$ and 
$\bar{\la}_{2\La}/\La=0.1$ leads to a weak first
order phase transition with small renormalized masses
and the system shows universal behavior
(cf.\ sect. \ref{ps}). For 
$\bar{\la}_{1\La}/\La=4$ and 
$\bar{\la}_{2\La}/\La=70$ one observes a 
relatively strong
first order phase transition. Nevertheless for 
both examples the coarse grained potential 
and the surface tension show 
a similarly high sensitivity on the scale $k$.
 
\begin{figure}[h]
\unitlength1.0cm
\begin{center}
\begin{picture}(13.,9.)
\put(-0.8,4.6){$\ds{\frac{U_k}{\La^3}}\, 10^3$}
\put(6.3,-0.5){$\ds{\frac{\vp_R}{\La^{1/2}}}$}
\put(-0.4,0.){
\epsfysize=13.cm
\epsfxsize=9.cm
\rotate[r]{\epsffile{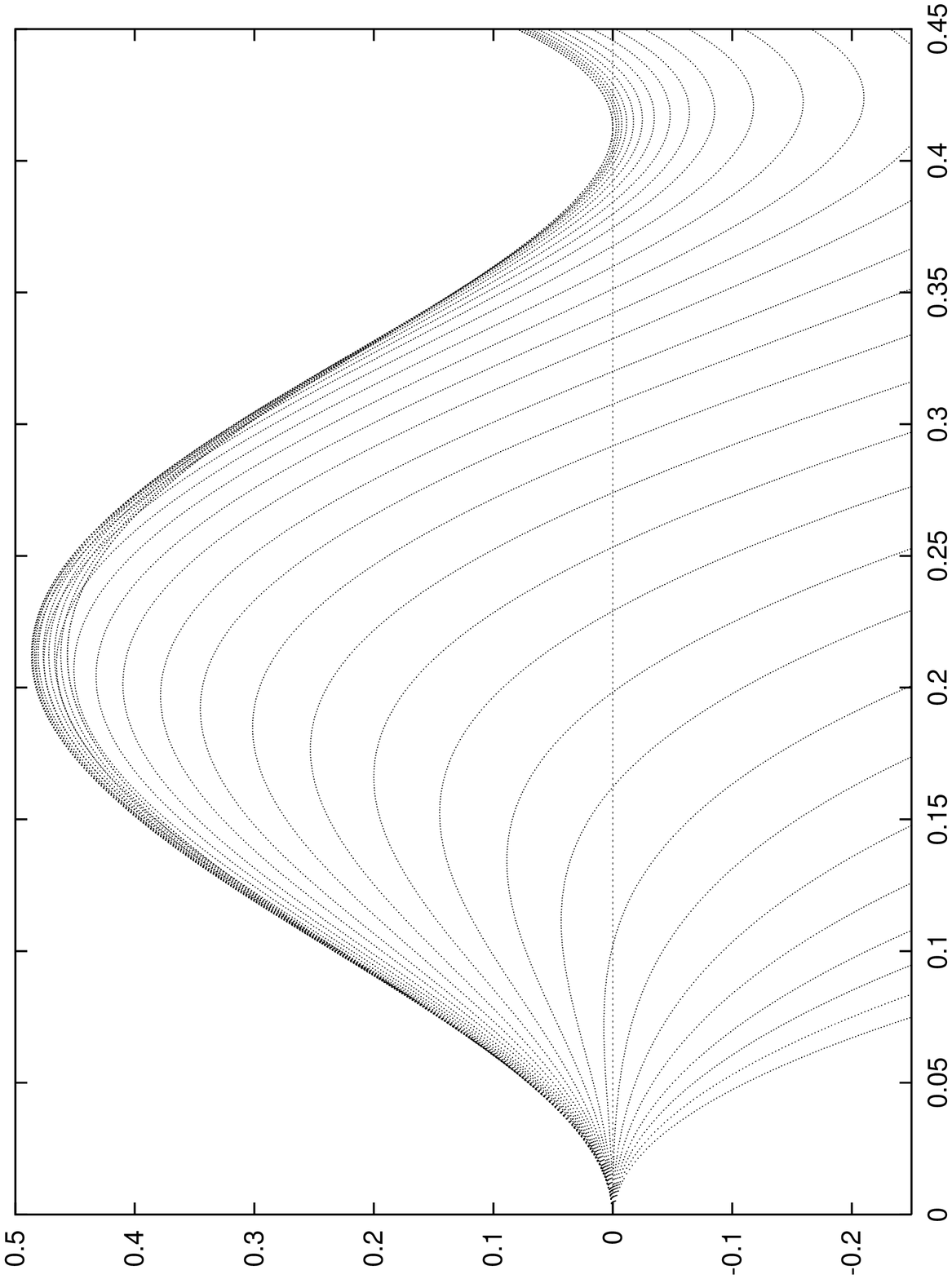}}
}
\end{picture}
\end{center}
\caption[]{\footnotesize The coarse grained   
potential in dependence on the coarse graining scale $k$
for fixed (almost critical) temperature.
The example corresponds to
$\bar{\la}_{1\La}/\La=0.1,\bar{\la}_{2\La}/\La=2$.
\label{innercoarse}
}
\end{figure}

\begin{figure}[h]
\unitlength1.0cm
\begin{center}
\begin{picture}(13.,9.)
\put(-0.8,4.6){$\ds{\frac{U_k}{\La^3}}\, 10^{15}$}
\put(6.3,-0.5){$\ds{\frac{\vp_R}{\La^{1/2}}}$}
\put(-0.4,0.){
\epsfysize=13.cm
\epsfxsize=9.cm
\rotate[r]{\epsffile{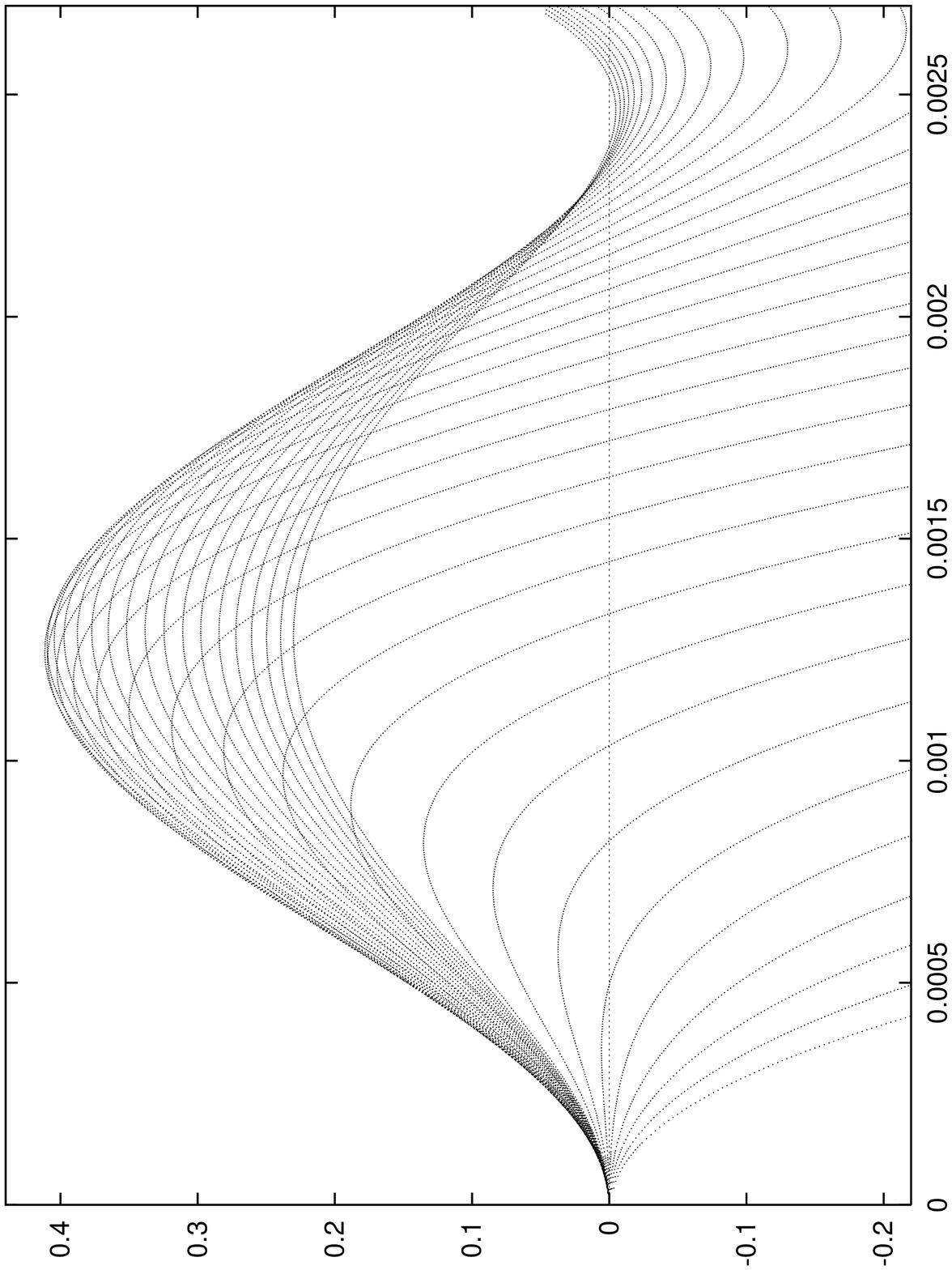}}
}
\end{picture}
\end{center}
\caption[]{\footnotesize The coarse grained potential for
$\bar{\la}_{1\La}/\La=2,\bar{\la}_{2\La}/\La=0.1$.
\label{innercoarse2}
}
\end{figure}

In figs.\  \ref{innercoarse}, \ref{innercoarse2}
the scale dependence of $U_k$
is shown for a fixed temperature in the 
vicinity of $T=T_{c}$ or $\dt \kp_{\La} =0$.
We plot $U_k$ in units of $\La^3$ as a function of 
$\vp_R/\La^{1/2} = (\rho_{R}/2\La)^{1/2}
= (Z_k \rho/2\La)^{1/2}$ 
for different values of $k$. Each curve differs
in $k=\La e^{t}$ 
by $\Dt t=1/18$ and
the first curve to
the left with a negative curvature at the origin
corresponds to $t \simeq -0.40$ 
for fig.\ \ref{innercoarse} 
($t \simeq -9.3$ for fig.\ \ref{innercoarse2}).
For $0 \ge t \, \gtap \, -0.40$ ($0 \ge t \, \gtap \, -9.3$)
there is only one minimum of the potential
away from the origin which lies below the plotted
range in fig.\ \ref{innercoarse}
(\ref{innercoarse2}). Lowering $k$ 
results in a successive
inclusion of fluctuations on larger length scales
and one observes the appearance
of a second local minimum at the origin of the
potential. The barrier between both minima increases
and reaches its maximum value. 
As $k$ is lowered further the potential 
barrier starts to
decrease whereas the location of
the minima becomes almost $k$-independent
and degenerate in height. 
In this region also the outer convex part of the
potential shows an almost stable profile due to the 
decoupling of the massive modes. 
We stop the integration at the scale
$k=k_f$ given by eq.\ (\ref{fix}).
The coarse grained effective potential $U_k$
with $k=k_f$ essentially includes all fluctuations
with squared momenta larger than the scale $|m_{B,R}^2|$
given by the curvature at the top of the potential barrier
(cf.\ eq.\ (\ref{mbr})). Successive inclusion of fluctuations
with momenta smaller than $k_f$ would result in a further
decrease of the potential 
barrier. The flattening of the barrier
is induced by the pole structure of the
'threshold' functions $l^3_n(\om)$ appearing
in the flow equations for the couplings (e.g.\ $l^3_1(\om)$ 
appearing in eq.\ (\ref{La1}) exhibits a pole at
$\om=-1$ according to $l^3_1(\om) \sim (\om+1)^{-1/2}$
for $\om$ near $-1$ (cf.\ appendix \ref{poles})).
The argument $\om$ corresponds
to dimensionless mass terms as $U_k^{\pri}(\rho_R)/k^2$ or
$(U_k^{\pri}(\rho_R)+2\rho_R U_k^{\pri\pri}(\rho_R))/k^2$. 
In the nonconvex region of the potential the curvature 
is negative. Since the pole
cannot be crossed, negative $U_k^{\pri}$ or 
$U_k^{\pri\pri}$ must go to zero with $k^2$ and
as a consequence the potential barrier flattens.
For the scalar 'vector' model the approach to convexity
in the limit $k\to 0$ has been studied analytically
previously \cite{RTW}. Here we are interested in the
potential for a nonzero coarse graining scale $k$.

\begin{figure}[h]
\unitlength1.0cm
\begin{center}
\begin{picture}(13.,9.)
\put(-0.7,4.6){$\ds{\frac{\si_k}{\si_{\max}}}$}
\put(6.3,-0.5){$\ds{\ln\left({k}/{k_f}\right)}$}
\put(6.9,2.){(1)}
\put(9.2,2.){(2)}
\put(11.3,2.){(3)}
\put(-0.4,0.){
\epsfysize=13.cm
\epsfxsize=9.cm
\rotate[r]{\epsffile{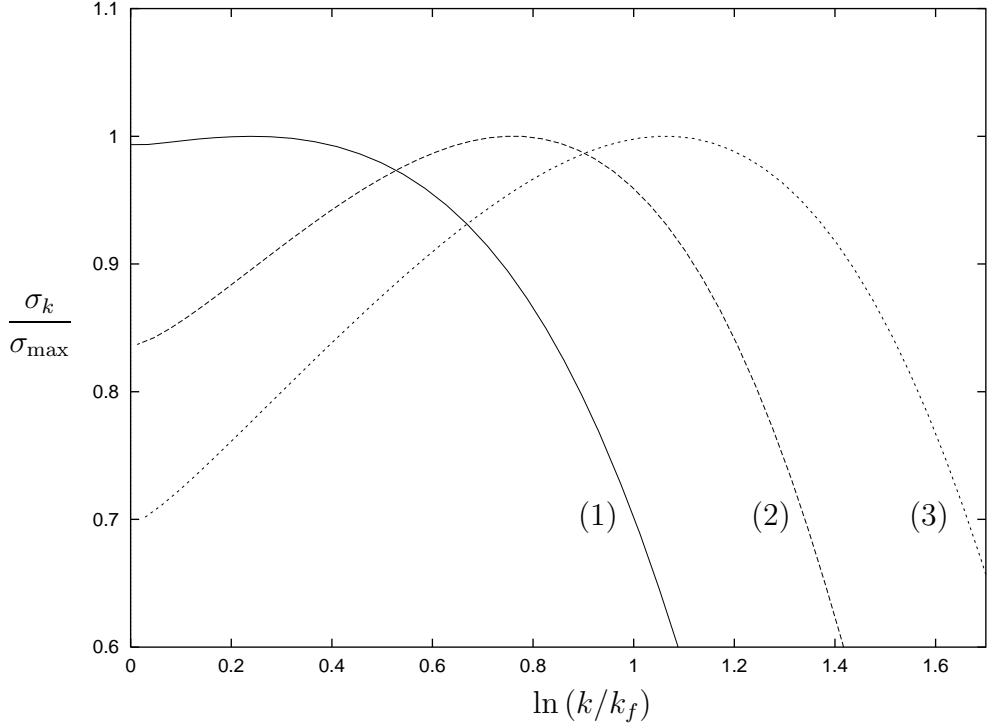}}
}
\end{picture}
\end{center}
\caption[]{\footnotesize The normalized surface tension
$\si_k/\si_{\max}$ as a function of $\ln (k/k_f)$.
The short distance parameters are (1)
$\bar{\la}_{1\La}/\La=0.1$, $\bar{\la}_{2\La}/\La=2$,
(2) $\bar{\la}_{1\La}/\La=2$, $\bar{\la}_{2\La}/\La=0.1$,
(3) $\bar{\la}_{1\La}/\La=4$, $\bar{\la}_{2\La}/\La=70$. 
\label{surface}
}
\end{figure}

The most significant
difference between fig.\ \ref{innercoarse} and 
fig.\ \ref{innercoarse2}
is the $k$-dependence of the potential barrier
in a region where the minima become degenerate
and almost independent of $k$. Fig.\ \ref{innercoarse}
shows a barrier with a weak scale dependence
in the range of $k$ where the location of the minima stabilizes.
In contrast, in fig.\ \ref{innercoarse2} one observes a barrier 
with a strong scale dependence in this region. 
Figs.\ \ref{scaledep1} and 
\ref{scaledep2} (cf.\ sect.\ \ref{rg}) show
the relative height $U_0(k)=U_k(\vp_0)-U_k(0)$ between the 
two local minima and the height of the
potential barrier $U_B(k)=U_k(\vp_B)-U_k(0)$ 
($(\prl U_k/\prl \vp)(\vp_B)=0$, $0 < \vp_B < \vp_0$) as a function
of $t=\ln(k/\La)$. Accordingly one observes that for
$\bar{\la}_{1\La}/\La=0.1$, $\bar{\la}_{2\La}/\La=2$ 
the top of the potential barrier shows a weak
scale dependence in a region of $k$ where $U_0(k)$ is small
whereas for
$\bar{\la}_{1\La}/\La=2$, $\bar{\la}_{2\La}/\La=0.1$
the top of the potential barrier
depends strongly on the coarse graining scale
in this region.
The surface tension $\si_k$ is displayed in
fig.\ \ref{surface} and shows a corresponding behavior.
Here we consider $\si_k$ also for the short distance 
parameters $\bar{\la}_{1\La}/\La=4$,
$\bar{\la}_{2\La}/\La=70$. In fig.\ \ref{surface}
the surface tension $\si_k$ is normalized to $\si_{\max}$
($\si_{\max}/\La^2=1.67 \times 10^{-2} 
(8.41 \times 10^{-11})
(1.01 \times 10^{-3})$ for 
$\bar{\la}_{1\La}/\La=0.1(2)(4)$, 
$\bar{\la}_{2\La}/\La=2(0.1)(70)$)
and given as a function of $\ln (k/k_f)$.\footnote{
The integration according to eq. (\ref{sur}) is performed  
between the two zeros $\vp=0$ and $\vp=\vp_0^{\pri} \, \ltap \vp_0$ 
of $U_k$.}
For $\bar{\la}_{1\La}/\La=0.1$,
$\bar{\la}_{2\La}/\La=2$ the curve exhibits a small curvature
around its maximum and $\si_{\max} \simeq \si_{k=k_f}$.
One observes for the second and the third example 
a comparably large curvature around $\si_{\max}$ and 
$\si_{k=k_f}$ becomes considerably smaller than the maximum.
One may consider what happens if this scale
dependence is not taken into account correctly.
If one takes the difference in $\sigma_k$
from its maximum value to $\sigma_{k=k_f}$ as a rough
measure for its uncertainty this is about 30\%
for the last example. Entering exponentially in eq.\
(\ref{nc}) this would lead to tremendous errors.

In order to quantify the differences between the three 
examples we have displayed some characteristic 
quantities in table \ref{table2}. 
\begin{table} [h]
\renewcommand{\arraystretch}{2.0}
\hspace*{\fill}
\begin{tabular}{|c|c|c|c|c|c|c|c|}     \hline

$\ds{\frac{\bar{\la}_{1\La}}{\La}}$
&$\ds{\frac{\bar{\la}_{2\La}}{\La}}$
&$\ds{\frac{\la_{1R}}{m_R^c}}$
&$\ds{\frac{\la_{2R}}{m_R^c}}$
&$\ds{\frac{m_R^c}{m_{2R}^c}}$
&$\ds{\frac{m_R^c}{\La}}$
&$\ds{\frac{m_{2R}^c}{\La}}$
&$\ds{\frac{k_f}{\La}}$
\\ \hline \hline
0.1
& 2
&0.228
&8.26
&0.235
&$1.55 \times 10^{-1}$
&$6.62 \times 10^{-1}$
&$1.011 \times 10^{-1}$
\\ \hline 
$2$
&0.1
&0.845
&15.0
&0.335
&$2.04 \times 10^{-5}$
&$6.10 \times 10^{-5}$
&$1.145 \times 10^{-5}$
\\ \hline 
4
&70
&0.980
&16.8
&0.341
&$6.96 \times 10^{-2}$
&$2.04 \times 10^{-1}$
&$3.781 \times 10^{-2}$
\\ \hline 
\end{tabular}
\hspace*{\fill}
\renewcommand{\arraystretch}{1}
\caption{\footnotesize The effective dimensionless couplings
$\la_{1R}/m_R^c$ and $\la_{2R}/m_R^c$. The couplings and the mass
terms $m_R^c$, $m_{2R}^c$ are evaluated at the scale $k_f$
and $\dt\kp_{\La}=0$. 
\label{table2}
}
\end{table}
The renormalized couplings
\be
\la_{1R}=U_{k_f}^{\pri\pri}(\rho_{0R}),\qquad
\la_{2R}=4\prl_{\tau}U_{k_f}(\rho_{0R})
\ee
are normalized to the mass term
\be
m_{R}^c=(2\rho_{0R} \la_{1R})^{1/2} \, .
\ee
In addition we give
the mass term
\be
m_{2R}^c=(\rho_{0R} \la_{2R})^{1/2}
\ee
corresponding to the curvature of the potential in the direction
of the second invariant $\tau$. In comparison with 
figs.\ \ref{innercoarse}--\ref{surface} one observes
that the smaller the effective couplings the weaker the scale 
dependence of $U_k$ and $\sigma_k$. In particular, a
reasonably weak scale dependence of $U_k$ and $\si_k$ requires
\be
\frac{\la_{1R}}{m_R^c}=\hal \frac{m_R^c}{\Dt \rho_{0R}}
\ll 1 \label{weakk}\, \, . 
\ee
This establishes a quantitative criterion for the range
where Langer's theory can be used without paying too much
attention to the precise definition of the coarse graining.
Comparison with fig.\ \ref{ratio} (cf.\ sect.\ \ref{phase}) 
shows that this condition is 
not realized for the range of couplings leading to
universal behavior and for large $\bar{\la}_{1\La}/\La$.
The only region where the saddle point expansion 
is expected to converge 
reasonably well is for small $\bar{\la}_{1\La}/\La$
and large $\bar{\la}_{2\La}/\La$.
The second and the third example 
given in fig.\ \ref{surface}, which exhibit a strong
$k$-dependence, show similar values for the effective 
couplings. More precisely, 
for the relatively strong phase transition
with slightly larger effective couplings 
one observes an increased scale 
dependence as compared to the weak phase transition.
In table \ref{table2} also the renormalized 
masses in units of $\La$ 
which indicate the strength of the phase transition and $k_f/\La$
are presented.

In summary, we have shown that 
the coarse grained free energy cannot be defined
without detailed information on the coarse graining scale
$k$ unless the effective dimensionless couplings are small. 
Only for small
couplings we observe a weak $k$-dependence 
of the surface tension in a range
where the location of the
minima of the potential remains almost fixed. There is
a close relation between the dependence of the coarse grained
free energy on the coarse graining scale and the reliability
of the saddle point approximation in Langer's theory
of bubble nucleation. For a strong $k$-dependence of $U_k$
a small variation in the coarse graining scale can induce 
large changes in the predicted nucleation rate
in lowest order in a saddle point approximation.
In this case the $k$-dependence of the pre\-factor
$A_k$ has also to be computed. For strong dimensionless
couplings a realistic estimate of the nucleation rate
therefore needs the capability to compute
$\ln A_k$ with the same accuracy as $16 \pi \sigma_k^3/3 \eps^2$
and a check of the cancelation of the $k$-dependence
in the combined expression (\ref{nc}). 
Our observation that the details of the coarse graining 
prescription become less important in the case of small
dimensionless couplings is consistent with the
fact that typically small couplings are needed for a
reliable saddle point approximation for $A_k$.
For the electroweak high temperature phase transition
a small $k$-dependence of $\sigma_k$ is found for a
small mass $M_H$ of the Higgs scalar whereas for 
$M_H$ near the W-boson mass the picture resembles our 
fig.\ \ref{innercoarse2} \cite{Tet}. This corresponds to
the observation \cite{La,Bu,Ba} that the saddle point approximation
around the critical bubble converges well only for a
small enough mass of the Higgs scalar.

\sect{Conclusions \label{con}}

We have presented in this paper a detailed investigation
of the phase transition in three dimensional models for 
complex $2 \times 2 $ matrices. They are 
characterized by two quartic couplings
$\bar{\la}_{1\La}$ and $\bar{\la}_{2\La}$. 
In the limit
$\bar{\la}_{1\La}\to \infty$, $\bar{\la}_{2\La}\to \infty$
this also covers the model of unitary matrices.
The picture arising from this study is unambiguous:

(1) One 
observes two symmetry breaking patterns for
$\bar{\la}_{2\La}>0$ and $\bar{\la}_{2\La}<0$
respectively. The case $\bar{\la}_{2\La}=0$
denotes the boundary between the two phases
with different symmetry breaking patterns. 
In this special case the theory
exhibits an enhanced $O(8)$ symmetry.  
The phase transition is always first order
for the investigated symmetry breaking
$U(2) \times U(2) \to U(2)$ ($\bar{\la}_{2 \La}>0$). 
For $\bar{\la}_{2 \La}=0$ 
the $O(8)$ symmetric Heisenberg model is recovered
and one finds a second order phase transition.

(2) The strength of the phase transition depends on the size
of the classical quartic couplings $\bar{\la}_{1\La}/\La$ and  
$\bar{\la}_{2\La}/\La$. They describe the short distance or
classical action at a momentum scale $\La$. The strength of the
transition can be parametrized by $m_R^c/\La$ with $m_R^c$ 
a characteristic inverse correlation length at the 
critical temperature.
For fixed $\bar{\la}_{2\La}$ the strength of the
transition decreases with increasing $\bar{\la}_{1\La}$.
This is analogous to the Coleman-Weinberg effect in 
four dimensions.

(3) For a wide range of classical couplings the critical
behavior near the phase transition is universal.
This means
that it becomes largely independent of the details of the 
classical action once everything is expressed in terms of the
relevant renormalized parameters. 
In particular, characteristic ratios like
$m_R^c/\Dt \rho_{0R}$ (critical inverse correlation length
in the ordered phase over discontinuity in the order parameter)
or $m_{0R}^c/\Dt \rho_{0R}$ (same for the disordered phase)
are not influenced by the addition of
new terms in the classical action
as far as the symmetries are respected.

(4) The range of short distance parameters
$\bar{\la}_{1\La}$, $\bar{\la}_{2\La}$ for 
which the phase transition exhibits universal behavior
is not only determined by the strength of the phase
transition as measured by $m_R^c/\La$.
For a given $\bar{\la}_{1\La}/\La$ and 
small enough $\bar{\la}_{2\La}/\La$ one always observes
universal behavior. In the range of small 
$\bar{\la}_{1\La}/\La$ the essential criterion for universal 
behavior is given by the size of 
$\bar{\la}_{2\La}/\bar{\la}_{1\La}$, with approximate 
universality for $\bar{\la}_{2\La} < \bar{\la}_{1\La}$.
For strong couplings universality extends to larger
$\bar{\la}_{2\La}/\bar{\la}_{1\La}$ and occurs for much larger
$m_R^c/\La$ (cf.\ table \ref{table1}).

(5) We have investigated how various characteristic 
quantities like  
the discontinuity in the order parameter
$\Dt \rho_{0}$ or the corresponding renormalized quantity
$\Dt \rho_{0R}$ or critical correlation lengths
depend on the classical parameters. 
In particular, at the critical temperature
one finds universal 
critical exponents for not too large $\bar{\la}_{2\La}$,
\bea
\Dt \rho_{0R} &\sim& (\bar{\la}_{2\La})^{\theta},
\qquad \theta=1.93\, ,\nnn
\Dt \rho_{0} &\sim& (\bar{\la}_{2\La})^{2\zeta},
\qquad \zeta=0.988 \, .
\eea 
These exponents are related by a scaling relation
to the critical 
correlation length and order parameter exponents
$\nu$ and $\beta$ of the $O(8)$ symmetric 
Heisenberg model according to 
$\theta/\zeta=\nu/\beta=1.95$ ($\nu=0.882$, 
$\beta=0.451$ in our calculation for 
$\bar{\la}_{2\La}=0$). Small values of $\bar{\la}_{2\La}$
can be associated with a perturbation of the 
$O(8)$ symmetric model and $\theta , \zeta$ are
related to the corresponding crossover exponents.
On the other hand, $\Dt \rho_{0R}$ ($\Dt \rho_{0}$)
becomes independent of $\bar{\la}_{2\La}$ in the 
infinite coupling limit. 

(6) We have computed the universal equation of state.
The equation of state relates the 
derivative of the free energy
$U$ to an external source, $\prl U/\prl \vp = j$.
From there one can extract universal ratios
e.g.\ for the jump in the order parameter
$(\Dt \rho_{0R}/m_R^c=0.592)$ or for the ratios of critical 
correlation lengths in the disordered (symmetric) and
ordered (spontaneously broken) phase
$(m_{0R}^c/m_{R}^c=0.746)$.
It specifies critical couplings
$(\la_{1R}/m_R^c=0.845,\la_{2R}/m_R^c=15.0)$. 
The universal behavior
of the potential for large field arguments 
$\rho_R \gg \rho_{0R}$ is
$U \sim \rho_R^3 \sim \rho^{3/(1+\eta)}$
provided $\rho_R$ is sufficiently 
small as compared to $\La$. Here the critical exponent
$\eta$ which characterizes the dependence of the potential
on the unrenormalized field $\rho$
is found to be $\eta=0.022$. For large $\rho$ the universal
equation of state equals the one for the $O(8)$ symmetric
Heisenberg model and $\eta$ specifies the anomalous
dimension or the critical exponent $\dt=(5-\eta)/(1+\eta)$.
The equation of state is computed for a nonzero coarse
graining scale $k$. It 
therefore contains information for quantities
like the 'classical' bubble surface tension in the
context of Langer's theory of bubble formation.
  
(7) We have investigated the dependence of the coarse
grained
effective potential $U_k(\rho)$ and the 'classical'
surface tension $\sigma_k$
on the coarse graining 
scale $k$ with special emphasis on the question
of the validity of Langer's 
theory of bubble formation. 
We find a strong scale
dependence of 
$U_k$ and $\sigma_k$
if the phase transition is characterized
by large dimensionless couplings.
A weak scale dependence is observed for small effective
couplings. 
There is
a close relation between the dependence of the coarse grained
free energy on the coarse graining scale and the reliability
of the saddle point approximation in Langer's theory
of bubble nucleation. A strong $k$-dependence of $\sigma_k$
is only compatible
with a large contribution from the higher orders of the
expansion.
We obtain a very consistent
picture: The validity of the saddle point
approximation typically requires small dimensionless couplings.
In this case also the details of the
coarse graining are not of crucial importance
within an appropriate range of $k$. The 
quantitative criterion for the validity of Langer's
formula is in our case $\la_{1R}/m_R^c \ll 1$.

(8) Our method is not restricted to the study of the universal
behavior. We can compute the effective potential for 
arbitrary values of the initial parameters and have done this
for particular examples.

The uncertainties of our results 
induced by the numerical integration
of the flow equations are well under control
and small. They
are negligible compared to
the expected error induced by our truncation.
For a significant improvement of our treatment one
would have to include higher order terms in the 
derivative expansion employed for the effective average
action. For weak first order or second order
phase transitions  we expect the error
to be related to the anomalous dimension
$\eta=0.022$. For the special case of the enhanced 
$O(8)$ symmetry one can compare e.g.\ with known values
for critical exponents obtained by other methods
\cite{ZJ,exp}. A comparison of our results 
for the critical exponents $\beta$ and $\nu$ with the
results of the most sophisticated calculations show
agreement within a few per cent. The anomalous dimension
is also well determined, even though it is most affected 
by our truncation.  

Finally, we should mention that our approach can be extended
in several directions. The generalization to complex
$N \times N$ matrices for arbitrary $N$ is straightforward.
For large $N$ this opens the possibility of a comparison with
$1/N$-expansions \cite{Fer,Nis}. 
Similarly, one may study systems with
symmetry $U(N,N)$ or $U(2N)$ and symmetry breaking
to the subgroup $U(N) \times U(N)$,
relevant for the study of the metal insulator transition
\cite{Weg}. 
Very interesting generalizations are the systems with
reduced $SU(N) \times SU(N)$ symmetry. They obtain by adding to
the classical potential a term involving the invariant
$\xi=\det \vp + \det \vp^{\dagger}$. ( Note that $\xi$
is not invariant with respect to $U(N) \times U(N)$).
This will give an even richer pattern of phase transitions
and permits a close contact to realistic meson models in
QCD where the axial anomaly is incorporated. Finally
one can extend the three dimensional treatment to a
four dimensional study of field theories at nonvanishing
temperature along the lines of ref.\ \cite{TetWet}.
We hope to gain in this way new information about
details of the chiral phase transition in QCD.

\newpage           

\appendix{Pole structure of the $l^d_n$ integrals \label{poles}}

The integrals
\bea
\ds{l^d_n(\om)} &=&
\ds{- n \int^{\infty}_0 dy y^{\frac{d}{2}+1} 
\frac{\partial r(y)}{\partial y}
\left[ y(1+r(y)) + \om \right]^{-(n+1)}}, 
\nnn 
\ds{\hat{l}^d_n(\om)}
&= &\ds{ \frac{n}{2} \int^{\infty}_0 dy y^{\frac{d}{2}} }
r(y)
\left[ y(1+r(y)) + \om \right]^{-(n+1)}
\label{threshldn} 
\eea
with
\be
r(y)=\ds{\frac{e^{-y}}{1-e^{-y}}}\quad , \quad
\ds{\frac{\partial r(y)}{\partial y}=
-\frac{e^{-y}}{(1-e^{-y})^2}}
\ee
exhibit for $d \le 2(n+1)$ a singularity at $\om=-1$.
The massless dimensionless average propagator
$y(1+r(y))$ is a monotonic function of $y$ that takes
on its minimum at $y=0$ with $\lim_{y \to 0}
y(1+r(y))=1$. We define new variables $\dt$ and $z$,
\be
\dt=\om + 1\quad,\quad z=y(1+r(y))-1
\label{deltaz}
\ee
and substitute in (\ref{threshldn}),
\bea
\ds{l^d_n(\dt)} &=&
\ds{\int^{\infty}_0 dz G^d_n(z) (z+\dt)^{-(n+1)}}, 
\nnn 
\ds{\hat{l}^d_n(\dt)}
&= &\ds{ \int^{\infty}_0 dz \hat{G}^d_n(z) (z+\dt)^{-(n+1)}}
\label{subst} 
\eea
with
\bea
G^d_n(z)&=&\ds{-\frac{n y^{\frac{d}{2}+1} 
\ds{\frac{\partial r(y)}{\partial y}}}{1+r(y)+
y\ds{\frac{\partial r(y)}{\partial y}}} },\nnn
\hat{G}^d_n(z)&=&\ds{\frac{n y^{\frac{d}{2}} 
r(y)}{2 \left(1+r(y)+
y\ds{\frac{\partial r(y)}{\partial y}}\right)} }
\eea
and $y=y(z)$. For $d < 2(n+1)$ the integrals (\ref{subst})
have a pole at $\dt = 0$. (The singularity becomes
logarithmic in $\dt$ for $d = 2(n+1)$). In this case
for $\dt \to 0$ the dominant contribution to the 
integral comes from the region $y \simeq 0$ or equivalently
$z \simeq 0$. To find an approximate expression for
$l^d_n$ and $\hat{l}^d_n$ near the pole we expand
the regular part of $G^d_n$ and $\hat{G}^d_n$ around
$z=0$. With
\bea
r(y)&=&\ds{\frac{1}{y}\left(1-\hal y+\frac{1}{12}y^2+O(y^4)
\right)},\nnn
\ds{\frac{\partial r(y)}{\partial y}}&=&\ds{-\frac{1}{y^2}
\left(1-\frac{1}{12} y^2 +O(y^4)\right)}
\eea
one obtains
\bea
G^d_n(z)&=&\ds{n y^{\frac{d}{2}-1}\left(
2-\frac{2}{3}y+\frac{1}{18}y^2+O(y^3)\right)},\nnn
\hat{G}^d_n(z)&=&\ds{\frac{n}{2} y^{\frac{d}{2}-1}
\left(2-\frac{5}{3}y+\frac{13}{18}y^2+O(y^3)\right)}.
\label{gy}
\eea
The inversion of $z(y)$ given by (\ref{deltaz})
can be done by expanding
\be
z=\ds{\hal y + \frac{1}{12} y^2 + O(y^4)}
\ee
and we find
\be
y=2 z - \frac{2}{3} z^2 + \frac{4}{9} z^3 + O(z^4).
\label{inv}
\ee
Insertion of (\ref{inv}) in (\ref{gy}) yields
\bea
G^d_n(z)&=&\ds{ 2 n (2 z)^{\frac{d}{2}-1}
\left(1-\frac{1}{3}\left(\frac{d}{2}+1\right)z
+ \left[\frac{1}{3}+\frac{1}{18}\left(\frac{d}{2}-1\right)
\left(\frac{d}{2}+6\right)  \right] z^2 +O(z^3)\right)},\nnn
\hat{G}^d_n(z)&=&\ds{ n (2 z)^{\frac{d}{2}-1}
\left(1-\frac{1}{3}\left(\frac{d}{2}+4\right)z
+ \left[2+\frac{1}{18}\left(\frac{d}{2}-1\right)
\left(\frac{d}{2}+12\right)  \right] z^2 +O(z^3)\right)}.
\label{gz}
\eea
We consider for $d=3$ and $n \ge 1$ the zeroth order
expression for $l^d_n$ and $\hat{l}^d_n$ that obtains from
the first term in (\ref{gz}) and the exchange of 
summation and integration in (\ref{subst}). Near the
pole one finds
\bea
l^3_n(\dt) &\simeq& \ds{2^{3/2} n \int^{\infty}_0 dz
z^{1/2}(z+\dt)^{-(n+1)}}\nnn
\hat{l}^3_n(\dt) &=& \ds{\hal l^3_n(\dt)}.
\eea
The leading contributions to $l^3_1$, $l^3_2$ and
$l^3_3$ are therefore given by
\bea
l^3_1(\dt)&=&2^{1/2} \pi \dt^{-1/2},\nnn
l^3_2(\dt)&=&2^{-1/2} \pi \dt^{-3/2},\nnn
l^3_3(\dt)&=&2^{-5/2} 3\pi \dt^{-5/2}.\nnn
\eea
We have verified this numerically.

\end{document}